\documentclass[preprint]{elsarticle}

\usepackage{hyperref}

\usepackage{hyperref}
\usepackage{amsmath} 
\usepackage{amssymb} 
\usepackage{tikz,pgfplots} 
\usepackage{caption} 
\usepackage{subcaption} 
\pgfplotsset{compat=newest} 
\usetikzlibrary{calc} 

\newlength\figureheight 
\newlength\figurewidth  

\usepackage{array} 
\usepackage{booktabs}
\usepgfplotslibrary{groupplots}

\usepackage[ruled,vlined]{algorithm2e}

\tikzset{
	font={\fontsize{9pt}{11}\selectfont}}
\usetikzlibrary{fit,positioning} 

\newtheorem{thm}{Theorem}

\newtheorem{mydef}{Definition}
\newdefinition{rmk}{Remark}
\newproof{pf}{Proof}

\newcommand{\real}{\mathbb{R}}
\newcommand{\model}{\mathcal{M}}
\newcommand{\data}{\mathcal{Y}}

\journal{arXiv}

\begin{document}

\begin{frontmatter}

\title{Hierarchical surrogate-based Approximate Bayesian Computation for an electric motor test bench}


\author[addressbosch,addressemcl]{David N. 
John\corref{mycorrespondingauthor}}
\cortext[mycorrespondingauthor]{Corresponding author}
\ead{david.john@de.bosch.com}
\author[addressmannheim]{Livia Stohrer}
\author[addressmannheim]{Claudia Schillings} 
\author[addressbosch]{Michael Schick}
\author[addressemcl]{Vincent Heuveline}

\address[addressbosch]{Bosch Research, Renningen, Germany}
\address[addressemcl]{Engineering Mathematics and Computing Lab (EMCL), Interdisciplinary Center for Scientific Computing (IWR), Heidelberg University, Germany}
\address[addressmannheim]{Mathematical Optimization Group, Institute of Mathematics, University of Mannheim, Germany}

\begin{abstract}
Inferring parameter distributions of complex industrial systems from noisy time series data requires methods to deal with the uncertainty of the underlying data and the used simulation model. Bayesian inference is well suited for these uncertain inverse problems. Standard methods used to identify uncertain parameters are Markov Chain Monte Carlo (MCMC) methods with explicit evaluation of a likelihood function. However, if the likelihood is very complex, such that its evaluation is computationally expensive, or even unknown in its explicit form,  Approximate Bayesian Computation (ABC) methods provide a promising alternative. 
In this work both methods are first applied to artificially generated data and second on a real world problem, by using data of an electric motor test bench. We show that both methods are able to infer the distribution of varying parameters with a Bayesian hierarchical approach. But the proposed ABC method is computationally much more efficient in order to achieve results with similar accuracy. We suggest to use summary statistics in order to reduce the dimension of the data which significantly increases the efficiency of the algorithm. Further the simulation model is replaced by a Polynomial Chaos Expansion (PCE) surrogate to speed up model evaluations. We proof consistency for the proposed surrogate-based ABC method with summary statistics under mild conditions on the (approximated) forward model.
\end{abstract}


\end{frontmatter}


\section{Introduction}\label{sec:intro}

In design processes, computer simulations are often used to virtually build a first set of prototypes  and to analyze their 
validity, efficiency or robustness. Typically mathematical models are used to 
describe complex processes in many applications, such as engineering, physical 
sciences, biology, finance and many others. Models however are only 
approximations of complex phenomena of the real world and thus, contain sources 
of uncertainty. This can be due to restrictive assumptions that have to be 
made, lack of knowledge, uncertain or even unknown parameter values or any kind 
of error arising from the use of numerical approximations or other 
simplifications. By quantifying these uncertainties, more refined predictions 
can be made and this is what Uncertainty Quantification (UQ) aims to achieve.

Measurement data can be used to improve the accuracy of the simulations by improving the knowledge of parameters in the mathematical models. This can be done by Bayesian inference which has its origin in 
the Bayes' theorem. The idea of Bayesian inference is to improve knowledge of 
uncertain parameters by incorporating all available information, namely 
information about the measurement process and prior knowledge of the uncertain 
parameters. The solution of this probabilistic approach to inverse 
problems is the so called posterior distribution which describes the updated 
knowledge of the uncertain parameters, see e.g. \cite{Stuart.2010,Dashti.2017,Kaipio.2005}.

Since the result of the Bayesian approach to inverse problems is a 
distribution rather than a single point estimate, one gains additional 
information about the uncertainty on the estimate in contrast to deterministic 
approaches. 
For complex models, the posterior distribution is usually not available in closed form solution and needs to be approximated. Markov 
Chain Monte Carlo (MCMC) methods which construct a Markov chain such that 
(under suitable conditions) its distribution converges to the posterior 
distribution, see e.g. \cite{Robert:2005:MCS:1051451, Kaipio.2005}, are often used in practice to compute samples from the posterior distribution. 
Another alternative is to approximate the posterior distribution using the  
Approximate Bayesian Computation (ABC) method, which is of advantage whenever the 
likelihood function is not known in explicit form or if its evaluation is too 
computationally expensive. The likelihood is then approximated based on the 
comparison of model simulations with the measurements \cite{Wilkinson}.
ABC  methods have been extensively used in population genetics and became known by 
one of the first publications in 1997 by Tavar{\'e} et al.\ \cite{Tavare505}. Besides 
the rejection algorithms, ABC methods have been 
further extended to "likelihood-free" MCMC by Marjoram et 
al.\ \cite{Marjoram15324}, sequential Monte Carlo by Sisson et al.\ \cite{Sisson1760}, probabilistic approximate rejection ABC by Wilkinson \cite{Wilkinson} and 
Gibbs Sampling by Wilkinson et al.\ \cite{10.1093/sysbio/syq054}. 
In ABC methods it is common practice to summarize the 
measurements by so-called summary statistics in order to improve the efficiency of such 
algorithms, see e.g. \cite{Najm2016,Prangle12}. However, the identification of proper summary statistics containing sufficient information for inference is a difficult task, see e.g. \cite{Cam} and \cite{Nunes,Prangle12,Barnes,Prangle2014}. For a comparative review of dimension reduction methods in ABC we refer to \cite{blum2013}.
A detailed introduction to ABC can be found in the book \cite{sisson2018handbook}.
A direct comparison between MCMC and ABC based on summary 
statistics is performed by Beaumont et al.\ \cite{Beaumont2025} in the context of 
population genetics.

In this work methods based on MCMC and ABC approaches are developed and applied to an electric motor in order 
to infer aleatoric parameters from noisy and varying measurement data.
These aleatoric parameter fluctuations are, for instance, due to manufacturing deviations and tolerances of motor components.
The variability and uncertainty of the aleatoric parameters is naturally modeled by random variables.
With hierarchical modelling the unknown random variables are then approximated by a parametric probability distribution. 
Note that the deterministic setting is a special case of this probabilistic setting where the distributions collapse to a Dirac distribution.
To validate and compare both methods 
they are first applied to artificially generated data and second to real world measurements from an electric motor test bench.
Since the underlying forward model is computationally intense, we propose a surrogate-based sampling strategy. The forward model w.r.t.~the uncertain parameters will be approximated by a polynomial chaos approach where the resulting high-dimensional integrals are approximated using sparse grids. Further, we introduce the Laplace approximation of the posterior distribution to efficiently construct proposals for MCMC. 

We suggest a novel approach to infer unknown parameter distributions from noisy observations in complex systems. The parameters are estimated in a hierarchical Bayesian setting.
In particular, we make the following main contributions: 
\begin{itemize}
\item We show that ABC is stable w.r.t.~approximations of the underlying forward model.
\item We suggest a very general methodology which leads to an immense speed-up of ABC by a combination of polynomial chaos approximations of the forward problem, sparse grid techniques and the use of summary statistics. The proposed methodology is application-neutral and can be applied to a wide range of problem classes.
\item We showcase the proposed methodology on an electric motor test case with artificial and real measurement data obtained from a test bench. The results of the surrogate-based ABC method are compared to an MCMC method using polynomial chaos approximations of the forward model and Laplace based preconditioners for the sampling. In particular, we demonstrate that the proposed surrogate-based ABC method leads to a performance comparable to MCMC in terms of accuracy while reducing the computational time. The speedup depends on the data size and increases with larger data.
\end{itemize}

This work is organized as follows: Section \ref{ch:application} introduces the electric motor model and  data. Further the considered mathematical problem is described in general. Section 
\ref{ch:methods} sets up the methodology and theoretical background needed 
for solving the problem and summarizes the proposed methodology.
Section \ref{ch:results} presents detailed
numerical results and Section~\ref{ch:conlusion} concludes this work.
\section{Application and problem description}\label{ch:application}

The methods considered in this work are applied to a direct current (DC) electric motor test bench. In particular we use measurements of current and angular velocity to infer system parameter distributions.
We describe the electric motor mathematically by a simplified and a more detailed model. The basic model consists of a system of linear ordinary differential equations (ODEs) which is introduced in Section~\ref{ssec:synthetic}. It is later used for synthetic data generation and detailed analysis of the considered methods.
Section~\ref{ssec:test bench} extends on the test bench and the detailed model. Later on the real world data from the test bench is used to test the considered methods on robustness.
Finally, Section~\ref{ssec:mathematical problem} sets the mathematical description of the parameter distribution identification problem.

\subsection{Basic electric motor model}
\label{ssec:synthetic}

\begin{figure}[htbp]
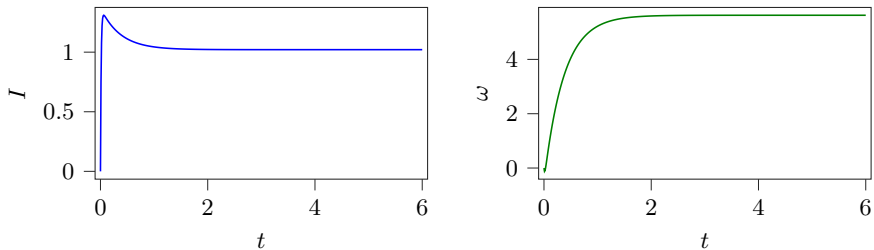

	\centering
	\begin{subfigure}{.5\textwidth}
		\centering
		\setlength\figureheight{0.2\textheight} 
		\setlength\figurewidth{0.99\textwidth}
		\input{Plots/synthetic/511_simulation_data_I.tex}
	\end{subfigure}%
	\begin{subfigure}{.5\textwidth}
		\centering
		\setlength\figureheight{0.2\textheight} 
		\setlength\figurewidth{0.99\textwidth}
		\input{Plots/synthetic/511_simulation_data_omega.tex}
	\end{subfigure}
	\caption[DC electric motor]{Numerical approximation of current $I$ and angular velocity $\omega$ of the basic electric motor model \eqref{eq:ODE}.}
	\label{fig:synth_simulation}
\end{figure}

An electric motor is a machine that converts electrical to mechanical energy and a DC motor runs on direct current instead of alternating current (AC). To construct a model describing a DC motor two equations are combined, one representing the electrical and one the mechanical side. The electrical part of a DC motor can be explained by the armature circuit. It is determined by an applied voltage $V$ that has to be compensated by a resistance $R$ ($V=I R$, where $I$ is the armature current\footnote{This relationship is called Ohm's law and was proven in 1826 by Ohm \cite{Ohm.1826}.}), a coil ($V=I' L$, where $L$ is the armature circuit inductance) and the internal voltage generated by the motor ($E_a=c_m \omega$, where $c_m$ is a motor constant and $\omega$ the rotational speed) \cite{Toliyat.2004}. Combining all parts of the armature loop results in the following equation
 \begin{equation}
 V = R I(t) + L I'(t) + c_m \omega(t).
\label{eq:electrical}
\end{equation}
The mechanical part of the DC motor is determined by its torque $T_m$. The torque is compensated by inertia of the motor $J$ which arises only when rotational speed changes,
by its friction $D$ which has a direct influence 
and by a (constant) motor load $T$. Altogether, it leads to the relationship
\begin{equation}
T_m = J \dot{\omega}(t) + D\omega(t) + T,
\label{eq:mechanical}
\end{equation}
where $T_m$ can be expressed by $c_g I(t)$ with another motor constant $c_g$ \cite{Toliyat.2004}.
Combining Equation \eqref{eq:electrical} and \eqref{eq:mechanical} the electric motor model is given by
\begin{align}
	\dot{I}(t) &= \frac{1}{L} \left( -R I(t) - c_m \omega(t) + V \right),\\
	\dot{\omega}(t) &= \frac{1}{J} \left( c_g I(t) - D \omega(t) - T \right).
\label{eq:ODE}
\end{align}
This linear ODE system with given constant coefficients can in principle be solved analytically. However, this is generally not the case for most problems. Hence, to be more general we use a numerical method to approximate the solution. In particular the ODE system is solved by an explicit Runge-Kutta Method of order 4 with $N_t=601$ equidistant time steps in the time interval $[0,6]$. We refer to this numerical approximation as simulation model $\model$. 
An approximation for the current and rotational speed for a motor starting from rest, i.e. with initial conditions $I(0)=0,\ \omega(0)=0$, and with a fixed set of parameters ($R=9 , L=0.11, c_m=0.5, c_g=3, D=0.1, J=0.1$ and $V=12$, $T=2.5$) is displayed in Figure~\ref{fig:synth_simulation}.

\subsection{Electric motor test bench and model}
\label{ssec:test bench}
In this section we first describe the electric motor test bench and how measurements are obtained. Second we describe the simulation model approximating the test bench. Both steps are based on previous work by Glaser et.al. \cite{Glaser.2016} and Glaser~\cite{Glaser.diss}.

\begin{figure}[htp]
	\centering
		\begin{subfigure}{.45\textwidth}
		\centering
		\includegraphics[width=\textwidth]{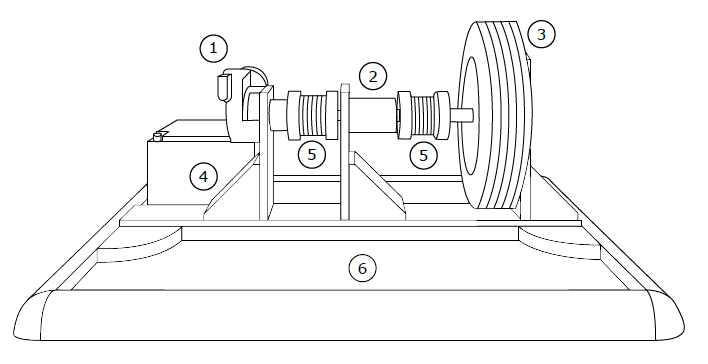}
	\end{subfigure}%
	\begin{subfigure}{.45\textwidth}
		\centering
		\includegraphics[width=0.9\textwidth]{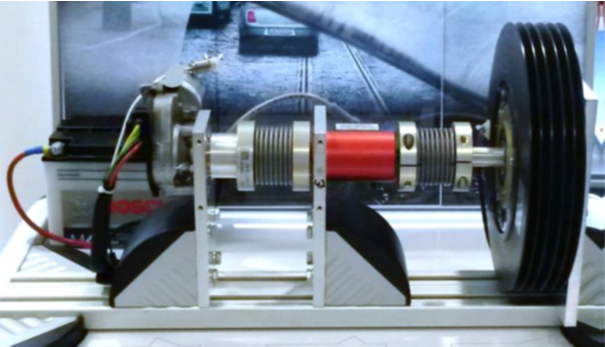}
	\end{subfigure}
	\caption[Schematic diagram and picture of the test bench hardware]{Schematic diagram and picture of the test bench hardware \cite{Glaser.2016}.} 
	\label{fig:test_bench}
\end{figure}

The test bench is based on a windshield wiper electric motor with an attached break to mimic mechanical loading.
Figure~\ref{fig:test_bench} represents a schematic diagram \cite{Glaser.2016}. The mounting \textcircled{6} builds the base of the test bench. Starting from left to right a power supply \textcircled{4} is needed to drive the engine. Then there is the electric drive \textcircled{1} composed of the motor and a worm gear. Metal couplings \textcircled{5} connect the shaft components and a torque sensor \textcircled{2} is mounted on the motor shaft. Besides the torque sensor additional sensors are connected with the electric drive to measure thermal characteristics. Furthermore, measurements of rotational speed can be obtained. On the right side of the shaft there is an electromagnetic powder brake \textcircled{3} that is needed for correctly dealing with a load by minimizing the back-drive ability. The brake opens the possibility to run the motor in different modes. Thereby, different measurements corresponding to the different operating modes of the test bench can be obtained to analyze motor characteristics.

Due to tolerances, material uncertainties and different suppliers, some components of electric motors have varying properties when coming from the production line. This yields varying characteristics of the motor. The uncertainty in the model is parametrized by the uncertain voltage $V$ and load torque $T$. Both quantities are the main factors on current and speed and therefore allow to capture the dominant source of uncertainty in both quantities of interest.
The special nature of the electric drive test bench is that some parameters can be varied on a single set-up in order to mimic some of those variations. This can be done in an automatic way and without replacing components of the test bench.
In particular the test bench easily allows to vary the voltage $V$ and the load torque $T$. To mimic aleatoric parameters we define reference distributions $\pi(V)$ and $\pi(T)$, draw samples from those distributions and use them as input for the test bench. In reality the load might vary largely, e.g. for a windshield wiper due to effects of wind, rain, snow, ice and dirt. And the voltage might vary due to different batteries, the battery age and usage and due to the consumption of other devices. For safety reasons it is important that the electric motor works reliable in all scenarios.

\begin{figure}[htbp]
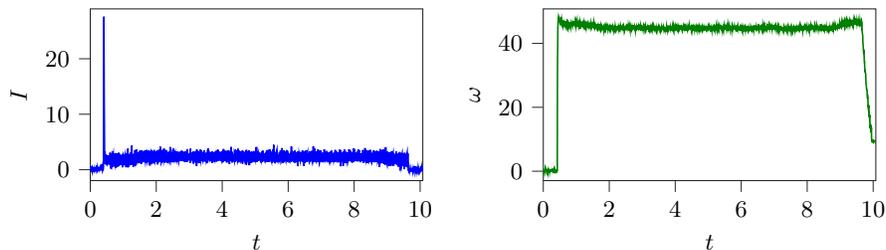

	\centering
	\begin{subfigure}{.5\textwidth}
		\centering
		\setlength\figureheight{0.2\textheight} 
		\setlength\figurewidth{0.99\textwidth}
		\input{Plots/testbench/401_raw_data_I.tex}
	\end{subfigure}%
	\begin{subfigure}{.5\textwidth}
		\centering
		\setlength\figureheight{0.2\textheight} 
		\setlength\figurewidth{0.99\textwidth}
		\input{Plots/testbench/401_raw_data_omega.tex}
	\end{subfigure}
	\caption[Test bench raw data]{Unprocessed test bench measurement data of current $I$ and angular velocity $\omega$.}
	\label{fig:raw_data}
\end{figure}

For a given sample of $V$ and $T$ we start the test bench and record measurement signals of current and angular velocity with the frequency 10 kHz for 10 seconds. 
Figure~\ref{fig:raw_data} displays these signals.
Now, based on the reference distributions $\pi(V)$ and $\pi(T)$ we draw 200 samples $(V_i, {T}_i), i=1,\dots,200$ and sequentially run the test bench on each of these samples to obtain a set of 200 measurements. From this set we discard the first 100 measurements in order to account for a warm-up phase of the electric motor. The remaining 100 measurements are now within a similar temperature range.
Then the test bench measurement recordings are preprocessed by aligning the starting point (discard the first few observations where nothing is happening), cut-off after 6 seconds, apply a smoothing filter 
to remove unwanted noise from test bench control devices 
(a second order Butterworth low pass filter \cite{butterworth1930theory,tuzlukov2018signal})
and sample down to 100 Hz. 
For details we refer to \cite[Ch. 2]{Glaser.diss} and \cite[Ch. 2]{John.diss}.

The resulting test bench measurement data set
\begin{align}
\label{eq:data}
\data := \{ y^{I}_i, y^{\omega}_i, i=1,\dots,N \}
\end{align}
contains $N=100$ noisy measurement series of current $I$ and rotational speed $\omega$. For a fixed $i$ the discrete observations 
$y^{I} :=[y^{I}(t_1), \dots, y^{I}(t_{N_t})] \in \mathbb{R}^{N_t}$ 
and 
$y^{\omega} :=[y^{\omega}(t_1), \dots, y^{\omega}(t_{N_t})] \in \mathbb{R}^{N_t}$ 
are each of size $N_t = 601$ in the time interval $[0,6]$ seconds with equidistant time points $(t_1, \dots, t_{N_t}) \in [0,6]$.
Overall, $\data$ contains $2\times N \times N_t$ data points.
Later, in Section~\ref{ch:results} an overview of $\data$ is given in Figure~\ref{fig:testbench_meas}.

To make things clear: In reality the distributions $\pi(V)$ and $\pi(T)$ might not be known. Also we might be interested in other distributions of the electric motor parameters. 
However, this test bench allows to test inference methods on real world data and additionally validate the results on reference distributions and even on reference samples.

The simulation model describing the test bench differs to the previous basic model~\eqref{eq:ODE} by modeling the worm gear and additionally considering a detailed thermal model interconnecting with the electrical and mechanical part. This leads to the introduction of additional parameters describing thermal characteristics and an overall very detailed and complex model. In the following we only give a rough overview on the model, for details see \cite{Glaser.2016,Glaser.diss}. The model equations are 
\begin{align}
	\dot{I}(t) &= 
	\frac{1}{L} \left( -R I(t) - c_m \omega(t) + (V_t(t)-V_{drop}) \right),\\
	\dot{\omega}(t) &= 
	\frac{1}{J} \left( (c_g I(t) - \tau_{loss} - \tau_{fric})\eta - D \omega(t) - {T}_t(t) i_g \right),
	\label{eq:ODE_tb}
\end{align}
where:
the resistance $R$ depends linear on the temperature of the coil;
the motor constant $c_m$ depends nonlinear on temperature of the magnet and on the current $I(t)$; 
$c_g \equiv c_m$;
$V_{drop}$ depends nonlinear on the current $I(t)$;
the gear meshing efficiency $\eta$ of the worm gear is a nonlinear function of the lead angle of the worm, the pressure angle and the friction of the worm (depending nonlinear on the worm temperature);
$\tau_{loss}$ summarizes the hysteresis loss and the eddy current loss, which both depend nonlinear on current $I(t)$ and temperature of the magnet;
$\tau_{fric} = \tau_{fric,air} + \tau_{fric,motor}$ summarizes the air friction loss $\tau_{fric,air}$ that depends on the angular velocity $\omega(t)$ and the friction loss of the motor $\tau_{fric,motor}$;
$\tau_{fric,motor}$ is a sum of the friction losses at the bearings, depending on the bearing temperatures and $\eta$ and the friction loss at the commutator depending on the temperature of the commutator;
$i_g$ is a constant proportional to the worm gear ratio.
For details on the thermal model equations we refer to \cite{Glaser.diss}.
Voltage $V_t(t)$ and load ${T}_t(t)$ for the test bench are considered as time depended, as they are delayed in the hardware until they reach stationary values $V$ and $T$.     

The model parameters are either known from expert knowledge, given by look-up tables or calibrated based on one reference measurement with known realization of $V$ and $T$ (solve a nonlinear optimization problem). An assumption for the model calibration is that the temperatures of the test bench motor are already in a stationary level. For this reason we discarded the first 100 measurements above. Albeit the model is very detailed it still does not describe the reality perfectly and parameter calibration is non-trivial. 

The test bench model is solved by an adaptive numerical integration scheme in the time interval $[0,6]$. The numerical approximation is then linearly interpolated to $N_t=601$ equidistant time steps, leading to similar outputs as in the basic model case above. 
As it will be clear from context which model is used, we overload notation and also refer to this numerical approximation as simulation model $\model$.

\subsection{Mathematical problem description}
\label{ssec:mathematical problem}

At the end of each of the two previous sections we denoted the numerical approximation of the two electric motor ODE systems as simulation model $\model$. 
To be more specific, for given deterministic model parameters $x \in \mathcal{X} \subseteq \mathbb{R}^n$, the simulation model 
\begin{align}
\model:\mathcal{X} \to \mathbb{R}^{k\times N_t}
\end{align}
is the operator that numerically approximates the solution of an ODE system of order $k \in \mathbb{N}$ and returns an approximation of the $k = 2$ states (current $I$ and angular velocity $\omega$) at $N_t$ discrete time points $(t_1, \dots, t_{N_t})$. 

Now, recall the definition and notation of the test bench measurement data $\data$ in \eqref{eq:data}.
The measurements in $\data$ have some variations that cannot be explained by measurement noise alone. Those variations are due to changing system parameters, so-called aleatoric parameters, i.e.~instead of taking a fixed single value they can fluctuate. Consequently, we need to reflect these variations in the simulation model parameters $x$ as well, by modeling them as random variables. 
Let $X_i, \,i=1,...,n$ be the aleatoric model parameters and $X=(X_1,...,X_n)$ the vector of all these random variables on an underlying probability space $(\Omega, \mathcal{A}, \mathbb{P})$ with $\Omega$ the underlying sample space, $\mathcal{A}$ the sigma algebra and $\mathbb{P}:\mathcal{A} \to [0,1]$ the probability measure, such that
$X:\Omega \to \mathcal{X}$.
A realization of $X$ is denoted by $x :=  X(\omega)=\left(X_1(\omega),...,X_n(\omega)\right)\in \mathcal{X}$ for $\omega \in \Omega$. 
With the random vector $X$ the simulation model $\model(X)$ becomes random as well with 
\begin{align}
\model(X):\Omega \to \mathbb{R}^{k\times N_t}.
\end{align} 
If $\model$ would be continuous in time, then $\model(X)$ would be a $k$ dimensional stochastic process.

Now, given the measurement data $\data$ and the simulation model $\model(X)$ the goal is to find the unknown probability distribution $\pi(X)$ of the aleatoric parameters $X$ such that the distribution of $\model(X)$ equals the distribution of the true underlying process, which is only known by the sample data $\data$. The problem to solve is an inverse problem. However, the parameters to infer are of stochastic nature and are not deterministic. In the next section we present methods in order to deal with this problem.

\section{Methods}\label{ch:methods}

We summarize in the following the main ingredients of the proposed algorithm:
starting with the Bayesian approach to inverse problems, followed by the  methods and approximations to solve the resulting problem.

\subsection{Bayesian inference}
In an abstract way, we can formulate the inverse problem, which consists of inferring the unknown parameters from noisy measurements $\data \in \mathbb{R}^{k \times N_t}$, as follows. The goal of computation is to find the unknown parameters $x^\dagger$ from 
\begin{equation}
\data  = \model(x^\dagger) + e\,,
\label{eq:RModel}
\end{equation}
where $e$ models the measurement and model error and $\model:\mathcal X \to \mathbb{R}^{k\times N_t}$ is the uncertainty-to-observation map, which consists of the solution operator of the underlying forward model and the observation operator. To quantify the uncertainty in the unknown parameters $x$, we adopt a Bayesian approach to inverse problems. We refer the reader to \cite{Kaipio.2005,Stuart.2010, Dashti.2017} for more details on Bayesian inverse problems. We model the unknown parameters $x\in\mathbb R^n$ as random variables, characterized according to a given prior distribution $\mu_0$ and assume that the measurement and model error $E$ is independent of $x$ and (for simplicity) normally distributed, i.e. $E\sim\mu_e=\mathcal N(0,\Gamma)$ with $\Gamma\in\mathbb R^{kN_t \times kN_t}$ symmetric, positive definite. The solution of the Bayesian inverse problem is then the posterior distribution, 
which can be characterized via Bayes' formula. 
\begin{thm}\cite{Stuart.2010}
Assume that $\Phi:\mathcal X \times \mathbb R^{k\times N_t}$ with $\Phi(x;\data)=\frac12\|\data-\model(x)\|_\Gamma^2$ is measurable w.r.t.~the product measure $\nu_0(\mathrm dx,\mathrm d\data)=\mu_0(\mathrm dx) \mu_e(\mathrm d\data)$ and that
\begin{align}
Z=\int_{\mathbb R^n} \exp(-\Phi(x;\data)) \mu_0(\mathrm dx)>0
\end{align}
for $\data \ \mu_e$-a.s..
Then, the conditional distribution $\mu^\data$ of $x|\data$ exists, is absolutely continuous w.r.t.\ $\mu_0$ and the Radon-Nikodym derivative is given by
\begin{align}
\mu^\data(\mathrm dx)=\frac1Z  \exp(-\Phi(x;\data)) \mu_0(\mathrm dx)
\end{align}
for $\data \ \mu_e$-a.s..
\end{thm}
In the finite dimensional setting it can be formulated as
\begin{align}
\pi(x\mid \data)=\frac{1}{Z}\pi_e\left(\data-\model(x)\right)\pi_{0}(x)\,,
\end{align}
where $\pi_0$ and $\pi_e$ denote the corresponding Lebesgue densities of $\mu_0$ and $\mu_e$. $\pi(x\mid \data)$ is denoted as posterior distribution of $x$ given $\data$. 

Due to the solution operator of the underlying forward map involved in the characterization of the posterior distribution, we usually rely on sampling methods to explore the posterior distribution. This is usually done by Markov Chain Monte Carlo (MCMC) methods, which generate a Markov chain with the posterior as limit distribution. As these methods are often seen as the gold standard for Bayesian inverse problems, we will compare our proposed algorithm to the Metropolis Hastings MCMC method. We refer the reader to  \cite{Robert:2005:MCS:1051451, Dashti.2017} and the references therein for more details on MCMC methods.

\subsection{Approximate Bayesian Computation and summary statistics}
\label{ssc:abc}
Approximate Bayesian Computation (ABC) methods aim to calculate an ap\-prox\-i\-ma\-tion $\hat{\pi}(x\mid \data)$ of the true posterior distribution $\pi(x\mid \data)$. The Approximate Rejection algorithm \cite{Wilkinson} is an intuitive member of this family. It is based on rejection sampling with a distance measure between simulated values and observed data $d(\cdot,\cdot): \mathbb{R}^{k\times N_t} \times \mathbb{R}^{k\times N_t} \rightarrow \mathbb{R}$ and an acceptance threshold $\delta \geq 0$.
The accepted prior samples are independent and identically distributed samples of $\hat{\pi}\left(x\mid \data,d(\data^{sim},\data)\le\delta\right)$.
Two key challenges of ABC methods are finding an appropriate $d(\cdot,\cdot)$ as well as a suitable value for $\delta$. 
With $\delta \rightarrow 0$ the accuracy increases, but simultaneously the overall acceptance rate decreases, leading to increased computational effort. Consequently, $\delta$ has to be chosen as trade-off between computational capacity and accuracy.
In practice, a
$\delta>0$ is necessary, since the probability that $\data^{sim}=\data$ is in most non-trivial models very low or even impossible due to model misspecification or measurement noise \cite{Sunnaker}.

One possibility to increase the acceptance rate and thus the efficiency is to use summary statistics $S(\cdot)$. This summary can be a set of statistics, i.e. $S(\cdot)=(S_1(\cdot),...,S_d(\cdot))$, $d \in \mathbb{N}$, where a $S_i(\cdot)$ can be any  statistical measure of the data.
The selection of statistical functions needs to be done carefully, such that $S(\data)$ is a sufficient representation of $\data$.
\begin{algorithm}[h]
	1. Sample $x$ from $\pi_0(\cdot).$ \\
	2. Sample $e$ from $\pi_e(\cdot)$ and set $\data^{sim}:= \model(x) +e$. Calculate $S(\data^{sim})$. \\
	3. Accept $x$ if $d\left(S(\data^{sim}),S(\data)\right)\le\delta$; return to 1.\\	
	\caption{Approximate Rejection with Summary Statistics \cite{barber2015}}
	\label{alg:rejection_sum}
\end{algorithm}
Algorithm~\ref{alg:rejection_sum} generates samples of an approximation of the posterior distribution 
$\pi(x\mid S(\data))$. The acceptance ratio is proportional to the probability 
that $S(\data^{sim})$ exactly fits $S(\data)$ and this is usually more probable than matching the whole 
data exactly.
Summary statistics are used in many 
practical applications, although it is hard to tell if the summaries are 
sufficient. See e.g. \cite{Najm2016} for more details.

\subsection{Sequential Monte Carlo method}
Sequential Monte Carlo (SMC) methods iteratively sample from a sequence of intermediate distributions, starting at the prior distribution and becoming increasingly similar to the true posterior $\pi(x\mid \data)$.
Algorithm~\ref{alg:abc_smc} is a version of such a SMC (also known as Population Monte Carlo (PMC)) method.
\begin{algorithm}[h]
	\textbf{Set iteration number} $p=1$\\
	1. Sample $x_i^{(1)}$ for $i=1,...,M$ from $\pi_0(\cdot).$ \\
	2. Sample $e_i$ from $\pi_e(\cdot)$ and set $\data_i^{sim}:= \model(x_i^{(1)}) +e_i$ for $i=1,...,M$. Calculate the summary $S(\data_i^{sim})$. \\
	3. Accept $x_i^{(1)}$ if $d\left(S(\data_i^{sim}),S(\data)\right)\le\delta^{(1)}$ and set $w_i^{(1)}=1/m$, where $m\leq M$ is the number of accepted samples.\\
	4. Calculate $\Sigma^{(1)}=2\cdot \text{Var}\left(\{x_i^{(1)}|i=1,...,m \}\right)$. \\
	\textbf{For iteration number} $p=2,...,P$ \textbf{do}\\
	5.1. Sample $x_i^{*}$ from $\{x_j^{(p-1)}|j=1,...,m \}$ with probabilities $w_j^{(p-1)}$ for $i=1,...,M$.\\
	5.2. Sample $x_i^{(p)}$ from $\mathcal{N}(x_i^{*},\Sigma^{(p-1)})$ for $i=1,...,M$.\\
	5.3. Sample $e_i$ from $\pi_e(\cdot)$ and set $\data_i^{sim}:= \model(x_i^{(p)}) +e_i$ for $i=1,...,M$. Calculate the summary $S(\data_i^{sim})$. \\
	5.4. Accept $x_i^{(p)}$ if $d\left(S(\data_i^{sim}),S(\data)\right)\le\delta^{(p)}$ and \\ 
	set $w_i^{(p)}\propto \pi_0(x_i^{(p)})/\sum_{j=1}^m w_j^{(p-1)}\phi\{(x_i^{(p)}-x_j^{(p-1)})/\sqrt{\Sigma^{(p-1)}}\}$.\\ 
	5.5. Calculate $\Sigma^{(p)}=2\cdot \text{Var}\left(\{x_i^{(p)}|i=1,...,m \}\right)$.	
	\caption{Sequential Monte Carlo ABC \cite{Beaumont2009,Marin2012}}
	\label{alg:abc_smc}
\end{algorithm}
In each iteration the acceptance rate increases, allowing for decreasing thresholds $\delta^{(p)}$ in iterations $p\leq P$ \cite{Sisson1760,Lintusaari2016}.
In Step~5.4.\ the previous samples $x_i^{(p-1)}$ are used to define a mixed density of normal distributions weighted by $w_j^{(p-1)}$ known from importance sampling. $\phi$ is most often a standardized Gaussian or a Student's $t$ density \cite{Beaumont2009}.
Consequently the mixed density approaches the true posterior distribution allowing to generate approximated samples from it directly.

\subsection{Polynomial chaos expansion of the forward map
\label{ssec:pce}}
Sampling from the posterior requires numerous evaluations of the simulation model $\model$. If $\model$ is computationally expensive to evaluate one might consider replacing it by a faster to evaluate surrogate. We choose a Polynomial Chaos expansion (PCE) to approximate $\model$ by orthogonal polynomials. 
Based on Wiener \cite{wiener1938} and Cameron and Martin \cite{cameron_martin}, Xiu and Karniadakis \cite{xiu_gpx} introduced the generalized Polynomial Chaos (gPC). Here a square-integrable random variable $X$ can be represented via 
\begin{equation}
X = \sum_{i=0}^\infty x_i\psi_i(\xi),
\end{equation}
with coefficients $x_i$, multi-dimensioanl orthogonal polynomials $\psi_i$ and random variables $\{\xi_i\}_{i=1}^\infty$.
For computational feasibility, the PCE must be truncated at some finite order $p$, which leaves $P + 1 = (p + d)!/(p! d!)$ terms in the expansion ($d$ denotes the dimension of $\xi$).
We approximate the coefficients $\{x_i\}_{i=0}^P$ via a Smolyak sparse pseudo-spectral projection method \cite{Smolyak,Constantine} using Sparse Grid numerical 
integration rules.
Here the level $L  \in \mathbb{N}$, a growth rule defining the maximum polynomial degree in each dimension $g(L) = 2 L +1$, and a truncation scheme (total order) determine the basis functionals and the Sparse Grid nodes. The accuracy of the surrogate increases with $L$, but also the associated cost as $\model$ needs to be evaluated at an increasing number of nodes.

\subsection{Stability analysis of surrogate models with Approximate Bayesian Computation}
The (truncated) polynomial chaos expansion leads to an approximation of the forward map, thus, to an approximation of the posterior measure. To ensure the convergence to the true solution of the Bayesian inverse problem, we shortly discuss a stability analysis in the following.
Assume that the goal of computation is to compute the mean of a quantity of interest $\varphi$ w.r.t.~the posterior $\mu^\data$ with Lebesgue density $\pi(x|\data)$. We denote by $\tilde \mu^\data_L$ the approximation of the posterior using the polynomial chaos surrogate model for the forward map and by $\tilde \pi_L(x|\data)$ the corresponding Lebesgue density. The subscript indicates the number of terms in the polynomial chaos expansion. Given the summary statistic $S:\mathbb R^{k \times N_t} \to \mathbb R^d$ with $d\le k N_t$, the inference using Algorithm~\ref{alg:rejection_sum} is based on $s^\dagger=S(\data)$ and constructs samples $\tilde x_{j}^{(\delta)}$ based on the polynomial chaos surrogate. Assuming  $S$ to be a sufficient statistic, i.e.
\begin{align}
\mathbb E^{\tilde \mu^\data}[\varphi(x)|\data]=\mathbb E^{\tilde \mu^\data}[\varphi(x)|s^\dagger]
\end{align}
for the surrogate-based model, we are interested in the mean-square-error
\begin{align}
\mathbb E^{\mu_0}\left[\left(\mathbb E^{\mu^\data}[\varphi(x)|\data]-\frac{1}{m}\sum_{j=1}^m \varphi(\tilde x_{j}^{(\delta)})\right)^2\right]\,.
\end{align}
We denote in the following the approximation of the least-squares potential $\Phi$ based on the truncated polynomial chaos expansion by $\Phi^L$, where $L$ is the truncation parameter of the expansion.
\begin{thm}\label{thm:consistency}
Let the quantity of interest $\varphi:\mathbb R^n\to \mathbb R$ with $\varphi\in L^1_{\mu_0}$. Assume that $\Phi\in C(\mathbb R^n;\mathbb R)$ and there are functions $M_i:\mathbb R_+\to\mathbb R_+,\ i=1,2$ independent of $L$ and monotonic non-decreasing separately in each argument, and with $M_2$ strictly positive, such that for all $x\in\mathbb R^n$
\begin{align}
\Phi(x) &\ge -M_1(\|x\|)\\
\Phi^L(x) &\ge -M_1(\|x\|)\\
\|\Phi(x)-\Phi^L(x)\| &\le M_2(\|x\|)\psi(L)\,,
\end{align}
where $\psi(L)\to 0$ as $L\to\infty$. Further assume that $\mu_0(\mathbb R^n\cap B)>0$ for some bounded set $B\subset \mathbb R^n$ and 
\begin{align}
\exp(M_1(\|x\|))(1+M_2(\|x\|)^2)\in L^1_{\mu_0}\,.
\end{align} 
In addition let $\varphi\in L^2_{\mu^\data}$ and $\varphi\in L^2_{\tilde \mu^\data_L}$, uniformly in $L$. 
Then, the mean square error converges to $0$ for $L,m\to \infty$ and $\delta\to 0$, i.e.
\begin{align}
\mathbb E^{\mu_0}\left[\left(\mathbb E^{\mu^\data}[\varphi(x)|\data]-\frac{1}{m}\sum_{j=1}^m \varphi(\tilde x_{j}^{(\delta)})\right)^2\right] \to 0\,,\quad m,L\to \infty, \ \delta\to 0\,.
\end{align}
\end{thm}
\begin{pf}
The triangle inequality gives the separate estimation of the individual errors
\begin{align}
&\mathbb E^{\mu_0}\left[\left(\mathbb E^{\mu^\data}[\varphi(x)|\data]-\frac{1}{m}\sum_{j=1}^m \varphi(\tilde x_{j}^{(\delta)})\right)^2\right]^{1/2}\\
&\le \left| E^{\mu^\data}[\varphi(x)|\data]-\mathbb E^{\tilde \mu^\data_L}[\varphi(x)|\data]\right|
+ \mathbb E^{\mu_0}\left[\left(\mathbb E^{\tilde \mu^\data_L}[\varphi(x)|\data]-\frac{1}{m}\sum_{j=1}^m \varphi(\tilde x_{j}^{(\delta)})\right)^2\right]^{1/2}\nonumber \\
&=: I_1+I_2. \nonumber
\end{align}
The assumptions on the least-squares potential ensure the closeness of the two measures $\mu^\data$ and $\tilde \mu^\data_L$ in the Hellinger distance, see \cite[Theorem 4.9]{Dashti.2017}, which implies
\begin{align}
I_1\le C\psi(L)\,.
\end{align}
The second summand $I_2$ corresponds to the ABC error, which converges to $0$ as $m\to \infty$ and $\delta\to 0$, see \cite[Proposition 3.1]{barber2015}.
\end{pf}

\subsection{Improvement with maximum a posteriori estimator}
\label{ssec:map}
For many real-world applications, the approximation of the whole posterior distribution is computationally infeasible. Point estimators or approximations via simpler distributions such as Gaussian distributions are very common in practice to reduce the overall computational effort. We will focus here on the maximum a posteriori (MAP) estimator, which is defined as the realization of the unknown $X$ with highest posterior density given prior assumption and given observed data $\data$,
\begin{equation}
x_{\text{MAP}}=\arg \max_{x \in \mathcal{X}}\pi(x\mid \data).
\label{eq:MAP}
\end{equation}
Note that there is the possibility that the MAP estimate is not unique or does not exist. For the definition of the MAP in the infinite dimensional setting, we refer to \cite{Dashti2013, Helin_2015}. 
In case of large data or informative data, the posterior distribution often shows a concentrated behavior in a small region of the parameter domain. In this setting, the posterior can be well represented by the Laplace approximation, a Gaussian distribution with mean $x_{\text{MAP}}$ and covariance $C=H^{-1}(x_{\text{MAP}})$, where $H(x_{\text{MAP}})$ denotes the Hessian of the log-posterior density at the MAP. We refer to \cite{schillings2019convergence} for more details. 

The Laplace approximation can be further used to accelerate sampling by increasing sampler efficiency, see e.g. \cite{refId0,schillings2019convergence,Rudolf2018}. 
To do so the estimation of the  MAP $x_{\text{MAP}}$ can be used as starting point for the sampler, this clearly reduces the burn-in phase of the 
algorithm. And in case of a Metropolis Hastings MCMC with Gaussian proposal distribution, $C$ can be used as initialization of the proposal distribution covariance.
I.e. replace the proposal distribution $\mathcal{N}(x_j,s^2\mathbf{I})$ with
$\mathcal{N}\left(x_j,s^2 C)\right)$,
where $x_j$ denotes 
the current state of all unknown variables that shall be inferred, 
$\mathbf{I}\in\mathbb{R}^{n\times n}$ the identity matrix and $s$ the step size 
parameter.
Using a Quasi-Newton method for optimization yields the MAP and additionally an approximation to the inverse Hessian at the MAP. By using the Symmetric-Rank-1 (SR1) Hessian update strategy theoretic results for convergence to the true Hessian are available, see e.g. \cite{refId0} and the references therein. 

In general Metropolis Hastings MCMC with Gaussian random walk proposal is not dimension-independent. We refer to \cite{Sprungk.2018} and \cite{Hu2017adaptive_pCN} for further details on methods to increase the performance of the algorithm in high dimensional spaces.

\subsection{Hierarchical estimation}
The need for hierarchical models in Bayesian statistics and in particular in the context of non-parametric methods in machine learning, is well established \cite{Bishop.2006}.
For state of the art in hierarchical Bayesian estimation we refer to \cite{robert2007bayesian} and for hierarchical Bayesian inverse problems to \cite{sraj2016coordinate,dunlop2017hierarchical,roininen2019hyperpriors,latz2019fast} and the references therein. Those methods deal with inference of continuous-parameter random fields both for priors and hyper-priors.
As this is an active research field in its own, we assume in the following parametric distributions for the unknown aleatoric parameters and thus constrain, i.e.\ discretize the function space. 
In particular, we assume Gaussian distributions and inference is then based on the hyper-parameters mean and standard deviation, following \cite{Glaser.2016, glaser2017modeling, Glaser.diss}.
This corresponds to the traditional way of dimensionality reduction in forward and inverse stochastic problems by using the truncated Karhunen-Lo\`{e}ve expansion (KLE) \cite{MARZOUK2009,lemaitre2010spectral,sraj2016coordinate}, 
with only two coefficients.
For theoretical error analysis, i.e.\ whether the reduced problem can well approximate the original one, we refer to \cite{LI2015KL}. 
Further several works on theoretical error analysis in the forward problem are mentioned. The effects of the truncated KLE on the Bayesian inverse problem solution is investigated in \cite{URIBE2020}.

For a summary of some hierarchical ABC approaches in the context of psychological models we refer to \cite{turner2014hierarchical} and the references therein. In case of high dimensional hierarchical problems, their introduced algorithm Gibbs ABC improves rejection rate for hierarchical ABC.

\subsection{Summary of the methods}\label{ssec:method_summary}
We summarize the two methods that will be applied to the electric 
motor later on and already use the notation of our problem for convenience. First, the common assumptions for both methods are presented and then 
each method is explained in detail.
Recall the problem description in Section~\ref{ssec:mathematical problem}. We have given data $\data$, the simulation model $\model(X)$ and the unknown aleatoric parameters $X$ with unknown probability distribution $\pi(X)$.
%
In order to estimate $\pi(X)$ with the hierarchical approach. 
we factorize and parameterize
\begin{align}
\pi(X \mid \theta) = \prod_{i=1}^{n} \pi(X_i \mid \theta),
\end{align}
where $\theta$ is a real-valued vector of hyper-parameters. 
For our case, where $X=(V,T)$, we assume 
$\pi(V \mid m_V, \sigma_V) = \mathcal{N}(m_V, \sigma_V^2)$ 
and $\pi(T \mid m_{T}, \sigma_{T}) = \mathcal{N}(m_{T}, \sigma_{T}^2)$ 
where $m_V, m_{T} \in \real$, $\sigma_V, \sigma_{T} > 0$, hence, $\theta := (m_V, m_{T},\sigma_V, \sigma_{T})$. Let $\pi_0(\theta)$ denote the prior for the hyper-parameters.
As the distribution of $X$ is now fully determined by the hyper-parameters $\theta$, the goal is to approximate $\pi(\theta \mid \data)$ instead of $\pi(X \mid \data)$.
Note that the assumption on the parametrization of the distribution is a modeling assumption which might introduce model error. The model error is not quantified in the following, i.e. we assume that the true distribution belongs to the class of parametrized distributions.
Further note that the distributions of $V$ and $T$ do not have a direct physical relation and can be considered as independent, which justifies the factorization assumption. For other use cases this might not be the case and  correlation needs be considered additionally.
	
Let $y_i^{(\cdot)} \in \data$ be one of the measurements, where $(\cdot)$ either stands for the current $I$ or angular velocity $\omega$. Let $x_i = X(\omega)$ denote a realization of the random vector $X$. Further, let $\model^{(\cdot)}(x_i)$ denote the output of the simulation for model parameters $x_i$ either for $I$ or $\omega$. We assume additive Gaussian measurement noise that is independent and identically distributed
\begin{align}
	y_i^{(\cdot)} = 
	\underbrace{\model^{(\cdot)}(x_i) + e_i}_{\textstyle =: \mathcal{G}^{(\cdot)}(x_i, \sigma_i^{(\cdot)})}, 
	\quad e_i \sim \mathcal{N}(0, \Sigma_i^{(\cdot)}), \quad \Sigma_i^{(\cdot)} = \sigma_i^{(\cdot)} \mathbf{I}_{N_t},
	\label{eq:noise_model}
\end{align}
with $\sigma_i^{(\cdot)} > 0$ for $i=1,\dots,N$. Here $\mathcal{G}^{(\cdot)}(x_i, \sigma_i^{(\cdot)})$ denotes the generative model that produces noisy simulations. 

As already motivated in Section~\ref{ssec:pce} we replace the original simulation model $\model(X)$ by a PCE surrogate $\model^{PCE}(X)$. For surrogate generation we assume $\pi(X)$ to be uniform within $D\subset \real^n$ with $D$ large enough to cover the whole range of the hierarchical prior distribution $\pi(X \mid \theta) \pi_0(\theta)$.

At this point the hierarchical MCMC and the hierarchical ABC method deviate in approximating the posterior distribution.

\subsubsection{Hierarchical surrogate-based MCMC}\label{sssec:mcmc}
Picking up from the noise model \eqref{eq:noise_model} the likelihood for one measurement is then given as 
\begin{align}
\pi(y_i^{(\cdot)} \mid x_i, \sigma_i^{(\cdot)}) = \mathcal{N}\left(y_i^{(\cdot)} - \model^{(\cdot)}(x_i), \Sigma_i^{(\cdot)} \right).
\end{align}
And the likelihood for all measurements in $\data$ as
\begin{align}
\pi(\data \mid X, \sigma) = 
\prod_{i=1}^{N} 
\mathcal{N}\left(y_i^{I} - \model^{I}(x_i), \Sigma_i^{I}\right)
\mathcal{N}\left(y_i^{\omega} - \model^{\omega}(x_i), \Sigma_i^{\omega}\right).
\end{align}
Here $\sigma$ denotes the vector of all $\sigma_i^{(\cdot)}$ and $\pi_0(\sigma)$ the corresponding prior. The posterior distribution is
\begin{align}
\pi(\theta , \sigma \mid \data) \propto \pi(\data \mid X, \sigma) \pi(X \mid \theta) \pi_0(\theta) \pi_0(\sigma). 
\end{align}

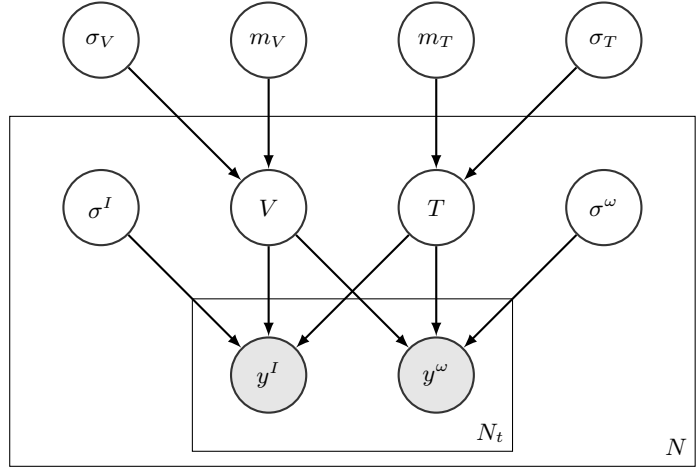
\begin{figure}[htbp]
	\flushright
	\begin{tikzpicture}
	\tikzstyle{main}=[circle, minimum size = 10mm, thick, draw =black!80, node distance = 12mm]
	\tikzstyle{connect}=[-latex, thick]
	\tikzstyle{box}=[rectangle, draw=black!100]
	\node[main, fill = white!100] (sigmaV) [label=center:$\sigma_V$] {};
	\node[main, fill = white!100] (mV) [right=of sigmaV, label=center:$m_V$] {};
	\node[main, fill = white!100] (mT) [right=of mV, label=center:$m_{T}$] {};
	\node[main, fill = white!100] (sigmaT) [right=of mT, label=center:$\sigma_{T}$] {};
	\node[main, fill = white!100] (sigmaI) [below=of sigmaV, label=center:$\sigma^{I}$] {};
	\node[main, fill = white!100] (sigmaO) [below=of sigmaT, label=center:$\sigma^{\omega}$] {};
	\node[main, fill = white!100] (V) [below=of mV, label=center:$V$] {};
	\node[main, fill = white!100] (T) [below=of mT, label=center:$T$] {};
	\node[main, fill = black!10] (yI) [below=of V, label=center:$y^{I}$] {};
	\node[main, fill = black!10] (yO) [below=of T, label=center:$y^{\omega}$] {};
	
	\path (mV) edge [connect] (V)
	(sigmaV) edge [connect] (V)
	(mT) edge [connect] (T)
	(sigmaT) edge [connect] (T)
	(sigmaI) edge [connect] (yI)
	(sigmaO) edge [connect] (yO)
	(V) edge [connect] (yI)
	(V) edge [connect] (yO)
	(T) edge [connect] (yI)
	(T) edge [connect] (yO);
	
	\node[rectangle, inner sep=0mm, fit= (yO) (yI), label=below right:$N_t$, xshift=10mm] {};
	\node[rectangle, inner sep=5mm, draw=black!100, fit= (yO) (yI)] {};
	\node[rectangle, inner sep=2mm, fit= (sigmaI) (sigmaO) (yO) (yI), label=below right:$N$, xshift=22mm] {};
	\node[rectangle, inner sep=7mm, draw=black!100, fit = (sigmaI) (sigmaO) (yO) (yI)] {};
	\end{tikzpicture}
	\caption{Graphical model of the hierarchical inference structure for MCMC.}
	\label{fig:graph}
\end{figure}

To approximate the posterior distribution $\pi(\theta, \sigma \mid \data)$ with Metropolis Hastings MCMC in the hierarchical setup one needs to sample the following parameters:
	the hyper-parameters $\theta$,
	the measurement noise standard deviations $\sigma$,
	and the realizations $x_i=X(\omega), i=1,\dots,N$.
In our case this is a total of $4+ 2N + 2N$ parameters to sample from. For a visualization of the inference structure see the 
graphical model\footnote{For details on graphical models we refer to \cite[Ch. 8]{Bishop.2006}.}
in Figure~\ref{fig:graph}.  

As stated in Section~\ref{ssec:map} the sampling might be highly inefficient in the case of concentrated posterior distributions due to highly informative data. Handing over MAP estimates for initialization and the inverse of the Hessian is an option to increase sampler efficiency. However computing the full inverse Hessian is to expensive, thus we suggest to estimate only the MAP and the inverse Hessian for the realizations $x_i=X(\omega), i=1,\dots,N$ by
\begin{align}
	(x_i) _{\text{MAP}} = \operatorname{arg} \operatornamewithlimits{min}_{x_i \in \mathcal{X}} 
	\| y_i^{I} - \model^{I}(x_i) \|_{\Sigma^{I}_i} 
	+ \| y_i^{\omega} - \model^{\omega}(x_i) \|_{\Sigma^{\omega}_i}
	+ log(\pi(X)).
\end{align}
And then use the $(x_i) _{\text{MAP}}$ for initialization and only a diagonal approximation $\text{diag}(\tilde{C})$ of $C$ for the proposal distribution.


\subsubsection{Hierarchical surrogate-based ABC with summary statistics}\label{sssec:abc}

Again, picking up from the noise model \eqref{eq:noise_model} we use summary statistics to summarize the data $y_i^{(\cdot)}, i=1,\dots,N$ and the generative model $\mathcal{G}^{(\cdot)}(X, \sigma_i^{(\cdot)})$, for $I$ and $\omega$ respectively. 

In general one could use any summary statistic that seems useful for the given data. Here we use the empirical mean and standard deviation, i.e. 
\begin{equation}
S_1^{(\cdot)}:=\frac{1}{N}\sum_{i=1}^{N}y_i^{(\cdot)} \quad \text{and} \quad 
S_2^{(\cdot)}:=\sqrt{\frac{1}{N-1}\sum_{i=1}^{N}(y_i^{(\cdot)}-S_1^{(\cdot)})^2}.
\end{equation}
The summary of the data $\data$ is then 
\begin{align}
	S(\data) := (S_1^{I}, S_1^{\omega}, S_2^{I}, S_2^{\omega} ) \in \real^{4 N_t}.
\end{align}
At this point we could do so with the generative model as well: generate a set of noisy simulation data $\data^\mathcal{G}$ depending on samples of $X \mid \theta$ for given $\theta$ and $\sigma_i^{(\cdot)}$, then compute $S(\data^\mathcal{G})$ and compare to $S(\data)$. The size $N^\mathcal{G}$ of $\data^\mathcal{G}$ does not necessarily need to correspond with the size $N$ of $\data$ as only the summary statistics are compared. However $N^\mathcal{G}$ needs to be sufficiently large to obtain accurate summary statistics. Considering the slow convergence rate of Monte Carlo integration, improving the quality of $S(\data^\mathcal{G})$ by increasing $N^\mathcal{G}$ might get prohibitively expensive as $S(\data^\mathcal{G})$ needs to be computed for each sample ($S(\data)$ only once!).

In the following we introduce a more efficient approach by exploiting the additive noise structure and the parametric distribution of $X$ further. Assume $\sigma_i^{(\cdot)} = \sigma^{(\cdot)}$ for $i=1,\dots,N$. This assumption makes sense since summarizing the data with mean and standard deviation makes it anyway impossible to infer the noise structure of individual measurements.
Then the mean of the generative model with respect to $\pi(X \mid \theta)$ is
\begin{align}
S_1^{(\cdot)}(\theta) :=& \mathbb{E}[\mathcal{G}^{(\cdot)}(X, \sigma^{(\cdot)})] 
= \mathbb{E}[\model(X) + e^{(\cdot)})] 
= \mathbb{E}[\model(X)] + \underbrace{\mathbb{E}[e^{(\cdot)})]}_{=0}\\
=& \int_\mathcal{X} \model(x) \pi(x \mid \theta) dx.
\end{align}
Since noise $e$ is assumed to be stochastic independent and Gaussian distributed with zero mean, it cancels by taking the expectation.
For the standard deviation we obtain
\begin{align}
S_2^{(\cdot)}(\theta, \sigma^{(\cdot)})^2 :=& V[\mathcal{G}^{(\cdot)}(X, \sigma^{(\cdot)})] 
= V[\model(X) + e^{(\cdot)})] 
= V[\model(X)] + \underbrace{V[e^{(\cdot)})]}_{=\operatorname{diag}(\Sigma^{(\cdot)})} \nonumber\\
=& \int_\mathcal{X} \left(\model(x) - S_1^{(\cdot)}(\theta)\right)^2 \pi(x \mid \theta) dx + \operatorname{diag}(\Sigma^{(\cdot)}).
\label{eq:sumstat_std}
\end{align}
In order to approximate the multidimensional integrals $S_1^{(\cdot)}(\theta)$ and $S_2^{(\cdot)}(\theta, \sigma^{(\cdot)})$ 
one can either do Monte Carlo sampling or exploit the assumption of the parametric distribution $\pi(X \mid \theta)$ further and utilize Gauss-Hermite sparse grid quadrature. Gauss-Hermite sparse grid quadrature decreases the number of simulation model evaluations (surrogate evaluations) further by simultaneously increase approximation quality. Denote the sparse grid approximations by $\hat{S}$.
The summary of the generative model is then 
\begin{align}
S(\theta, \sigma^{I}, \sigma^{\omega}) := (\hat{S}_1^{I}(\theta), \hat{S}_1^{\omega}(\theta), \hat{S}_2^{I}(\theta, \sigma^{I}), \hat{S}_2^{\omega}(\theta, \sigma^{\omega}) ) \in \real^{4 N_t}.
\end{align}

As the current and the angular velocity have different magnitude it is important to normalize the summary statistics. To do so we divide each component in $S(\data)$ by its $L^1$ norm, i.e. $\frac{S_{(\cdot)}}{\|S_{(\cdot)}\|_1}$ and so on, and use the exact same scaling also for the components of $S(\theta, \sigma^{I}, \sigma^{\omega})$. 

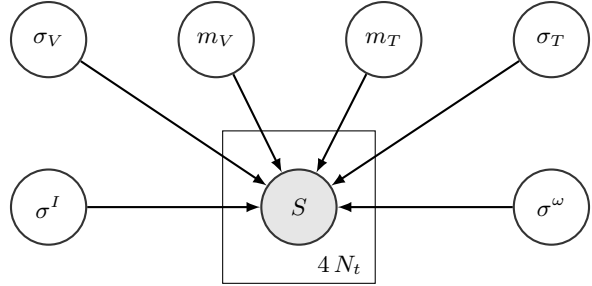
\begin{figure}[htbp]
	\centering
	\begin{tikzpicture}
	\tikzstyle{main}=[circle, minimum size = 10mm, thick, draw =black!80, node distance = 12mm]
	\tikzstyle{main2}=[circle, minimum size = 10mm, thick, draw =black!80, node distance = 23mm]
	\tikzstyle{connect}=[-latex, thick]
	\tikzstyle{box}=[rectangle, draw=black!100]
	\node[main, fill = white!100] (sigmaV) [label=center:$\sigma_V$] {};
	\node[main, fill = white!100] (mV) [right=of sigmaV, label=center:$m_V$] {};
	\node[main, fill = white!100] (mT) [right=of mV, label=center:$m_{T}$] {};
	\node[main, fill = white!100] (sigmaT) [right=of mT, label=center:$\sigma_{T}$] {};
	\node[main, fill = white!100] (sigmaI) [below=of sigmaV, label=center:$\sigma^{I}$] {};
	\node[main, fill = white!100] (sigmaO) [below=of sigmaT, label=center:$\sigma^{\omega}$] {};
	\node[main2, fill = black!10] (S) [right=of sigmaI, label=center:$S$] {};
	
	\path (mV) edge [connect] (S)
	(sigmaV) edge [connect] (S)
	(mT) edge [connect] (S)
	(sigmaT) edge [connect] (S)
	(sigmaI) edge [connect] (S)
	(sigmaO) edge [connect] (S);
	
	\node[rectangle, inner sep=0mm, fit= (S), label=below right:$4\,N_t$, xshift=-4mm] {};
	\node[rectangle, inner sep=5mm, draw=black!100, fit= (S) ] {};
	\end{tikzpicture}
	\caption{Graphical model of the hierarchical inference structure for ABC.}
	\label{fig:graph_abc}
\end{figure}

Let $d(\cdot, \cdot)$ be the $L^2$ norm. Finally, for a given threshold $\delta > 0$ and priors $\pi_{0}(\theta), \pi_{0}(\sigma)$, we sample from the posterior $\pi(\theta, \sigma \mid \data)$ with Algorithm~\ref{alg:rejection_sum} and the normalized summary statistics $S(\data)$ and $S(\theta, \sigma^{I}, \sigma^{\omega})$.  
For a visualization of the inference structure see the graphical model in Figure~\ref{fig:graph_abc}.


The noise standard deviations $\sigma^{I}$ and $\sigma^{\omega}$ are difficult to infer with the chosen summary statistics, as they are only additive terms in $S_2^{(\cdot)}(\theta, \sigma^{(\cdot)})$. Thus we estimate them a priori and keep them fixed during inference (Another option would be to define a highly informative prior centered on the a priori estimation, to allow the sampler to deviate a bit from the fixed value).
Let $J \subset \{1,\dots, N_t\}$ be an index set of time points $t_j \in [t_a, t_b]$ for $j \in J$ in a stationary time domain $[t_a, t_b]$.
We estimate the noise standard deviation of each measurement (for $ i=1,\dots, N$) in this stationary time domain by
\begin{align}
	\overline{\sigma_i^{(\cdot)}} =
	\sqrt{\frac{1}{|J|-1}\sum_{j\in J}\left(y_i^{(\cdot)}(t_j)-\overline{y_i^{(\cdot)}}\right)^2}, 
	\label{eq:noise_est}
\end{align}
where the mean in the stationary time domain is estimated by
\begin{align}
	\overline{y_i^{(\cdot)}} = 
	\frac{1}{|J|}\sum_{j\in J}y_i^{(\cdot)}(t_j).
\end{align}
Then we use either the mean or median of $\overline{\sigma_i^{(\cdot)}}, i=1,\dots,N$ as estimate for $\sigma^{I}$ and $\sigma^{\omega}$. In cases where the measurement data is non-stationary a moving average could be used to remove the trend.

\paragraph*{Remark} Note that similar to the MCMC case one could use MAP $\theta_{\text MAP}$ and inverse Hessian estimates $C$ of the parameters to accelerate the ABC sampler efficiency, e.g. by modifying the prior to $\pi_{0}(\theta) = \mathcal{N}(\theta_{\text MAP}, s C)$ for $s>1$. Provided that the estimates are correct this would lead to less rejected samples. But assume these estimates are distorted, then the modified prior (in case it is more informative than the original) might prevent us from sampling in regions where the posterior is non-zero, leading to a biased posterior. Consequently, in this work we are not using MAP and inverse Hessian estimates for ABC.

\section{Numerical experiments}\label{ch:results}
The numerical experiments are presented with increasing complexity in following 
order: First, we present the results of the methods introduced in Section~\ref{ssec:method_summary} applied to the basic electric motor model (see Section~\ref{ssec:synthetic}) with synthetic data. Second, the results for the real world test bench measurement data and the complex electric motor model (see Section~\ref{ssec:test bench}) are shown.
In both cases noisy measurement data $\data$ for the electric current $I$ and angular velocity $\omega$ is given to infer the distributions of parameters  voltage $V$ and load torque $T$. In  Section~\ref{ssec:method_summary} both are modeled as Gaussian random variables with unknown hyper-parameters mean $m_v, m_T$ and standard deviations $\sigma_V, \sigma_{T}$.


In the following we abbreviate the method \textit{hierarchical surrogate-based MCMC} summarized in Section~\ref{sssec:mcmc} by {MCMC} and in the case of initialization with MAP estimates by {MCMC(MAP)}. 
Further we abbreviate the method \textit{hierarchical surrogate-based ABC with summary statistics} summarized in Section~\ref{sssec:abc} by {ABC} and in the case of SMC by {SMC ABC}.

\subsection*{Inference setup}\label{ssec:inference_setup}
The prior distributions for $m_V, m_{T}$ are uniform with bounds $-30\% / +50\%$ based on the reference values and for $\sigma_V, \sigma_{T}$ with bounds $-75\% / +125\%$ based on the reference values. For MCMC the prior distributions for $\sigma^{I}$ and $\sigma^{\omega}$ are inverse Gamma distributions with mean and standard deviation $0.1$. Due to the summary statistics the identifiability of the noise is difficult, thus for ABC estimates of $\sigma^{I}$ and $\sigma^{\omega}$ based on the available data and Equation~\eqref{eq:noise_est} are used. 

For ABC and SMC ABC we use an implementation based on the Python package ELFI \cite{Elfi}. And for the the Metropolis-Hastings MCMC algorithm an implementation based on the Python package PyMC3 \cite{salvatier2016probabilistic}, with Gaussian random walk proposal.
All computations are performed on a standard Lenovo ThinkPad T460 with an Intel® Core™ i5-6300U CPU with 2 cores, 4 logical processors and 16 GB RAM. Multiprocessing on 3 cores is used for all method, i.e. for MCMC three parallel Markov Chains are generated and for ABC sampling is also carried out in parallel.

\subsection{Basic model with synthetic data}
\label{ssec:synt_results}
In the following we first describe the synthetic data generation, comment on the PCE surrogate, present detailed numerical results and finally draw conclusions on posterior consistency. 

\subsubsection{Synthetic data generation}

\begin{figure}[htbp]
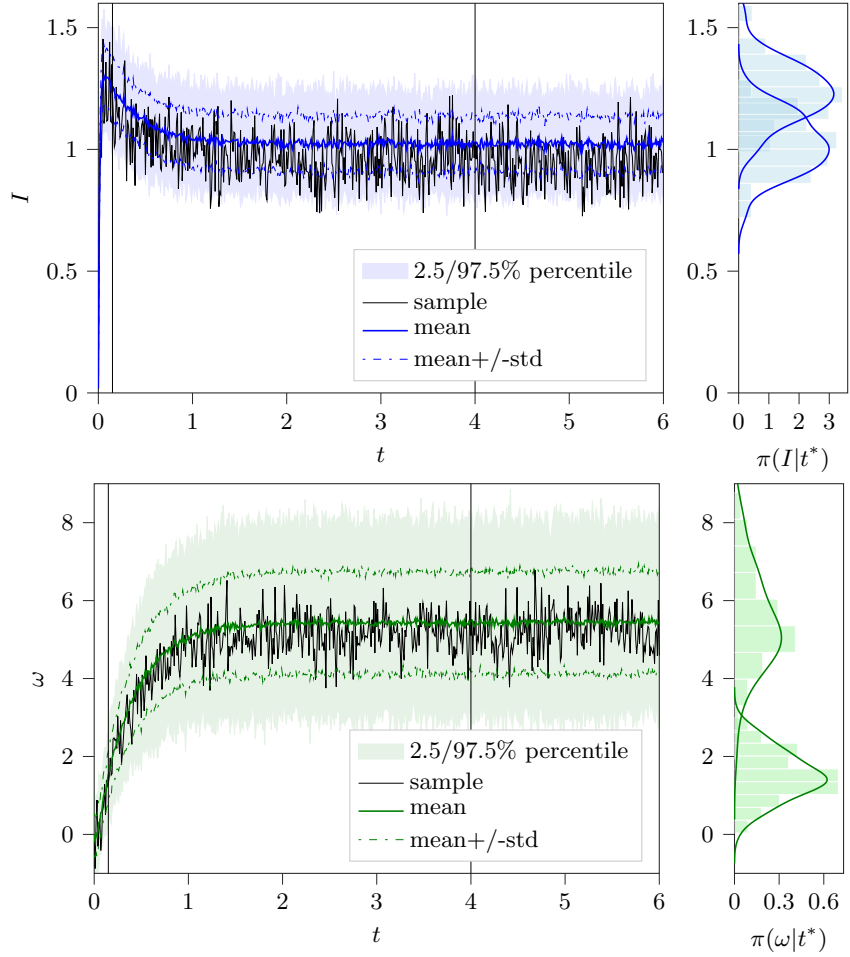

	\centering
	\begin{subfigure}{1\textwidth}
		\raggedleft
		\setlength\figureheight{0.35\textheight} 
		\setlength\figurewidth{1\textwidth}
		\input{Plots/synthetic/511_posterior_consistency_delta_data_I.tex}
	\end{subfigure}\\
	\begin{subfigure}{1\textwidth}
		\raggedleft
		\setlength\figureheight{0.35\textheight} 
		\setlength\figurewidth{1\textwidth}
		\input{Plots/synthetic/511_posterior_consistency_delta_data_omega.tex}
	\end{subfigure}
	\caption[Artificial measurements]{This figures 
		show the artificially generated noisy measurements $\data$ of the current $I$ and the rotational speed $\omega$ for $N = 100$. The area between the $2.5\%$ and $97.5\%$ percentile (shaded), mean+/-standard deviation (dash-dotted) and the mean (solid) of all $N$ measurements are depicted. The black lines show an exemplary noisy sample measurement series. Further, at two time points (vertical lines at $t^*=0.15$ and $t^*=4$ seconds) histograms and kernel density plots are displayed on the right hand side. }
	\label{fig:synthetic_meas}
\end{figure}

To generate synthetic measurement data $\data$ we use the generative model $\mathcal{G}(X, \sigma) = \model(X) + e$, where the simulation model $\model$ is the electric motor model defined in Section~\ref{ssec:synthetic}.
We define reference distributions for the aleatoric parameters $X=(V,T)$ by
$\pi(V \mid m_V, \sigma_V) = \mathcal{N}(m_V, \sigma_V^2)$ 
and $\pi(T \mid m_{T}, \sigma_{T}) = \mathcal{N}(m_{T}, \sigma_{T}^2)$, 
where the hyper-parameters are defined as $m_V = 12, m_{T} = 2.5$, $\sigma_V = 0.7, \sigma_{T} = 0.2$.
Further, reference values for the noise standard deviations are $\sigma^{I} = 0.1$ and $\sigma^{\omega} = 0.5$. 
%
\begin{figure}[htbp]
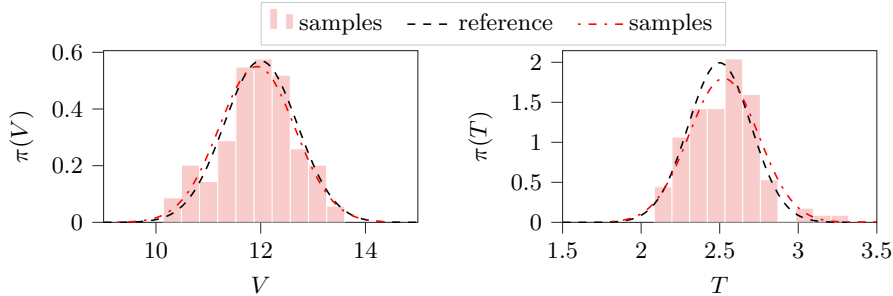

	\centering
	\begin{subfigure}[t]{.5\textwidth}
		\centering
		\setlength\figureheight{0.2\textheight} 
		\setlength\figurewidth{0.95\textwidth}
		\input{Plots/synthetic/511_samples_U.tex}
	\end{subfigure}%
	\begin{subfigure}[t]{.5\textwidth}
		\centering
		\setlength\figureheight{0.2\textheight} 
		\setlength\figurewidth{0.95\textwidth}
		\input{Plots/synthetic/511_samples_load.tex}
	\end{subfigure}
	\caption[Samples synthetic case]{This figures 
		show the Gaussian reference distributions (black dashed) of the parameters $V$ and $T$, histograms of $N=100$ samples and Gaussian distributions fitted to the samples (red dash-dotted).}
	\label{fig:synth_samples}
\end{figure}
With this setting, we sample $N$ times from $\mathcal{G}(X, \sigma)$ in order to generate the synthetic measurement data $\data$. 
Basically, the simulation model $\model$ is evaluated at $N$ independent samples of $X$, resulting in $N$ model outputs containing discrete time series of current $I$ and rotational speed $\omega$ each of size $N_t$. Then for each output independent Gaussian noise is added according to $e^{I}$ and $e^{\omega}$. 
Figure~\ref{fig:synth_samples} shows the reference distributions, the samples and Gaussian distributions fitted to the samples for $V$ and $T$, respectively. 
An overview of the resulting data $\data$ for $N = 100$ is visualized in Figure~\ref{fig:synthetic_meas}. 
Figure~\ref{fig:MCMC_noise} displays a histogram, mean and median of the noise standard deviation estimations $\overline{\sigma_i^{(\cdot)}}, i=1,\dots,N$ based on Equation~\eqref{eq:noise_est}. To obtain a correct estimate of $\sigma_V$ and $\sigma_{T}$ it is important to estimate the 
noise standard deviations correctly. An underestimation of the noise standard deviation results in an overestimation of $\sigma_V$ or $\sigma_{T}$ and vice versa, this is due to the additive structure in the summary statistic, see Equation~\eqref{eq:sumstat_std}.

\subsubsection{PCE surrogate}\label{sssec:synthetic_pce}
A PCE surrogate as introduced in Section~\ref{ssec:pce} is used. Validation is carried out with a set of 100 random samples and corresponding simulations.
The RMSE scaled by the standard deviation of the validation set for a level $L=2$ PCE (leads to 17 sparse grid points) is in the range of $10^{-7}$. I.e. the surrogate for this example can be seen as almost exact.  
The speed up is approximately factor 100. 
I.e. evaluation of one sample with the original model $\model$ takes $4.5 ms \pm 0.412 ms$ and with the surrogate $\model^{PCE}$ only $0.0446 ms \pm 0.007 ms$ (mean $\pm$ std. dev. of 7000 runs). 
Further the surrogate model, in particular the polynomial evaluation, can be vectorized to evaluate a batch of samples simultaneously, which increases efficiency additionally.

\subsubsection{Results}
\begin{figure}[htbp]
	\centering\tiny
	\begin{subfigure}{0.9\textwidth}
		\centering
		\setlength\figureheight{0.225\textheight} 
		\setlength\figurewidth{1\textwidth}
\begin{tikzpicture}

\definecolor{color0}{rgb}{0.901982184391939,0.507089934991626,0.577991751552348}
\definecolor{color1}{rgb}{0.732074288267711,0.54808023353899,0.271747550010192}
\definecolor{color2}{rgb}{0.548486066615094,0.585804228120484,0.249549387016444}
\definecolor{color3}{rgb}{0.260050252892592,0.633337204647202,0.411321010000812}

\begin{axis}[
axis line style={white!15.0!black},
height=\figureheight,
legend cell align={left},
legend style={at={(0.97,0.03)}, anchor=south east, draw=white!80.0!black},
tick align=outside,
tick pos=left,
width=\figurewidth,
x grid style={white!80.0!black},
xlabel={\(\displaystyle m_V\)},
xmin=8.16503912238725, xmax=15.9461898858467,
xtick style={color=white!15.0!black},
y grid style={white!80.0!black},
ymin=-0.5, ymax=6.5,
ytick style={color=white!15.0!black},
ytick={0,1,2,3,4,5,6},
y dir=reverse,
yticklabels={ABC(\(\displaystyle \delta=10.0\)),ABC(\(\displaystyle \delta=7.5\)),ABC(\(\displaystyle \delta=5.0\)),ABC(\(\displaystyle \delta=2.5\)),SMC ABC,MCMC,MCMC(MAP)}
]
\path [draw=white!25.098039215686274!black, fill=color0, opacity=0.5, semithick]
(axis cs:10.9593643320156,-0.4)
--(axis cs:10.9593643320156,0.4)
--(axis cs:13.3533909800314,0.4)
--(axis cs:13.3533909800314,-0.4)
--(axis cs:10.9593643320156,-0.4)
--cycle;
\path [draw=white!25.098039215686274!black, fill=color0, opacity=0.5, semithick]
(axis cs:11.0061885990045,0.6)
--(axis cs:11.0061885990045,1.4)
--(axis cs:12.8720030671941,1.4)
--(axis cs:12.8720030671941,0.6)
--(axis cs:11.0061885990045,0.6)
--cycle;
\path [draw=white!25.098039215686274!black, fill=color0, opacity=0.5, semithick]
(axis cs:11.2917316484164,1.6)
--(axis cs:11.2917316484164,2.4)
--(axis cs:12.5513791164348,2.4)
--(axis cs:12.5513791164348,1.6)
--(axis cs:11.2917316484164,1.6)
--cycle;
\path [draw=white!25.098039215686274!black, fill=color0, opacity=0.5, semithick]
(axis cs:11.7566209869239,2.6)
--(axis cs:11.7566209869239,3.4)
--(axis cs:12.1019575144298,3.4)
--(axis cs:12.1019575144298,2.6)
--(axis cs:11.7566209869239,2.6)
--cycle;
\path [draw=white!25.098039215686274!black, fill=color1, opacity=0.5, semithick]
(axis cs:11.7472867309733,3.6)
--(axis cs:11.7472867309733,4.4)
--(axis cs:12.1030296540029,4.4)
--(axis cs:12.1030296540029,3.6)
--(axis cs:11.7472867309733,3.6)
--cycle;
\path [draw=white!25.098039215686274!black, fill=color2, opacity=0.5, semithick]
(axis cs:12.0443848194668,4.6)
--(axis cs:12.0443848194668,5.4)
--(axis cs:12.1549793955253,5.4)
--(axis cs:12.1549793955253,4.6)
--(axis cs:12.0443848194668,4.6)
--cycle;
\path [draw=white!25.098039215686274!black, fill=color3, opacity=0.5, semithick]
(axis cs:11.8817743959457,5.6)
--(axis cs:11.8817743959457,6.4)
--(axis cs:11.9817881849732,6.4)
--(axis cs:11.9817881849732,5.6)
--(axis cs:11.8817743959457,5.6)
--cycle;
\addplot [thick, black, dashed]
table {%
12 6.5
12 -0.5
};
\addlegendentry{reference}
\addplot [thick, red, dash pattern=on 1pt off 3pt on 3pt off 3pt]
table {%
11.9273812062822 6.5
11.9273812062822 -0.5
};
\addlegendentry{samples}
\addplot [very thin, black, forget plot]
table {%
8.16503912238725 3.5
15.9461898858467 3.5
};
\addplot [very thin, black, forget plot]
table {%
8.16503912238725 4.5
15.9461898858467 4.5
};
\addplot [semithick, white!25.098039215686274!black, forget plot]
table {%
10.9593643320156 0
8.51872779345359 0
};
\addplot [semithick, white!25.098039215686274!black, forget plot]
table {%
13.3533909800314 0
15.5925012147804 0
};
\addplot [semithick, white!25.098039215686274!black, forget plot]
table {%
8.51872779345359 -0.2
8.51872779345359 0.2
};
\addplot [semithick, white!25.098039215686274!black, forget plot]
table {%
15.5925012147804 -0.2
15.5925012147804 0.2
};
\addplot [semithick, white!25.098039215686274!black, forget plot]
table {%
11.0061885990045 1
9.31374711578081 1
};
\addplot [semithick, white!25.098039215686274!black, forget plot]
table {%
12.8720030671941 1
14.7144286950256 1
};
\addplot [semithick, white!25.098039215686274!black, forget plot]
table {%
9.31374711578081 0.8
9.31374711578081 1.2
};
\addplot [semithick, white!25.098039215686274!black, forget plot]
table {%
14.7144286950256 0.8
14.7144286950256 1.2
};
\addplot [semithick, white!25.098039215686274!black, forget plot]
table {%
11.2917316484164 2
10.1926030169174 2
};
\addplot [semithick, white!25.098039215686274!black, forget plot]
table {%
12.5513791164348 2
13.6895206457363 2
};
\addplot [semithick, white!25.098039215686274!black, forget plot]
table {%
10.1926030169174 1.8
10.1926030169174 2.2
};
\addplot [semithick, white!25.098039215686274!black, forget plot]
table {%
13.6895206457363 1.8
13.6895206457363 2.2
};
\addplot [semithick, white!25.098039215686274!black, forget plot]
table {%
11.7566209869239 3
11.3760224890473 3
};
\addplot [semithick, white!25.098039215686274!black, forget plot]
table {%
12.1019575144298 3
12.4858698760571 3
};
\addplot [semithick, white!25.098039215686274!black, forget plot]
table {%
11.3760224890473 2.8
11.3760224890473 3.2
};
\addplot [semithick, white!25.098039215686274!black, forget plot]
table {%
12.4858698760571 2.8
12.4858698760571 3.2
};
\addplot [semithick, white!25.098039215686274!black, forget plot]
table {%
11.7472867309733 4
11.3516700639997 4
};
\addplot [semithick, white!25.098039215686274!black, forget plot]
table {%
12.1030296540029 4
12.4937041066287 4
};
\addplot [semithick, white!25.098039215686274!black, forget plot]
table {%
11.3516700639997 3.8
11.3516700639997 4.2
};
\addplot [semithick, white!25.098039215686274!black, forget plot]
table {%
12.4937041066287 3.8
12.4937041066287 4.2
};
\addplot [semithick, white!25.098039215686274!black, forget plot]
table {%
12.0443848194668 5
11.8802188564376 5
};
\addplot [semithick, white!25.098039215686274!black, forget plot]
table {%
12.1549793955253 5
12.3196069297959 5
};
\addplot [semithick, white!25.098039215686274!black, forget plot]
table {%
11.8802188564376 4.8
11.8802188564376 5.2
};
\addplot [semithick, white!25.098039215686274!black, forget plot]
table {%
12.3196069297959 4.8
12.3196069297959 5.2
};
\addplot [semithick, white!25.098039215686274!black, forget plot]
table {%
11.8817743959457 6
11.7322294007263 6
};
\addplot [semithick, white!25.098039215686274!black, forget plot]
table {%
11.9817881849732 6
12.1246140973822 6
};
\addplot [semithick, white!25.098039215686274!black, forget plot]
table {%
11.7322294007263 5.8
11.7322294007263 6.2
};
\addplot [semithick, white!25.098039215686274!black, forget plot]
table {%
12.1246140973822 5.8
12.1246140973822 6.2
};
\addplot [semithick, white!25.098039215686274!black, forget plot]
table {%
12.129747951835 -0.4
12.129747951835 0.4
};
\addplot [semithick, white!25.098039215686274!black, forget plot]
table {%
11.9441864854189 0.6
11.9441864854189 1.4
};
\addplot [semithick, white!25.098039215686274!black, forget plot]
table {%
11.9036459830063 1.6
11.9036459830063 2.4
};
\addplot [semithick, white!25.098039215686274!black, forget plot]
table {%
11.9270611941875 2.6
11.9270611941875 3.4
};
\addplot [semithick, white!25.098039215686274!black, forget plot]
table {%
11.92344637855 3.6
11.92344637855 4.4
};
\addplot [semithick, white!25.098039215686274!black, forget plot]
table {%
12.0994078544141 4.6
12.0994078544141 5.4
};
\addplot [semithick, white!25.098039215686274!black, forget plot]
table {%
11.9321915956255 5.6
11.9321915956255 6.4
};
\end{axis}

\end{tikzpicture}
	\end{subfigure}\\
	\begin{subfigure}{0.9\textwidth}
		\centering
		\setlength\figureheight{0.225\textheight} 
		\setlength\figurewidth{1\textwidth}
\begin{tikzpicture}

\definecolor{color0}{rgb}{0.901982184391939,0.507089934991626,0.577991751552348}
\definecolor{color1}{rgb}{0.732074288267711,0.54808023353899,0.271747550010192}
\definecolor{color2}{rgb}{0.548486066615094,0.585804228120484,0.249549387016444}
\definecolor{color3}{rgb}{0.260050252892592,0.633337204647202,0.411321010000812}

\begin{axis}[
axis line style={white!15.0!black},
height=\figureheight,
legend cell align={left},
legend style={at={(0.97,0.03)}, anchor=south east, draw=white!80.0!black},
tick align=outside,
tick pos=left,
width=\figurewidth,
x grid style={white!80.0!black},
xlabel={\(\displaystyle \sigma_V\)},
xmin=0.116797350389084, xmax=1.4232572240768,
xtick style={color=white!15.0!black},
y grid style={white!80.0!black},
ymin=-0.5, ymax=6.5,
ytick style={color=white!15.0!black},
ytick={0,1,2,3,4,5,6},
y dir=reverse,
yticklabels={ABC(\(\displaystyle \delta=10.0\)),ABC(\(\displaystyle \delta=7.5\)),ABC(\(\displaystyle \delta=5.0\)),ABC(\(\displaystyle \delta=2.5\)),SMC ABC,MCMC,MCMC(MAP)}
]
\path [draw=white!25.098039215686274!black, fill=color0, opacity=0.5, semithick]
(axis cs:0.441093020008162,-0.4)
--(axis cs:0.441093020008162,0.4)
--(axis cs:0.900734315713368,0.4)
--(axis cs:0.900734315713368,-0.4)
--(axis cs:0.441093020008162,-0.4)
--cycle;
\path [draw=white!25.098039215686274!black, fill=color0, opacity=0.5, semithick]
(axis cs:0.454743127057692,0.6)
--(axis cs:0.454743127057692,1.4)
--(axis cs:0.890598517545172,1.4)
--(axis cs:0.890598517545172,0.6)
--(axis cs:0.454743127057692,0.6)
--cycle;
\path [draw=white!25.098039215686274!black, fill=color0, opacity=0.5, semithick]
(axis cs:0.444049764360887,1.6)
--(axis cs:0.444049764360887,2.4)
--(axis cs:0.856210494114311,2.4)
--(axis cs:0.856210494114311,1.6)
--(axis cs:0.444049764360887,1.6)
--cycle;
\path [draw=white!25.098039215686274!black, fill=color0, opacity=0.5, semithick]
(axis cs:0.547076983731485,2.6)
--(axis cs:0.547076983731485,3.4)
--(axis cs:0.804072741297231,3.4)
--(axis cs:0.804072741297231,2.6)
--(axis cs:0.547076983731485,2.6)
--cycle;
\path [draw=white!25.098039215686274!black, fill=color1, opacity=0.5, semithick]
(axis cs:0.571743461583112,3.6)
--(axis cs:0.571743461583112,4.4)
--(axis cs:0.807212444395342,4.4)
--(axis cs:0.807212444395342,3.6)
--(axis cs:0.571743461583112,3.6)
--cycle;
\path [draw=white!25.098039215686274!black, fill=color2, opacity=0.5, semithick]
(axis cs:0.693249406102705,4.6)
--(axis cs:0.693249406102705,5.4)
--(axis cs:0.769857243296237,5.4)
--(axis cs:0.769857243296237,4.6)
--(axis cs:0.693249406102705,4.6)
--cycle;
\path [draw=white!25.098039215686274!black, fill=color3, opacity=0.5, semithick]
(axis cs:0.695754480284834,5.6)
--(axis cs:0.695754480284834,6.4)
--(axis cs:0.767686647800876,6.4)
--(axis cs:0.767686647800876,5.6)
--(axis cs:0.695754480284834,5.6)
--cycle;
\addplot [thick, black, dashed]
table {%
0.7 6.5
0.7 -0.5
};
\addlegendentry{reference}
\addplot [thick, red, dash pattern=on 1pt off 3pt on 3pt off 3pt]
table {%
0.725925067398919 6.5
0.725925067398919 -0.5
};
\addlegendentry{samples}
\addplot [very thin, black, forget plot]
table {%
0.116797350389084 3.5
1.4232572240768 3.5
};
\addplot [very thin, black, forget plot]
table {%
0.116797350389084 4.5
1.4232572240768 4.5
};
\addplot [semithick, white!25.098039215686274!black, forget plot]
table {%
0.441093020008162 0
0.176592956800082 0
};
\addplot [semithick, white!25.098039215686274!black, forget plot]
table {%
0.900734315713368 0
1.36387268436373 0
};
\addplot [semithick, white!25.098039215686274!black, forget plot]
table {%
0.176592956800082 -0.2
0.176592956800082 0.2
};
\addplot [semithick, white!25.098039215686274!black, forget plot]
table {%
1.36387268436373 -0.2
1.36387268436373 0.2
};
\addplot [semithick, white!25.098039215686274!black, forget plot]
table {%
0.454743127057692 1
0.177016106491189 1
};
\addplot [semithick, white!25.098039215686274!black, forget plot]
table {%
0.890598517545172 1
1.2284136240443 1
};
\addplot [semithick, white!25.098039215686274!black, forget plot]
table {%
0.177016106491189 0.8
0.177016106491189 1.2
};
\addplot [semithick, white!25.098039215686274!black, forget plot]
table {%
1.2284136240443 0.8
1.2284136240443 1.2
};
\addplot [semithick, white!25.098039215686274!black, forget plot]
table {%
0.444049764360887 2
0.177554046881201 2
};
\addplot [semithick, white!25.098039215686274!black, forget plot]
table {%
0.856210494114311 2
1.1482754384484 2
};
\addplot [semithick, white!25.098039215686274!black, forget plot]
table {%
0.177554046881201 1.8
0.177554046881201 2.2
};
\addplot [semithick, white!25.098039215686274!black, forget plot]
table {%
1.1482754384484 1.8
1.1482754384484 2.2
};
\addplot [semithick, white!25.098039215686274!black, forget plot]
table {%
0.547076983731485 3
0.176181890102162 3
};
\addplot [semithick, white!25.098039215686274!black, forget plot]
table {%
0.804072741297231 3
0.981468937662682 3
};
\addplot [semithick, white!25.098039215686274!black, forget plot]
table {%
0.176181890102162 2.8
0.176181890102162 3.2
};
\addplot [semithick, white!25.098039215686274!black, forget plot]
table {%
0.981468937662682 2.8
0.981468937662682 3.2
};
\addplot [semithick, white!25.098039215686274!black, forget plot]
table {%
0.571743461583112 4
0.238205753697079 4
};
\addplot [semithick, white!25.098039215686274!black, forget plot]
table {%
0.807212444395342 4
0.984478273914091 4
};
\addplot [semithick, white!25.098039215686274!black, forget plot]
table {%
0.238205753697079 3.8
0.238205753697079 4.2
};
\addplot [semithick, white!25.098039215686274!black, forget plot]
table {%
0.984478273914091 3.8
0.984478273914091 4.2
};
\addplot [semithick, white!25.098039215686274!black, forget plot]
table {%
0.693249406102705 5
0.583254490520628 5
};
\addplot [semithick, white!25.098039215686274!black, forget plot]
table {%
0.769857243296237 5
0.884507995143805 5
};
\addplot [semithick, white!25.098039215686274!black, forget plot]
table {%
0.583254490520628 4.8
0.583254490520628 5.2
};
\addplot [semithick, white!25.098039215686274!black, forget plot]
table {%
0.884507995143805 4.8
0.884507995143805 5.2
};
\addplot [semithick, white!25.098039215686274!black, forget plot]
table {%
0.695754480284834 6
0.588583503611931 6
};
\addplot [semithick, white!25.098039215686274!black, forget plot]
table {%
0.767686647800876 6
0.875412291989593 6
};
\addplot [semithick, white!25.098039215686274!black, forget plot]
table {%
0.588583503611931 5.8
0.588583503611931 6.2
};
\addplot [semithick, white!25.098039215686274!black, forget plot]
table {%
0.875412291989593 5.8
0.875412291989593 6.2
};
\addplot [semithick, white!25.098039215686274!black, forget plot]
table {%
0.676056172836893 -0.4
0.676056172836893 0.4
};
\addplot [semithick, white!25.098039215686274!black, forget plot]
table {%
0.682349737978765 0.6
0.682349737978765 1.4
};
\addplot [semithick, white!25.098039215686274!black, forget plot]
table {%
0.678666921459459 1.6
0.678666921459459 2.4
};
\addplot [semithick, white!25.098039215686274!black, forget plot]
table {%
0.682514736477524 2.6
0.682514736477524 3.4
};
\addplot [semithick, white!25.098039215686274!black, forget plot]
table {%
0.697077427842012 3.6
0.697077427842012 4.4
};
\addplot [semithick, white!25.098039215686274!black, forget plot]
table {%
0.72861779340607 4.6
0.72861779340607 5.4
};
\addplot [semithick, white!25.098039215686274!black, forget plot]
table {%
0.730545609488483 5.6
0.730545609488483 6.4
};
\end{axis}

\end{tikzpicture}
	\end{subfigure}\\
	\begin{subfigure}{0.9\textwidth}
		\centering
		\setlength\figureheight{0.225\textheight} 
		\setlength\figurewidth{1\textwidth}
\begin{tikzpicture}

\definecolor{color0}{rgb}{0.901982184391939,0.507089934991626,0.577991751552348}
\definecolor{color1}{rgb}{0.732074288267711,0.54808023353899,0.271747550010192}
\definecolor{color2}{rgb}{0.548486066615094,0.585804228120484,0.249549387016444}
\definecolor{color3}{rgb}{0.260050252892592,0.633337204647202,0.411321010000812}

\begin{axis}[
axis line style={white!15.0!black},
height=\figureheight,
legend cell align={left},
legend style={at={(0.97,0.03)}, anchor=south east, draw=white!80.0!black},
tick align=outside,
tick pos=left,
width=\figurewidth,
x grid style={white!80.0!black},
xlabel={\(\displaystyle m_{T}\)},
xmin=1.65145825496547, xmax=3.82053296265821,
xtick style={color=white!15.0!black},
y grid style={white!80.0!black},
ymin=-0.5, ymax=6.5,
ytick style={color=white!15.0!black},
ytick={0,1,2,3,4,5,6},
y dir=reverse,
yticklabels={ABC(\(\displaystyle \delta=10.0\)),ABC(\(\displaystyle \delta=7.5\)),ABC(\(\displaystyle \delta=5.0\)),ABC(\(\displaystyle \delta=2.5\)),SMC ABC,MCMC,MCMC(MAP)}
]
\path [draw=white!25.098039215686274!black, fill=color0, opacity=0.5, semithick]
(axis cs:2.21725446646073,-0.4)
--(axis cs:2.21725446646073,0.4)
--(axis cs:2.97429230950612,0.4)
--(axis cs:2.97429230950612,-0.4)
--(axis cs:2.21725446646073,-0.4)
--cycle;
\path [draw=white!25.098039215686274!black, fill=color0, opacity=0.5, semithick]
(axis cs:2.24032187168964,0.6)
--(axis cs:2.24032187168964,1.4)
--(axis cs:2.83392801172108,1.4)
--(axis cs:2.83392801172108,0.6)
--(axis cs:2.24032187168964,0.6)
--cycle;
\path [draw=white!25.098039215686274!black, fill=color0, opacity=0.5, semithick]
(axis cs:2.31927876074221,1.6)
--(axis cs:2.31927876074221,2.4)
--(axis cs:2.72939051520112,2.4)
--(axis cs:2.72939051520112,1.6)
--(axis cs:2.31927876074221,1.6)
--cycle;
\path [draw=white!25.098039215686274!black, fill=color0, opacity=0.5, semithick]
(axis cs:2.47264719011167,2.6)
--(axis cs:2.47264719011167,3.4)
--(axis cs:2.58314439214584,3.4)
--(axis cs:2.58314439214584,2.6)
--(axis cs:2.47264719011167,2.6)
--cycle;
\path [draw=white!25.098039215686274!black, fill=color1, opacity=0.5, semithick]
(axis cs:2.46657823365433,3.6)
--(axis cs:2.46657823365433,4.4)
--(axis cs:2.58773913061117,4.4)
--(axis cs:2.58773913061117,3.6)
--(axis cs:2.46657823365433,3.6)
--cycle;
\path [draw=white!25.098039215686274!black, fill=color2, opacity=0.5, semithick]
(axis cs:2.56617972475227,4.6)
--(axis cs:2.56617972475227,5.4)
--(axis cs:2.60191006658335,5.4)
--(axis cs:2.60191006658335,4.6)
--(axis cs:2.56617972475227,4.6)
--cycle;
\path [draw=white!25.098039215686274!black, fill=color3, opacity=0.5, semithick]
(axis cs:2.51227195792712,5.6)
--(axis cs:2.51227195792712,6.4)
--(axis cs:2.54363189901267,6.4)
--(axis cs:2.54363189901267,5.6)
--(axis cs:2.51227195792712,5.6)
--cycle;
\addplot [thick, black, dashed]
table {%
2.5 6.5
2.5 -0.5
};
\addlegendentry{reference}
\addplot [thick, red, dash pattern=on 1pt off 3pt on 3pt off 3pt]
table {%
2.52564823025527 6.5
2.52564823025527 -0.5
};
\addlegendentry{samples}
\addplot [very thin, black, forget plot]
table {%
1.65145825496547 3.5
3.82053296265821 3.5
};
\addplot [very thin, black, forget plot]
table {%
1.65145825496547 4.5
3.82053296265821 4.5
};
\addplot [semithick, white!25.098039215686274!black, forget plot]
table {%
2.21725446646073 0
1.75005255986059 0
};
\addplot [semithick, white!25.098039215686274!black, forget plot]
table {%
2.97429230950612 0
3.72193865776309 0
};
\addplot [semithick, white!25.098039215686274!black, forget plot]
table {%
1.75005255986059 -0.2
1.75005255986059 0.2
};
\addplot [semithick, white!25.098039215686274!black, forget plot]
table {%
3.72193865776309 -0.2
3.72193865776309 0.2
};
\addplot [semithick, white!25.098039215686274!black, forget plot]
table {%
2.24032187168964 1
1.75907135155905 1
};
\addplot [semithick, white!25.098039215686274!black, forget plot]
table {%
2.83392801172108 1
3.43848832722377 1
};
\addplot [semithick, white!25.098039215686274!black, forget plot]
table {%
1.75907135155905 0.8
1.75907135155905 1.2
};
\addplot [semithick, white!25.098039215686274!black, forget plot]
table {%
3.43848832722377 0.8
3.43848832722377 1.2
};
\addplot [semithick, white!25.098039215686274!black, forget plot]
table {%
2.31927876074221 2
1.94415078404568 2
};
\addplot [semithick, white!25.098039215686274!black, forget plot]
table {%
2.72939051520112 2
3.10867303075975 2
};
\addplot [semithick, white!25.098039215686274!black, forget plot]
table {%
1.94415078404568 1.8
1.94415078404568 2.2
};
\addplot [semithick, white!25.098039215686274!black, forget plot]
table {%
3.10867303075975 1.8
3.10867303075975 2.2
};
\addplot [semithick, white!25.098039215686274!black, forget plot]
table {%
2.47264719011167 3
2.3465287335028 3
};
\addplot [semithick, white!25.098039215686274!black, forget plot]
table {%
2.58314439214584 3
2.70520481915191 3
};
\addplot [semithick, white!25.098039215686274!black, forget plot]
table {%
2.3465287335028 2.8
2.3465287335028 3.2
};
\addplot [semithick, white!25.098039215686274!black, forget plot]
table {%
2.70520481915191 2.8
2.70520481915191 3.2
};
\addplot [semithick, white!25.098039215686274!black, forget plot]
table {%
2.46657823365433 4
2.34396427241902 4
};
\addplot [semithick, white!25.098039215686274!black, forget plot]
table {%
2.58773913061117 4
2.70458249333046 4
};
\addplot [semithick, white!25.098039215686274!black, forget plot]
table {%
2.34396427241902 3.8
2.34396427241902 4.2
};
\addplot [semithick, white!25.098039215686274!black, forget plot]
table {%
2.70458249333046 3.8
2.70458249333046 4.2
};
\addplot [semithick, white!25.098039215686274!black, forget plot]
table {%
2.56617972475227 5
2.51285618177214 5
};
\addplot [semithick, white!25.098039215686274!black, forget plot]
table {%
2.60191006658335 5
2.65550530908369 5
};
\addplot [semithick, white!25.098039215686274!black, forget plot]
table {%
2.51285618177214 4.8
2.51285618177214 5.2
};
\addplot [semithick, white!25.098039215686274!black, forget plot]
table {%
2.65550530908369 4.8
2.65550530908369 5.2
};
\addplot [semithick, white!25.098039215686274!black, forget plot]
table {%
2.51227195792712 6
2.46702160477067 6
};
\addplot [semithick, white!25.098039215686274!black, forget plot]
table {%
2.54363189901267 6
2.59055060704397 6
};
\addplot [semithick, white!25.098039215686274!black, forget plot]
table {%
2.46702160477067 5.8
2.46702160477067 6.2
};
\addplot [semithick, white!25.098039215686274!black, forget plot]
table {%
2.59055060704397 5.8
2.59055060704397 6.2
};
\addplot [semithick, white!25.098039215686274!black, forget plot]
table {%
2.58834847408894 -0.4
2.58834847408894 0.4
};
\addplot [semithick, white!25.098039215686274!black, forget plot]
table {%
2.52129699550317 0.6
2.52129699550317 1.4
};
\addplot [semithick, white!25.098039215686274!black, forget plot]
table {%
2.51566279520812 1.6
2.51566279520812 2.4
};
\addplot [semithick, white!25.098039215686274!black, forget plot]
table {%
2.52598524999656 2.6
2.52598524999656 3.4
};
\addplot [semithick, white!25.098039215686274!black, forget plot]
table {%
2.52900915929044 3.6
2.52900915929044 4.4
};
\addplot [semithick, white!25.098039215686274!black, forget plot]
table {%
2.58438811064585 4.6
2.58438811064585 5.4
};
\addplot [semithick, white!25.098039215686274!black, forget plot]
table {%
2.52707371054569 5.6
2.52707371054569 6.4
};
\end{axis}

\end{tikzpicture}
	\end{subfigure}\\
	\begin{subfigure}{0.9\textwidth}
		\centering
		\setlength\figureheight{0.225\textheight} 
		\setlength\figurewidth{1\textwidth}
\begin{tikzpicture}

\definecolor{color0}{rgb}{0.901982184391939,0.507089934991626,0.577991751552348}
\definecolor{color1}{rgb}{0.732074288267711,0.54808023353899,0.271747550010192}
\definecolor{color2}{rgb}{0.548486066615094,0.585804228120484,0.249549387016444}
\definecolor{color3}{rgb}{0.260050252892592,0.633337204647202,0.411321010000812}

\begin{axis}[
axis line style={white!15.0!black},
height=\figureheight,
legend cell align={left},
legend style={at={(0.97,0.03)}, anchor=south east, draw=white!80.0!black},
tick align=outside,
tick pos=left,
width=\figurewidth,
x grid style={white!80.0!black},
xlabel={\(\displaystyle \sigma_{T}\)},
xmin=0.0311265336034002, xmax=0.446561030475415,
xtick style={color=white!15.0!black},
y grid style={white!80.0!black},
ymin=-0.5, ymax=6.5,
ytick style={color=white!15.0!black},
ytick={0,1,2,3,4,5,6},
y dir=reverse,
yticklabels={ABC(\(\displaystyle \delta=10.0\)),ABC(\(\displaystyle \delta=7.5\)),ABC(\(\displaystyle \delta=5.0\)),ABC(\(\displaystyle \delta=2.5\)),SMC ABC,MCMC,MCMC(MAP)}
]
\path [draw=white!25.098039215686274!black, fill=color0, opacity=0.5, semithick]
(axis cs:0.138641659992098,-0.4)
--(axis cs:0.138641659992098,0.4)
--(axis cs:0.288811837490816,0.4)
--(axis cs:0.288811837490816,-0.4)
--(axis cs:0.138641659992098,-0.4)
--cycle;
\path [draw=white!25.098039215686274!black, fill=color0, opacity=0.5, semithick]
(axis cs:0.138865798806006,0.6)
--(axis cs:0.138865798806006,1.4)
--(axis cs:0.279346820003955,1.4)
--(axis cs:0.279346820003955,0.6)
--(axis cs:0.138865798806006,0.6)
--cycle;
\path [draw=white!25.098039215686274!black, fill=color0, opacity=0.5, semithick]
(axis cs:0.150665218320407,1.6)
--(axis cs:0.150665218320407,2.4)
--(axis cs:0.279147903477331,2.4)
--(axis cs:0.279147903477331,1.6)
--(axis cs:0.150665218320407,1.6)
--cycle;
\path [draw=white!25.098039215686274!black, fill=color0, opacity=0.5, semithick]
(axis cs:0.181925741261672,2.6)
--(axis cs:0.181925741261672,3.4)
--(axis cs:0.264611895963908,3.4)
--(axis cs:0.264611895963908,2.6)
--(axis cs:0.181925741261672,2.6)
--cycle;
\path [draw=white!25.098039215686274!black, fill=color1, opacity=0.5, semithick]
(axis cs:0.179884232378861,3.6)
--(axis cs:0.179884232378861,4.4)
--(axis cs:0.258220282162752,4.4)
--(axis cs:0.258220282162752,3.6)
--(axis cs:0.179884232378861,3.6)
--cycle;
\path [draw=white!25.098039215686274!black, fill=color2, opacity=0.5, semithick]
(axis cs:0.201757514680868,4.6)
--(axis cs:0.201757514680868,5.4)
--(axis cs:0.225186112714768,5.4)
--(axis cs:0.225186112714768,4.6)
--(axis cs:0.201757514680868,4.6)
--cycle;
\path [draw=white!25.098039215686274!black, fill=color3, opacity=0.5, semithick]
(axis cs:0.217518834910545,5.6)
--(axis cs:0.217518834910545,6.4)
--(axis cs:0.238382212689476,6.4)
--(axis cs:0.238382212689476,5.6)
--(axis cs:0.217518834910545,5.6)
--cycle;
\addplot [thick, black, dashed]
table {%
0.2 6.5
0.2 -0.5
};
\addlegendentry{reference}
\addplot [thick, red, dash pattern=on 1pt off 3pt on 3pt off 3pt]
table {%
0.22186759239743 6.5
0.22186759239743 -0.5
};
\addlegendentry{samples}
\addplot [very thin, black, forget plot]
table {%
0.0311265336034002 3.5
0.446561030475415 3.5
};
\addplot [very thin, black, forget plot]
table {%
0.0311265336034002 4.5
0.446561030475415 4.5
};
\addplot [semithick, white!25.098039215686274!black, forget plot]
table {%
0.138641659992098 0
0.0500099198248554 0
};
\addplot [semithick, white!25.098039215686274!black, forget plot]
table {%
0.288811837490816 0
0.42767764425396 0
};
\addplot [semithick, white!25.098039215686274!black, forget plot]
table {%
0.0500099198248554 -0.2
0.0500099198248554 0.2
};
\addplot [semithick, white!25.098039215686274!black, forget plot]
table {%
0.42767764425396 -0.2
0.42767764425396 0.2
};
\addplot [semithick, white!25.098039215686274!black, forget plot]
table {%
0.138865798806006 1
0.0500099198248554 1
};
\addplot [semithick, white!25.098039215686274!black, forget plot]
table {%
0.279346820003955 1
0.404825879824549 1
};
\addplot [semithick, white!25.098039215686274!black, forget plot]
table {%
0.0500099198248554 0.8
0.0500099198248554 1.2
};
\addplot [semithick, white!25.098039215686274!black, forget plot]
table {%
0.404825879824549 0.8
0.404825879824549 1.2
};
\addplot [semithick, white!25.098039215686274!black, forget plot]
table {%
0.150665218320407 2
0.050672693451734 2
};
\addplot [semithick, white!25.098039215686274!black, forget plot]
table {%
0.279147903477331 2
0.371879405135157 2
};
\addplot [semithick, white!25.098039215686274!black, forget plot]
table {%
0.050672693451734 1.8
0.050672693451734 2.2
};
\addplot [semithick, white!25.098039215686274!black, forget plot]
table {%
0.371879405135157 1.8
0.371879405135157 2.2
};
\addplot [semithick, white!25.098039215686274!black, forget plot]
table {%
0.181925741261672 3
0.0599381303086468 3
};
\addplot [semithick, white!25.098039215686274!black, forget plot]
table {%
0.264611895963908 3
0.317413574268423 3
};
\addplot [semithick, white!25.098039215686274!black, forget plot]
table {%
0.0599381303086468 2.8
0.0599381303086468 3.2
};
\addplot [semithick, white!25.098039215686274!black, forget plot]
table {%
0.317413574268423 2.8
0.317413574268423 3.2
};
\addplot [semithick, white!25.098039215686274!black, forget plot]
table {%
0.179884232378861 4
0.0648242466189489 4
};
\addplot [semithick, white!25.098039215686274!black, forget plot]
table {%
0.258220282162752 4
0.313217389024652 4
};
\addplot [semithick, white!25.098039215686274!black, forget plot]
table {%
0.0648242466189489 3.8
0.0648242466189489 4.2
};
\addplot [semithick, white!25.098039215686274!black, forget plot]
table {%
0.313217389024652 3.8
0.313217389024652 4.2
};
\addplot [semithick, white!25.098039215686274!black, forget plot]
table {%
0.201757514680868 5
0.167427413069686 5
};
\addplot [semithick, white!25.098039215686274!black, forget plot]
table {%
0.225186112714768 5
0.260015181031197 5
};
\addplot [semithick, white!25.098039215686274!black, forget plot]
table {%
0.167427413069686 4.8
0.167427413069686 5.2
};
\addplot [semithick, white!25.098039215686274!black, forget plot]
table {%
0.260015181031197 4.8
0.260015181031197 5.2
};
\addplot [semithick, white!25.098039215686274!black, forget plot]
table {%
0.217518834910545 6
0.187273343899176 6
};
\addplot [semithick, white!25.098039215686274!black, forget plot]
table {%
0.238382212689476 6
0.269562150011827 6
};
\addplot [semithick, white!25.098039215686274!black, forget plot]
table {%
0.187273343899176 5.8
0.187273343899176 6.2
};
\addplot [semithick, white!25.098039215686274!black, forget plot]
table {%
0.269562150011827 5.8
0.269562150011827 6.2
};
\addplot [semithick, white!25.098039215686274!black, forget plot]
table {%
0.216301853595793 -0.4
0.216301853595793 0.4
};
\addplot [semithick, white!25.098039215686274!black, forget plot]
table {%
0.214442100963087 0.6
0.214442100963087 1.4
};
\addplot [semithick, white!25.098039215686274!black, forget plot]
table {%
0.228930954099614 1.6
0.228930954099614 2.4
};
\addplot [semithick, white!25.098039215686274!black, forget plot]
table {%
0.22738297579241 2.6
0.22738297579241 3.4
};
\addplot [semithick, white!25.098039215686274!black, forget plot]
table {%
0.223352070749534 3.6
0.223352070749534 4.4
};
\addplot [semithick, white!25.098039215686274!black, forget plot]
table {%
0.213140671923298 4.6
0.213140671923298 5.4
};
\addplot [semithick, white!25.098039215686274!black, forget plot]
table {%
0.228081395304223 5.6
0.228081395304223 6.4
};
\end{axis}

\end{tikzpicture}
	\end{subfigure}%
	\caption[Marginal posterior distributions]{
		Boxplots of the samples approximating the hyper-parameters marginal posterior distributions in the synthetic case obtained with ABC, SMC ABC, MCMC and MCMC(MAP) in comparison to the reference values and sample mean and standard deviation.}
	\label{fig:posterior_consistency_delta}
\end{figure}
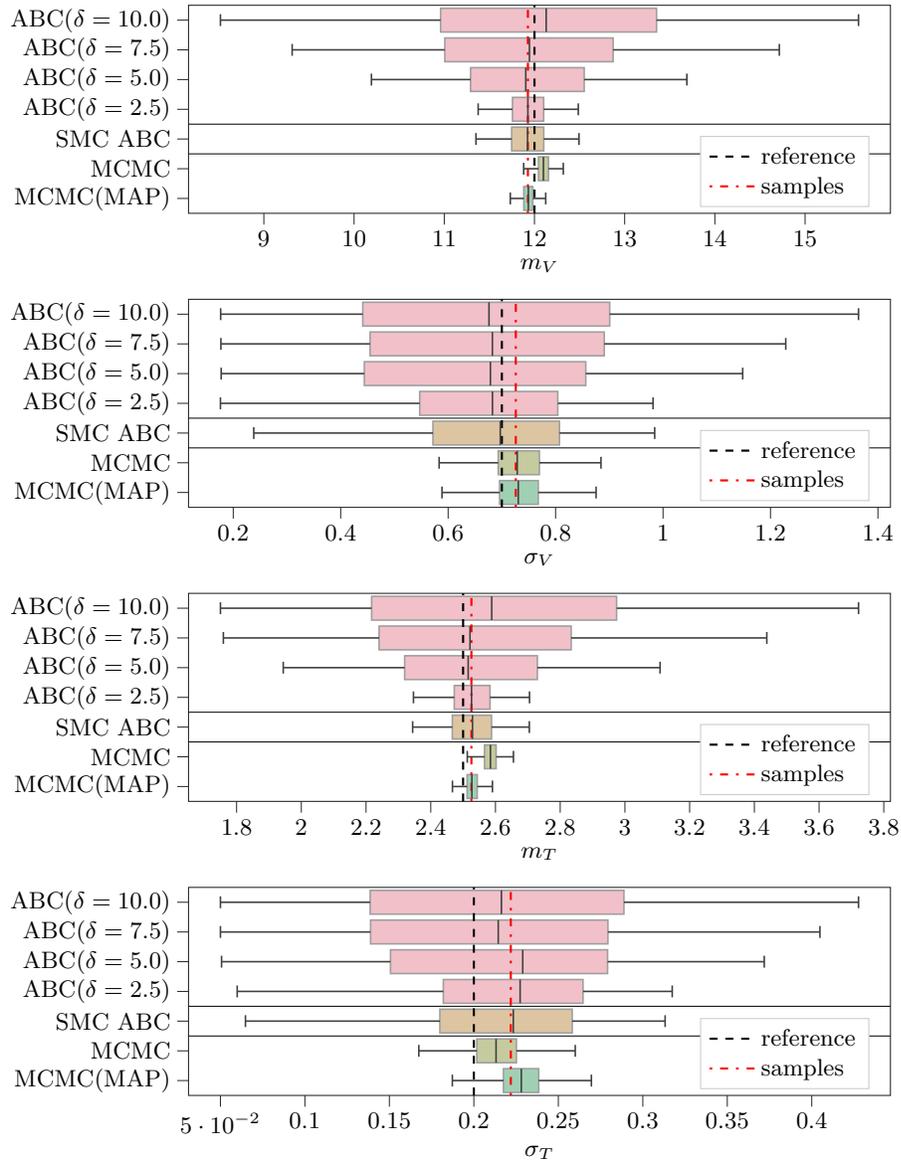

Figure~\ref{fig:posterior_consistency_delta} displays boxplots of the samples approximating the marginal posterior distributions of the hyper-parameters, obtained with 
ABC, SMC ABC,  MCMC and MCMC(MAP). Adding on this Table~\ref{tab:runtimes} lists the corresponding runtimes, number of proposals and samples of all methods.
%
%
\begin{table}[htbp]
	\centering\small
	\begin{tabular}{lrrr}
\toprule
{method} & runtime [s] & samples $m$ & proposals $M$ \\
\midrule
ABC($\delta=10.0$) &           3 &        1500 &         12000 \\
ABC($\delta=7.5$)  &           7 &        1500 &         28000 \\
ABC($\delta=5.0$)  &          27 &        1500 &        106000 \\
ABC($\delta=2.5$)  &        1536 &        1500 &       5718000 \\
SMC ABC            &         211 &        1500 &        626000 \\
MCMC               &         400 &   $n_{eff}$ &       3*11000 \\
MCMC(MAP)          &         387 &   $n_{eff}$ &       3*11000 \\
\bottomrule
\end{tabular}

	\caption{Runtimes in seconds, number of samples $m$ and proposals $M$ of the considered methods. For MCMC the number of effective samples $n_{eff}$ is given in detail in Table~\ref{tab:mcmc_gelman_rubin_neff}.}
	\label{tab:runtimes}
\end{table}
For ABC four decreasing values of $\delta \in \{10,\, 7.5,\, 5,\, 2.5\}$ show the concentration of the marginal posterior. The number $M$ of ABC sample proposals to obtain a fixed number of $m=1500$ samples that satisfy $d(\cdot,\cdot) \leq  \delta$ increases exponentially, see Figure~\ref{fig:posterior_consistency_delta_proposals}.
The SMC ABC results are obtained with the exact same $\delta$'s and are comparable to the ABC results for the smallest $\delta$. However, by reusing the information of sample-discrepancy pairs from previous populations, SMC ABC reduces the overall number of proposals by approximately factor 10 in order to obtain a similar result. 
%
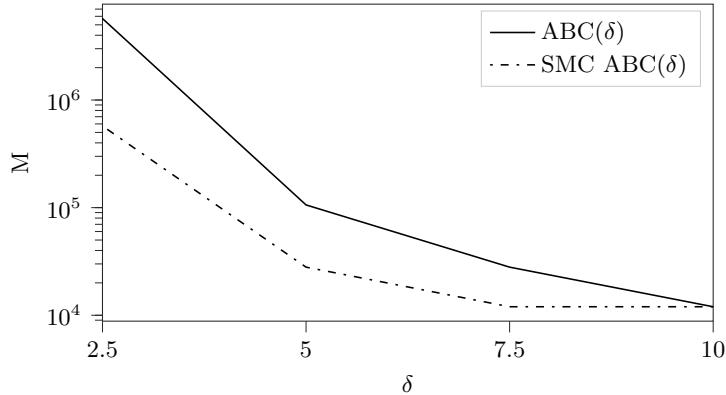
\begin{figure}[htbp]
	\centering
	\begin{subfigure}{0.8\textwidth}
		\centering
		\setlength\figureheight{0.3\textheight} 
		\setlength\figurewidth{1\textwidth}
\begin{tikzpicture}

\begin{axis}[
axis line style={white!15.0!black},
height=\figureheight,
legend cell align={left},
legend style={draw=white!80.0!black},
log basis y={10},
tick align=outside,
tick pos=left,
width=\figurewidth,
x grid style={white!80.0!black},
xlabel={\(\displaystyle \delta\)},
xmin=2.5, xmax=10,
xtick={2.5,5,7.5,10},
xtick style={color=white!15.0!black},
y grid style={white!80.0!black},
ylabel={M},
ymin=8816.13233993287, ymax=7783004.76380128,
ymode=log,
ytick style={color=white!15.0!black}
]
\addplot [semithick, black]
table {%
10 12000
7.5 28000
5 106000
2.5 5718000
};
\addlegendentry{ABC$(\delta)$}
\addplot [semithick, black, dash pattern=on 1pt off 3pt on 3pt off 3pt]
table {%
10 12000
7.5 12000
5 28000
2.5 574000
};
\addlegendentry{SMC ABC$(\delta)$}
\end{axis}

\end{tikzpicture}
	\end{subfigure}%
	\caption[Posterior consistency with respect to $\delta$ - number of proposals]{This figure shows the number $M$ of proposals to obtain $m=1500$ samples that satisfy $d(\cdot,\cdot) \leq \delta$, in order to obtain the results in Figure~\ref{fig:posterior_consistency_delta}. SMC ABC$(\delta)$ denotes the proposals that are required in each population, given the information of the previous population.}
	\label{fig:posterior_consistency_delta_proposals}
\end{figure}
For MCMC and MCMC(MAP) we sample three parallel Markov chains each of length 11.000. With a burn-in phase of 6.000 samples and a modest thinning (discard every second sample) due to autocorrelation we use three times 2.500 samples to approximate the posterior.  
Table~\ref{tab:mcmc_gelman_rubin_neff} lists the Gelman-rubin statistics $\hat{R}$ and the number of effective samples $n_{eff}$ of the samples (without thinning). 
The statistics show that MCMC has clearly difficulties to converge and an increase of sample size would be recommended. It is a bit better for the hyper-parameters and noise standard deviations better, but poor for $V_i$ and $T_i$ for $i=1,\dots,N$, which is due to the concentration effect of the posterior and the high dimensions. 
Over all this sampling difficulties lead to a biased (w.r.t.~the other methods) posterior for $m_V$ and $m_T$, displayed in Figure~\ref{fig:posterior_consistency_delta}.
By initializing with MAP estimates MCMC(MAP) shows clear improvements in sampler statistics and posterior distribution. However the sampler has still difficulties, e.g. the number of effective samples is still less than 15\% of the considered samples.
%
\begin{table}[htbp]
	\centering\small
	\begin{tabular}{lrrrrrrrr}
	\toprule
	{} & \multicolumn{2}{c}{mean($\hat{R}$)} & \multicolumn{6}{c}{$\hat{R}$} \\
	\cmidrule(lr){2-3} \cmidrule(lr){4-9}
	{} &             $V$ &   $T$ &     $m_V$ & $\sigma_V$ & $m_{T}$ & $\sigma_{T}$ & $\sigma_{I}$ & $\sigma_{\omega}$ \\
	\midrule
	MCMC      &           3.446 & 3.515 &     1.107 &      1.105 &   1.185 &        1.227 &        2.765 &             1.684 \\
	MCMC(MAP) &           1.328 & 1.319 &     1.000 &      1.005 &   1.005 &        1.001 &        1.003 &             1.002 \\

	\midrule
	{} & \multicolumn{2}{c}{mean($n_{eff}$)} & \multicolumn{6}{c}{$n_{eff}$} \\
	\cmidrule(lr){2-3} \cmidrule(lr){4-9}
	{} &             $V$ & $T$ &     $m_V$ & $\sigma_V$ & $m_{T}$ & $\sigma_{T}$ & $\sigma_{I}$ & $\sigma_{\omega}$ \\
	\midrule
	MCMC      &               2 &   2 &        13 &         12 &       7 &            5 &            2 &                 3 \\
	MCMC(MAP) &               8 &   8 &     2,215 &      2,114 &   1,679 &        2,353 &        1,297 &             1,521 \\
	\bottomrule
\end{tabular}

	\caption{Gelman-rubin statistics $\hat{R}$ and number of effective samples $n_{eff}$ for MCMC and MCMC(MAP), based on the last 5000 samples of 3 parallel sampled chains, each of total length 11.000, for $N=100$. Note that the values presented for $V$ and $T$ are 
		$mean(\hat{R}(V)) := \frac{1}{N} \sum_{i=1}^{N} \hat{R}(V_i)$, 
		respectively for $T$ (analog for $n_{eff}$).}
	\label{tab:mcmc_gelman_rubin_neff}
\end{table}
In addition to the posterior distributions of the hyper-parameters marginal posterior distributions of $V_i$ and $T_{i}$ for $i=1,\dots,N$ obtained via MCMC(MAP) are visualized in Figure \ref{fig:MCMC_samples_100}.
%
\begin{figure}[htbp]
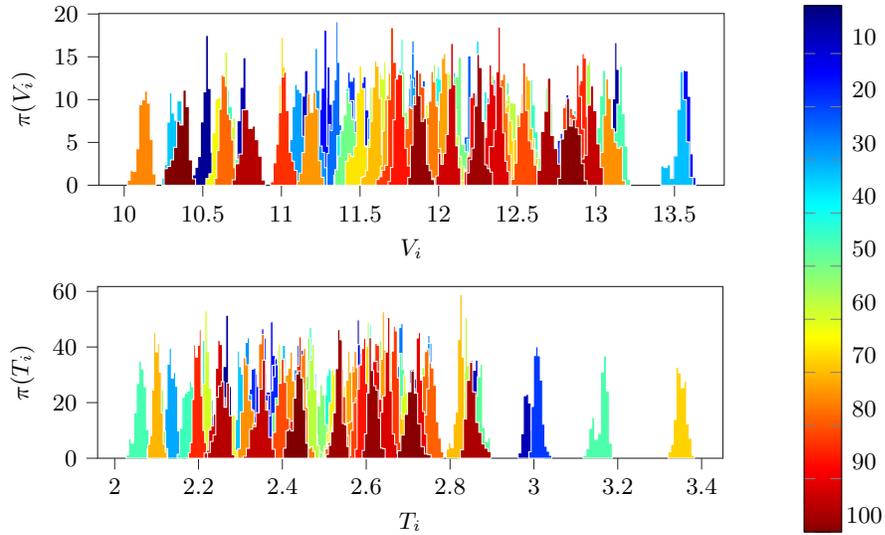

	\centering
	\begin{tabular}[c]{cc}
		\begin{tabular}[c]{c}
			\smallskip
			\begin{subfigure}[c]{0.85\textwidth}
				\centering
				\setlength\figureheight{0.2\textheight} 
				\setlength\figurewidth{0.96\textwidth}
				\input{Plots/synthetic/511_posterior_100_U.tex}
			\end{subfigure}\\
			\begin{subfigure}[c]{0.85\textwidth}
				\centering
				\setlength\figureheight{0.2\textheight} 
				\setlength\figurewidth{0.96\textwidth}
				\input{Plots/synthetic/511_posterior_100_load.tex}
			\end{subfigure}
		\end{tabular}
		&
		\begin{subfigure}[c]{0.15\textwidth}
			\setlength\figureheight{0.2\textheight} 
			\setlength\figurewidth{0.89\textwidth}
			\begin{tikzpicture}
\begin{axis}[
    hide axis,
    scale only axis,
    height=\figureheight,
    width=\figurewidth,
    colormap/jet,
    colorbar horizontal,
    point meta min=1,
    point meta max=100,
    colorbar style={
        xtick={10,20,30,...,100},
        xticklabel pos=upper,
        rotate=270,
        width=7cm,
        x tick label style={xshift=0.3cm}
    }]
    \addplot [draw=none] coordinates {(0,0)};
\end{axis}
\end{tikzpicture}
		\end{subfigure}
	\end{tabular}
	\caption[MCMC(MAP) estimates for the samples $V_i$ and ${T}_i$ based on syntehtic data]{This figure shows the marginal MCMC(MAP) samples for $V_i$ and 
		${T}_i$ for $i=1,...,100$. The color bar indicates the index $i$.}
	\label{fig:MCMC_samples_100}
\end{figure}
%
This offers, in contrast to the ABC methods, the opportunity to verify (presuming the posterior distributions are correct), if the underlying parameter distributions are actually Gaussian or if another parametric distribution would be more suitable.
A possibility to get an intuition on the distribution of $V$ or $T$, one could for example plot a histogram of the MAP estimates.
If the marginal posterior densities do not contain much 
uncertainty a parametric distribution fitted to the MAP estimates leads already to a reliable estimate of the distribution of $V$ or $T$. However if uncertainty is present,  a hierarchical method is better suited to capture the overall uncertainty.

\begin{figure}[htbp]
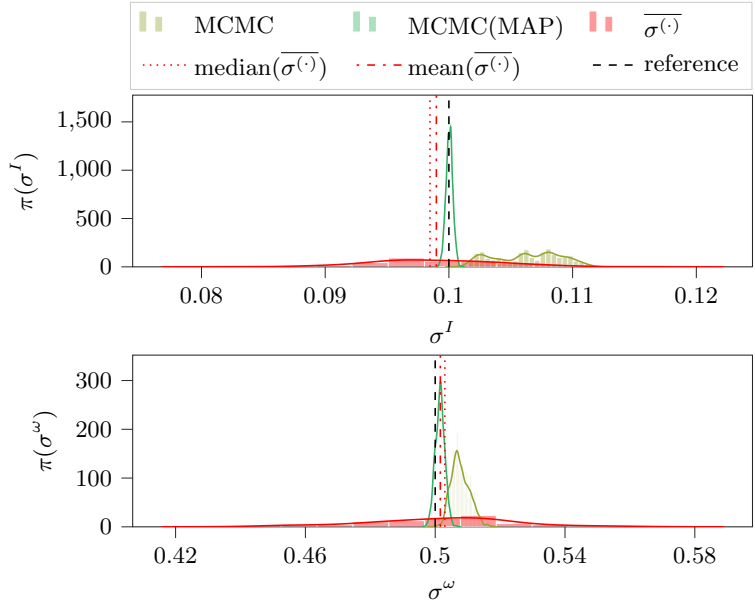

	\flushleft
	\begin{subfigure}{0.9\textwidth}
		\flushright
		\setlength\figureheight{0.2\textheight} 
		\setlength\figurewidth{0.9\textwidth}
		\input{Plots/synthetic/511_posterior_noise_sigma_I.tex}
	\end{subfigure}\\
	\begin{subfigure}{0.9\textwidth}
		\flushright
		\setlength\figureheight{0.2\textheight} 
		\setlength\figurewidth{0.9\textwidth}
		\input{Plots/synthetic/511_posterior_noise_sigma_omega.tex}
	\end{subfigure}
	\caption[MCMC posterior for noise standard deviations]{This figure shows the marginal MCMC and MCMC(MAP) samples for the noise standard deviations $\sigma^{I}$ and $\sigma^{\omega}$. Also the reference values, a histogram of the estimates 
		$\overline{\sigma^{(\cdot)}} := [\overline{\sigma_i^{(\cdot)}}]_{i=1,\dots,N}$ 
		via Equation~\eqref{eq:noise_est} and the corresponding mean and median are displayed.}
	\label{fig:MCMC_noise}
\end{figure}

Figure~\ref{fig:MCMC_noise} shows the MCMC and MCMC(MAP) marginal posterior distributions for the noise standard deviations $\sigma^{I}$ and $\sigma^{\omega}$. For MCMC they overestimate a bit, which corresponds to the fact that the Markov Chain did not converge properly. For MCMC(MAP) they are almost identical to the reference values, which is expected in this synthetic case. 

\begin{figure}[htbp]
	\centering
	\begin{subfigure}[b]{.5\textwidth}
		\centering
		\setlength\figureheight{0.3\textheight} 
		\setlength\figurewidth{0.95\textwidth}
\begin{tikzpicture}

\definecolor{color0}{rgb}{0.967797559291991,0.441274560091574,0.53581031550587}
\definecolor{color1}{rgb}{0.808795411310631,0.563470005005669,0.195026426967273}
\definecolor{color2}{rgb}{0.59208915296397,0.641846701637824,0.193506913499104}
\definecolor{color3}{rgb}{0.19783576093349,0.695551696606304,0.39953010374445}

\begin{axis}[
axis line style={white!15.0!black},
height=\figureheight,
legend cell align={left},
legend columns=4,
legend style={at={(0.03,1.03)}, anchor=south west, draw=white!80.0!black, 
	/tikz/every even column/.append style={column sep=0.25cm},},
log basis x={10},
log basis y={10},
tick align=outside,
tick pos=left,
width=\figurewidth,
x grid style={white!80.0!black},
xlabel={Model evaluations},
xmin=149874.249778859, xmax=132311080.984622,
xmode=log,
xtick style={color=white!15.0!black},
y grid style={white!80.0!black},
ylabel={\(\displaystyle \Sigma_i Bias(\theta_i)\)},
ymin=5.87702417108365e-05, ymax=0.204080831092211,
ymode=log,
ytick style={color=white!15.0!black}
]
\addplot [very thin, color0, mark=*, mark size=3, mark options={solid}]
table {%
204000 0.0474015043756714
476000 0.00225529790784776
1802000 0.00294629317320805
97206000 0.00191509227474545
};
\addlegendentry{ABC($\delta$)}
\addplot [very thin, color1, mark=square*, mark size=3, mark options={solid}]
table {%
408000 0.0283039604500902
884000 0.0015364209940172
10642000 0.000861168697244169
};
\addlegendentry{SMC ABC}
\addplot [very thin, color2, mark=triangle*, mark size=3, mark options={solid,rotate=180}]
table {%
1200000 0.140884695305936
3300000 0.0331269511301216
};
\addlegendentry{MCMC}
\addplot [very thin, color3, mark=triangle*, mark size=3, mark options={solid}]
table {%
1203348 0.000134300993269861
3303348 8.51325954589494e-05
};
\addlegendentry{MCMC(MAP)}
\end{axis}

\end{tikzpicture}
	\end{subfigure}%
	\begin{subfigure}[b]{.5\textwidth}
		\centering
		\setlength\figureheight{0.3\textheight} 
		\setlength\figurewidth{0.95\textwidth}
\begin{tikzpicture}

\definecolor{color0}{rgb}{0.967797559291991,0.441274560091574,0.53581031550587}
\definecolor{color1}{rgb}{0.808795411310631,0.563470005005669,0.195026426967273}
\definecolor{color2}{rgb}{0.59208915296397,0.641846701637824,0.193506913499104}
\definecolor{color3}{rgb}{0.19783576093349,0.695551696606304,0.39953010374445}

\begin{axis}[
axis line style={white!15.0!black},
height=\figureheight,
legend cell align={left},
legend style={draw=white!80.0!black},
log basis x={10},
log basis y={10},
tick align=outside,
tick pos=left,
width=\figurewidth,
x grid style={white!80.0!black},
xlabel={Model evaluations},
xmin=149874.249778859, xmax=132311080.984622,
xmode=log,
xtick style={color=white!15.0!black},
y grid style={white!80.0!black},
ylabel={\(\displaystyle \Sigma_i MSE(\theta_i)\)},
ymin=0.00700898221390617, ymax=3.67379839705604,
ymode=log,
ytick style={color=white!15.0!black}
]
\addplot [very thin, color0, mark=*, mark size=3, mark options={solid}]
table {%
204000 2.76378010134708
476000 1.73858272709997
1802000 0.765443619227399
97206000 0.0917295418738479
};
\addlegendentry{ABC($\delta$)}
\addplot [very thin, color1, mark=square*, mark size=3, mark options={solid}]
table {%
408000 1.53781364175105
884000 0.682064035164544
10642000 0.0927540527042589
};
\addlegendentry{SMC ABC}
\addplot [very thin, color2, mark=triangle*, mark size=3, mark options={solid,rotate=180}]
table {%
1200000 0.153638660072564
3300000 0.0442807290106134
};
\addlegendentry{MCMC}
\addplot [very thin, color3, mark=triangle*, mark size=3, mark options={solid}]
table {%
1203348 0.00931680042485736
3303348 0.00944905100625661
};
\addlegendentry{MCMC(MAP)}
\legend{}
\end{axis}

\end{tikzpicture}
	\end{subfigure}
	\caption[Bias and MSE]{Sum of Bias and MSE summed over all parameters $\theta$ versus the number of simulation model $\model$ (or metamodel $\model^{PCE}$) evaluations. The Bias is calculated w.r.t.~the median of the marginal posterior distribution and the empirical moments of the reference samples. Note that due to parallelization model evaluations do not directly correspond to runtimes. 
	}
	\label{fig:bias_mse_sum}
\end{figure}
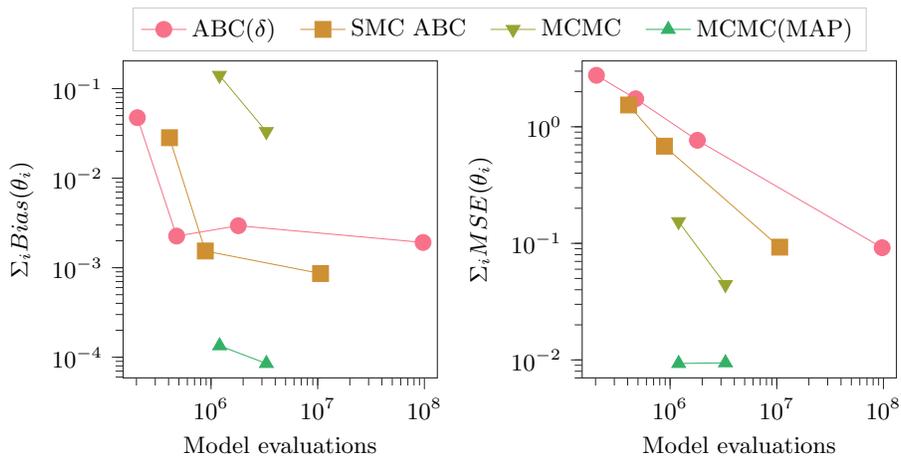

\begin{mydef}
	For estimator $\vartheta$ of a parameter $\varTheta \in \real$ define  $Bias(\vartheta) = \left( \hat{\vartheta} - \varTheta \right)^2$, where $\hat{\vartheta}$ is a point estimate of $\vartheta$, e.g. the mean, median or MAP.
	Further define the Mean (Median or MAP) Square Error by $MSE(\vartheta) = Bias(\vartheta) + V[\vartheta]$. 
\end{mydef}
In order to put the considered methods into perspective with respect to accuracy, uncertainty and computational effort, Figure~\ref{fig:bias_mse_sum} shows the summed Bias and MSE of all hyper-parameters versus the number of model evaluations. The Bias is calculated w.r.t.~the median of the marginal posterior distribution of $m_V, \sigma_V, m_T, \sigma_{T}$ and the reference samples empirical moments.
The number of model evaluations for ABC are determined by the product of number of proposals $M$ and the number of grid points of the Gauss-Hermite quadrature. Here for two dimensions 17 sparse grid nodes are used. 
For MCMC the number of model evaluations is the number of chains times number of samples $m$ times number of observations $N$. Sampling in one chain is only possible sequentially. 
Important to note is that the model evaluations do not directly translate to runtime. For ABC and also for SMC ABC in each population they can be easily be executed in parallel, which is not the case for MCMC.
This can be already seen in Table~\ref{tab:runtimes}, where SMC ABC is twice as fast as MCMC(MAP) albeit requiring a larger number of model evaluations.
Please note that computational times in Table~\ref{tab:runtimes} strongly depend on the implementation, however they already give an impression on each methods efficiency. 
As expected the summed MSE for ABC, SMC ABC and MCMC reduce with the number of model evaluations, i.e. corresponding number of samples, where the major part is due to a reduction of the variance, but also due to a reduction in the summed Bias. For MCMC(MAP) initialized already with the MAP estimates no further reduction of the summed MSE can be achieved, however the summed Bias reduces a bit.
Overall, w.r.t.~the summed Bias ABC and SMC ABC achieve results that are only a bit worse than MCMC(MAP) and way better than MCMC, in much less time.

\begin{figure}[htbp]
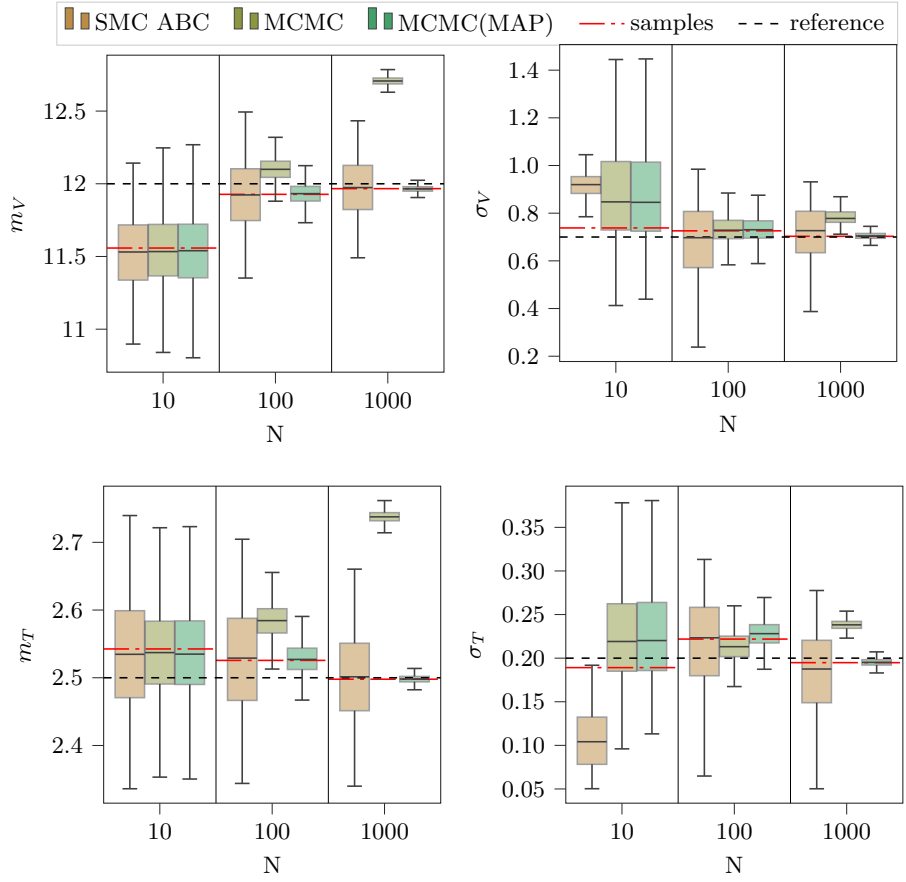

	\centering
	\begin{subfigure}{0.5\textwidth}
		\centering
		\setlength\figureheight{0.3\textheight} 
		\setlength\figurewidth{\textwidth}
		\input{Plots/synthetic/513_posterior_N_0.tex}
	\end{subfigure}%
	\begin{subfigure}{0.5\textwidth}
		\centering
		\setlength\figureheight{0.3\textheight} 
		\setlength\figurewidth{\textwidth}
		\input{Plots/synthetic/513_posterior_N_1.tex}
	\end{subfigure}\\
	\begin{subfigure}{0.5\textwidth}
		\centering
		\setlength\figureheight{0.3\textheight} 
		\setlength\figurewidth{\textwidth}
		\input{Plots/synthetic/513_posterior_N_2.tex}
	\end{subfigure}%
	\begin{subfigure}{0.5\textwidth}
		\centering
		\setlength\figureheight{0.3\textheight} 
		\setlength\figurewidth{\textwidth}
		\input{Plots/synthetic/513_posterior_N_3.tex}
	\end{subfigure}		
	\caption[Posterior for $N = 10, 100, 1000$]{This figure shows histograms of the marginal posterior distributions of $m_V, \ \sigma_V,\ m_T,\ \sigma_T$ obtained with SMC ABC, MCMC and MCMC(MAP) for data size $N = 10, 100, 1000$. Also the reference value and sample values depending on $N$ are displayed. Table~\ref{tab:mcmc_n_obs} lists the corresponding computational time, $\hat{R}$ and $n_{eff}$.}
	\label{fig:posterior_n_obs_all}
\end{figure}

So far the results presented above were for a fixed data size $N=100$. In the following we compare the MCMC, MCMC(ABC) and SMC ABC methods for $N = 10, 100, 1000$. 
Table~\ref{tab:mcmc_n_obs} contains measures for the MCMC and MCMC(MAP) sampling efficiency for increasing values of $N =10,100,1000$ but constant length of the three parallel sampled Markov chains of length $11,000$. 
\begin{table}[htbp]
	\centering
	\begin{tabular}{rrrrrrrr}
\toprule
{} & \multicolumn{3}{c}{MCMC} & \multicolumn{3}{c}{MCMC(MAP)} & {SMC ABC} \\
\cmidrule(lr){2-4} \cmidrule(lr){5-7} \cmidrule(lr){8-8}
N & $\hat{R}$ & $n_{eff}$ &  time [s] & $\hat{R}$ & $n_{eff}$ &  time [s] &  time [s] \\
\midrule
10   &     1.13 &   453 &   158 &     1.03 &   486 &   151 & 217 \\
100  &     3.42 &     2 &   400 &     1.31 &    62 &   387 & 211 \\
1000 &     5.48 &     2 & 3,562 &     2.23 &     9 & 3,536 & 205 \\
\bottomrule
\end{tabular}

	\caption{Dimension dependency of MH-MCMC sampler efficiency with respect to the number of observations $N$. Note that the actual number of inferred parameters is $2N+6$. For concise presentation, the values for $\hat{R}$ and $n_{eff}$ are the mean of all of those parameters statistics. For comparison the runtime of SMC ABC is added.}
	\label{tab:mcmc_n_obs}
\end{table}
For increasing $N$ the values for $\hat{R}$, $n_{eff}$ and the computational time show that the efficiency decreases drastically, which was already discussed in Section~\ref{ssec:map}. This is not only due to the high dimensions $N$ but also due to the concentration effect of the posterior distributions as a consequence of highly informative data. Note that already for the low dimensional case with $N=10$ the $n_{eff}$ is below $10\%$ of the considered samples.
$\hat{R}$ increases in higher dimensional state spaces indicating that the distribution of the Markov chain has not yet converged and the number of samples 
should be increased. Furthermore, the time needed to execute the algorithm 
strongly increases with $N$. We observe a similar behavior for MCMC(MAP), however in a smaller scale. This confirms that the statistical efficiency of 
the Metropolis-Hastings Algorithm with Gaussian random walk proposal does not 
perform dimension-independent and justifies the ABC and in particular the SMC ABC approach that improves in performance with increasing $N$, since the summary statistics improve with increasing $N$. Also the runtime stays almost constant. For $N= 1000$ SMC ABC is roughly 17 times faster than MCMC(MAP). 
Figure~\ref{fig:posterior_n_obs_all} adds on Table~\ref{tab:mcmc_n_obs} by presenting the marginal posterior distribution for varying $N$. For $N=1000$ the MCMC posterior of $m_V$ and $m_T$ are largely biased and only the SMC ABC and MCMC(MAP) results deliver proper estimates. 
We also observe that the SMC ABC posterior only slightly concentrates with increasing data. This might be due to the fact that the summary statistics only represent part of the original information.
%

\subsubsection{Posterior consistency}
Our proposed method has several layers of approximation, thus we comment in the following on posterior consistency w.r.t.~Theorem~\ref{thm:consistency}.  
In order to show posterior consistency we present in Figure~\ref{fig:posterior_consistency_delta} results of the method for small ABC threshold limit ($\delta \rightarrow 0$). Corresponding to the theory the posterior concentrates for decreasing values of $\delta$, which is also summarized in Figure~\ref{fig:bias_mse_sum}.

For the considered example the surrogate already has an very high approximation quality, see Section~\ref{sssec:synthetic_pce} Thus we do not consider posterior consistency with respect to the level $L$ of the surrogate.  

Posterior consistency in the small noise limit ($\sigma^{I}, \sigma^{\omega} \rightarrow 0$) is difficult to show. With decreasing noise in the single measurements also the noise in the summary statistics reduces. Thus the minimal discrepancy value $d(\cdot,\cdot)$ decreases as well. If we now fix the threshold $\delta$ for several noise levels the posterior distribution does not concentrate, but rather smear out a bit as more samples are accepted. 
Consequently in order to observe posterior consistency in the small noise limit we additionally need to decrease $\delta$. By decreasing $\delta$ in the order as the minimal discrepancy value $d(\cdot,\cdot)$ decreases (which is known due to the reference values) we also observe posterior concentration on the empirical moments of the parameter samples.

In the case of large data limit, i.e. number of observations $N \rightarrow \infty$, we observe in Figure~\ref{fig:posterior_n_obs_all} that the empirical moments of the parameter samples converge to the reference values (due to the law of large numbers) and with this also the posterior distribution shifts more to the reference values. However a concentration can hardly be observed, this might be a consequence of the information loss due to the summary statistics. With increasing $N$ the summary statistics get more and more accurate (again due to the law of large numbers), however they still represent only part of the whole data $\data$.
A second parameter determining the data size is the number of discrete time steps $N_t$. As it is already very large, $N_t \rightarrow \infty$ would not give much insight and is thus not considered in this work.

\subsection{Test bench with real world measurement data}
\label{ssec:tb_results}
The test bench was already introduced in detail in Section~\ref{ssec:test bench}. It allows a predefinition of parameter distributions in order to generate measurement data $\data$. This is used to validate the methods and test them for robustness. In comparison to the artificial setting, inference based on real measurements introduces 
additional challenges, like model discrepancies and complex structured 
measurement noise. 


\subsubsection{Test bench data generation}
\begin{figure}[htbp]
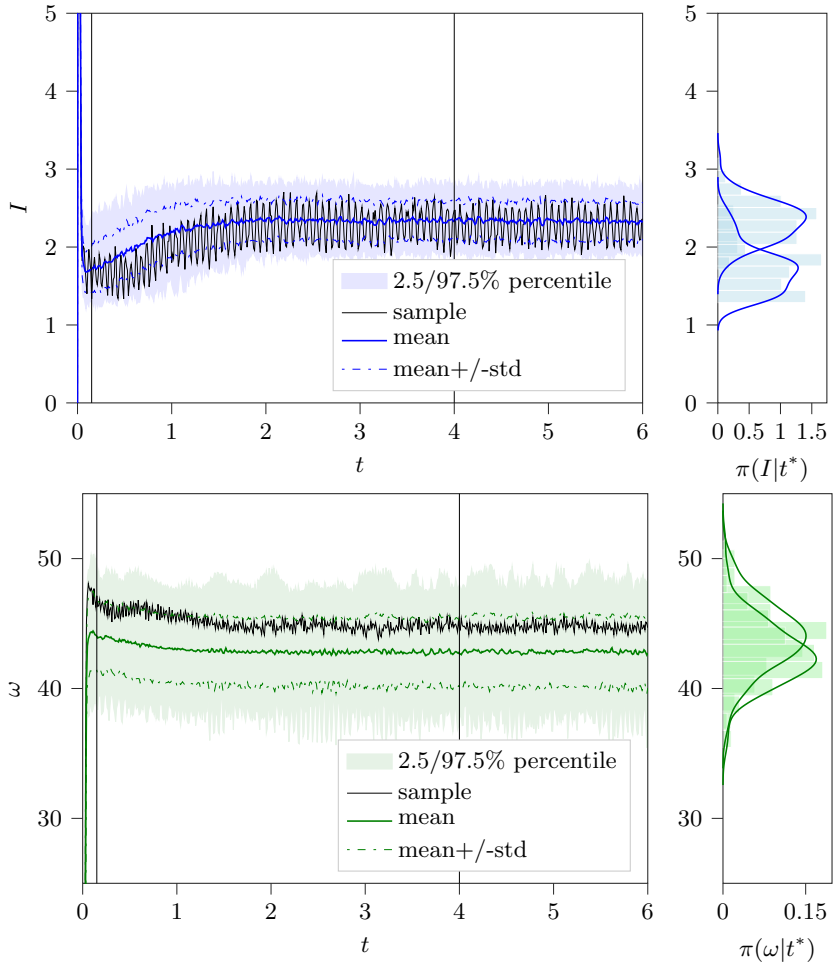

	\centering
	\begin{subfigure}{1\textwidth}
		\raggedleft
		\setlength\figureheight{0.35\textheight} 
		\setlength\figurewidth{1\textwidth}
		\input{Plots/testbench/511_posterior_consistency_delta_data_I.tex}
	\end{subfigure}\\
	\begin{subfigure}{1\textwidth}
		\raggedleft
		\setlength\figureheight{0.35\textheight} 
		\setlength\figurewidth{1\textwidth}
		\input{Plots/testbench/511_posterior_consistency_delta_data_omega.tex}
	\end{subfigure}
	\caption[Test bench measurements]{This figures 
		show the test bench measurements $\data$ of the current $I$ and the rotational speed $\omega$ for $N = 100$. The area between the $2.5\%$ and $97.5\%$ percentile (shaded), mean+/-standard deviation (dash-dotted) and the mean (solid) of all $N$ measurements are depicted. The black lines show an exemplary noisy sample measurement series. Further, at two time points (vertical lines at $t^*=0.15$ and $t^*=4$ seconds) histograms and kernel density plots are displayed on the right hand side.}
	\label{fig:testbench_meas}
\end{figure}

We define reference distributions for the aleatoric parameters $X=(V,T)$ by
$\pi(V \mid m_V, \sigma_V) = \mathcal{N}(m_V, \sigma_V^2)$ 
and $\pi(T \mid m_{T}, \sigma_{T}) = \mathcal{N}(m_{T}, \sigma_{T}^2)$, 
where the hyper-parameters are defined as $m_V = 13.5, m_{T} = 2.5$, $\sigma_V = 0.7, \sigma_{T} = 0.2$. 
%
With this setting, we generate the test bench data $\data$ as described in Section~\ref{ssec:test bench}. $\data$ contains discrete time series of current $I$ and rotational speed $\omega$ each of size $N_t = 601$ in the time interval $[0,6]$ seconds. Later the entire time interval is used for inference. 
An overview of the resulting data $\data$ is visualized in Figure~\ref{fig:testbench_meas}. 

The the noise standard deviations $\sigma^{I}$ and $\sigma^{\omega}$ are estimated by taking the median of the estimations $\overline{\sigma_i^{(\cdot)}}$ for each measurement obtained via Equation~\eqref{eq:noise_est} in the stationary time domain $[4,6]$ seconds. The measurement noise varies for different measurements, which can be observed in Figure~\ref{fig:MCMC_noise_tb}, where a histogram of $\overline{\sigma_i^{(\cdot)}}$ is plotted.

\subsubsection{PCE surrogate}\label{ssec:tb_pce}

The simulation model $\model$ of the the electric motor test bench defined in Section~\ref{ssec:test bench} is computationally too expensive to be used for sampling with a reasonable number of samples. Thus a cheaper to evaluate PCE
surrogate model is used as introduced in Section~\ref{ssec:pce}.
The RMSE scaled by the standard deviation of the validation set for a level $L=5$ PCE (leads to 181 sparse grid points) is in the range of $10^{-3}$ for the current and $10^{-4}$ for the angular velocity. This is sufficient for the following analysis. 
The speed up by using the surrogate for this example is approximately factor 2440. I.e. evaluation of the original model $\model$ for one sample takes $ 1.78 s \pm 192 ms$ (mean $\pm$ std. dev. of 7 runs) and evaluation of the surrogate $\model^{PCE}$ for one sample takes $ 729 \mu s \pm 76.4 \mu s$ (mean $\pm$ std. dev. of 7000 runs).
Note that depending on the implementation vectorized evaluation of the surrogate is possible. E.g. evaluation of the surrogate $\model^{PCE}$ for 100 samples takes $ 1.78 ms \pm 39.3 \mu s$ (mean $\pm$ std. dev. of 700 runs). The speed up compared to the original model is then approximately factor $10^5$.

\subsubsection{Results}
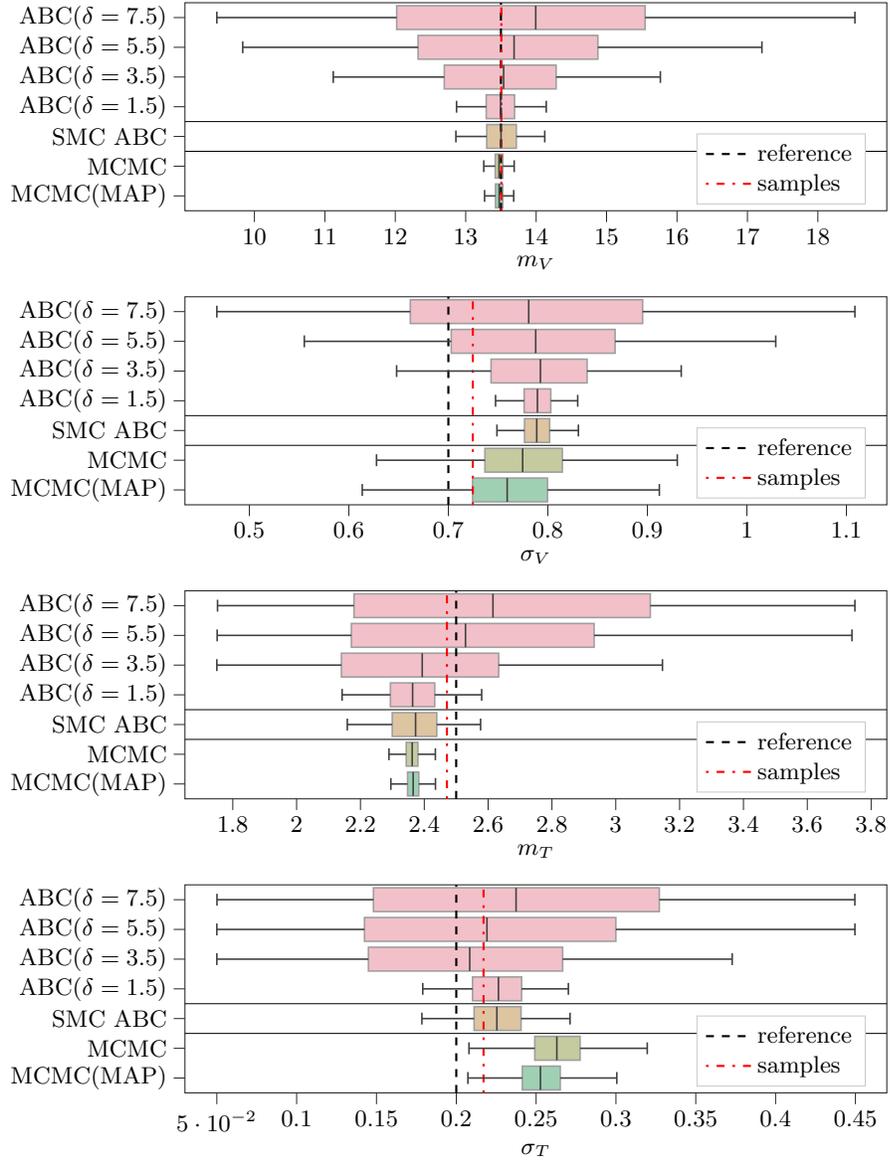
\begin{figure}[htbp]
	\centering\tiny
	\begin{subfigure}{0.9\textwidth}
		\centering
		\setlength\figureheight{0.225\textheight} 
		\setlength\figurewidth{1\textwidth}
\begin{tikzpicture}

\definecolor{color0}{rgb}{0.901982184391939,0.507089934991626,0.577991751552348}
\definecolor{color1}{rgb}{0.732074288267711,0.54808023353899,0.271747550010192}
\definecolor{color2}{rgb}{0.548486066615094,0.585804228120484,0.249549387016444}
\definecolor{color3}{rgb}{0.260050252892592,0.633337204647202,0.411321010000812}

\begin{axis}[
axis line style={white!15.0!black},
height=\figureheight,
legend cell align={left},
legend style={at={(0.97,0.03)}, anchor=south east, draw=white!80.0!black},
tick align=outside,
tick pos=left,
width=\figurewidth,
x grid style={white!80.0!black},
xlabel={\(\displaystyle m_V\)},
xmin=9.01336561833055, xmax=18.9792385625388,
xtick style={color=white!15.0!black},
y grid style={white!80.0!black},
ymin=-0.5, ymax=6.5,
ytick style={color=white!15.0!black},
ytick={0,1,2,3,4,5,6},
y dir=reverse,
yticklabels={ABC(\(\displaystyle \delta=7.5\)),ABC(\(\displaystyle \delta=5.5\)),ABC(\(\displaystyle \delta=3.5\)),ABC(\(\displaystyle \delta=1.5\)),SMC ABC,MCMC,MCMC(MAP)}
]
\path [draw=white!25.098039215686274!black, fill=color0, opacity=0.5, semithick]
(axis cs:12.0202148511582,-0.4)
--(axis cs:12.0202148511582,0.4)
--(axis cs:15.5474113173655,0.4)
--(axis cs:15.5474113173655,-0.4)
--(axis cs:12.0202148511582,-0.4)
--cycle;
\path [draw=white!25.098039215686274!black, fill=color0, opacity=0.5, semithick]
(axis cs:12.3241516326214,0.6)
--(axis cs:12.3241516326214,1.4)
--(axis cs:14.8765127045755,1.4)
--(axis cs:14.8765127045755,0.6)
--(axis cs:12.3241516326214,0.6)
--cycle;
\path [draw=white!25.098039215686274!black, fill=color0, opacity=0.5, semithick]
(axis cs:12.6958345899983,1.6)
--(axis cs:12.6958345899983,2.4)
--(axis cs:14.283628713528,2.4)
--(axis cs:14.283628713528,1.6)
--(axis cs:12.6958345899983,1.6)
--cycle;
\path [draw=white!25.098039215686274!black, fill=color0, opacity=0.5, semithick]
(axis cs:13.2915193092773,2.6)
--(axis cs:13.2915193092773,3.4)
--(axis cs:13.6929146845732,3.4)
--(axis cs:13.6929146845732,2.6)
--(axis cs:13.2915193092773,2.6)
--cycle;
\path [draw=white!25.098039215686274!black, fill=color1, opacity=0.5, semithick]
(axis cs:13.2988472896131,3.6)
--(axis cs:13.2988472896131,4.4)
--(axis cs:13.7184774209351,4.4)
--(axis cs:13.7184774209351,3.6)
--(axis cs:13.2988472896131,3.6)
--cycle;
\path [draw=white!25.098039215686274!black, fill=color2, opacity=0.5, semithick]
(axis cs:13.4191073674098,4.6)
--(axis cs:13.4191073674098,5.4)
--(axis cs:13.5275631606398,5.4)
--(axis cs:13.5275631606398,4.6)
--(axis cs:13.4191073674098,4.6)
--cycle;
\path [draw=white!25.098039215686274!black, fill=color3, opacity=0.5, semithick]
(axis cs:13.423216215347,5.6)
--(axis cs:13.423216215347,6.4)
--(axis cs:13.52668568293,6.4)
--(axis cs:13.52668568293,5.6)
--(axis cs:13.423216215347,5.6)
--cycle;
\addplot [thick, black, dashed]
table {%
13.5 6.5
13.5 -0.5
};
\addlegendentry{reference}
\addplot [thick, red, dash pattern=on 1pt off 3pt on 3pt off 3pt]
table {%
13.5060881971486 6.5
13.5060881971486 -0.5
};
\addlegendentry{samples}
\addplot [very thin, black, forget plot]
table {%
9.01336561833055 3.5
18.9792385625388 3.5
};
\addplot [very thin, black, forget plot]
table {%
9.01336561833055 4.5
18.9792385625388 4.5
};
\addplot [semithick, white!25.098039215686274!black, forget plot]
table {%
12.0202148511582 0
9.46635984306729 0
};
\addplot [semithick, white!25.098039215686274!black, forget plot]
table {%
15.5474113173655 0
18.5262443378021 0
};
\addplot [semithick, white!25.098039215686274!black, forget plot]
table {%
9.46635984306729 -0.2
9.46635984306729 0.2
};
\addplot [semithick, white!25.098039215686274!black, forget plot]
table {%
18.5262443378021 -0.2
18.5262443378021 0.2
};
\addplot [semithick, white!25.098039215686274!black, forget plot]
table {%
12.3241516326214 1
9.83398253693355 1
};
\addplot [semithick, white!25.098039215686274!black, forget plot]
table {%
14.8765127045755 1
17.2056164268554 1
};
\addplot [semithick, white!25.098039215686274!black, forget plot]
table {%
9.83398253693355 0.8
9.83398253693355 1.2
};
\addplot [semithick, white!25.098039215686274!black, forget plot]
table {%
17.2056164268554 0.8
17.2056164268554 1.2
};
\addplot [semithick, white!25.098039215686274!black, forget plot]
table {%
12.6958345899983 2
11.1193563946231 2
};
\addplot [semithick, white!25.098039215686274!black, forget plot]
table {%
14.283628713528 2
15.7639236951488 2
};
\addplot [semithick, white!25.098039215686274!black, forget plot]
table {%
11.1193563946231 1.8
11.1193563946231 2.2
};
\addplot [semithick, white!25.098039215686274!black, forget plot]
table {%
15.7639236951488 1.8
15.7639236951488 2.2
};
\addplot [semithick, white!25.098039215686274!black, forget plot]
table {%
13.2915193092773 3
12.8700004667686 3
};
\addplot [semithick, white!25.098039215686274!black, forget plot]
table {%
13.6929146845732 3
14.143727556755 3
};
\addplot [semithick, white!25.098039215686274!black, forget plot]
table {%
12.8700004667686 2.8
12.8700004667686 3.2
};
\addplot [semithick, white!25.098039215686274!black, forget plot]
table {%
14.143727556755 2.8
14.143727556755 3.2
};
\addplot [semithick, white!25.098039215686274!black, forget plot]
table {%
13.2988472896131 4
12.8607551948336 4
};
\addplot [semithick, white!25.098039215686274!black, forget plot]
table {%
13.7184774209351 4
14.1224732375054 4
};
\addplot [semithick, white!25.098039215686274!black, forget plot]
table {%
12.8607551948336 3.8
12.8607551948336 4.2
};
\addplot [semithick, white!25.098039215686274!black, forget plot]
table {%
14.1224732375054 3.8
14.1224732375054 4.2
};
\addplot [semithick, white!25.098039215686274!black, forget plot]
table {%
13.4191073674098 5
13.2564386854954 5
};
\addplot [semithick, white!25.098039215686274!black, forget plot]
table {%
13.5275631606398 5
13.6882787518575 5
};
\addplot [semithick, white!25.098039215686274!black, forget plot]
table {%
13.2564386854954 4.8
13.2564386854954 5.2
};
\addplot [semithick, white!25.098039215686274!black, forget plot]
table {%
13.6882787518575 4.8
13.6882787518575 5.2
};
\addplot [semithick, white!25.098039215686274!black, forget plot]
table {%
13.423216215347 6
13.2680694707177 6
};
\addplot [semithick, white!25.098039215686274!black, forget plot]
table {%
13.52668568293 6
13.6813793538537 6
};
\addplot [semithick, white!25.098039215686274!black, forget plot]
table {%
13.2680694707177 5.8
13.2680694707177 6.2
};
\addplot [semithick, white!25.098039215686274!black, forget plot]
table {%
13.6813793538537 5.8
13.6813793538537 6.2
};
\addplot [semithick, white!25.098039215686274!black, forget plot]
table {%
13.9941231519209 -0.4
13.9941231519209 0.4
};
\addplot [semithick, white!25.098039215686274!black, forget plot]
table {%
13.6867010444932 0.6
13.6867010444932 1.4
};
\addplot [semithick, white!25.098039215686274!black, forget plot]
table {%
13.5382877296476 1.6
13.5382877296476 2.4
};
\addplot [semithick, white!25.098039215686274!black, forget plot]
table {%
13.4949802923416 2.6
13.4949802923416 3.4
};
\addplot [semithick, white!25.098039215686274!black, forget plot]
table {%
13.5015210635666 3.6
13.5015210635666 4.4
};
\addplot [semithick, white!25.098039215686274!black, forget plot]
table {%
13.4731665906114 4.6
13.4731665906114 5.4
};
\addplot [semithick, white!25.098039215686274!black, forget plot]
table {%
13.4751780990662 5.6
13.4751780990662 6.4
};
\end{axis}

\end{tikzpicture}
	\end{subfigure}\\
	\begin{subfigure}{0.9\textwidth}
		\centering
		\setlength\figureheight{0.225\textheight} 
		\setlength\figurewidth{1\textwidth}
\begin{tikzpicture}

\definecolor{color0}{rgb}{0.901982184391939,0.507089934991626,0.577991751552348}
\definecolor{color1}{rgb}{0.732074288267711,0.54808023353899,0.271747550010192}
\definecolor{color2}{rgb}{0.548486066615094,0.585804228120484,0.249549387016444}
\definecolor{color3}{rgb}{0.260050252892592,0.633337204647202,0.411321010000812}

\begin{axis}[
axis line style={white!15.0!black},
height=\figureheight,
legend cell align={left},
legend style={at={(0.97,0.03)}, anchor=south east, draw=white!80.0!black},
tick align=outside,
tick pos=left,
width=\figurewidth,
x grid style={white!80.0!black},
xlabel={\(\displaystyle \sigma_V\)},
xmin=0.435252329469937, xmax=1.14063034363784,
xtick style={color=white!15.0!black},
y grid style={white!80.0!black},
ymin=-0.5, ymax=6.5,
ytick style={color=white!15.0!black},
ytick={0,1,2,3,4,5,6},
y dir=reverse,
yticklabels={ABC(\(\displaystyle \delta=7.5\)),ABC(\(\displaystyle \delta=5.5\)),ABC(\(\displaystyle \delta=3.5\)),ABC(\(\displaystyle \delta=1.5\)),SMC ABC,MCMC,MCMC(MAP)}
]
\path [draw=white!25.098039215686274!black, fill=color0, opacity=0.5, semithick]
(axis cs:0.661748665957165,-0.4)
--(axis cs:0.661748665957165,0.4)
--(axis cs:0.895199059414287,0.4)
--(axis cs:0.895199059414287,-0.4)
--(axis cs:0.661748665957165,-0.4)
--cycle;
\path [draw=white!25.098039215686274!black, fill=color0, opacity=0.5, semithick]
(axis cs:0.702881026376353,0.6)
--(axis cs:0.702881026376353,1.4)
--(axis cs:0.867531318111347,1.4)
--(axis cs:0.867531318111347,0.6)
--(axis cs:0.702881026376353,0.6)
--cycle;
\path [draw=white!25.098039215686274!black, fill=color0, opacity=0.5, semithick]
(axis cs:0.742904662879729,1.6)
--(axis cs:0.742904662879729,2.4)
--(axis cs:0.839398832484503,2.4)
--(axis cs:0.839398832484503,1.6)
--(axis cs:0.742904662879729,1.6)
--cycle;
\path [draw=white!25.098039215686274!black, fill=color0, opacity=0.5, semithick]
(axis cs:0.776167947395,2.6)
--(axis cs:0.776167947395,3.4)
--(axis cs:0.802927055726145,3.4)
--(axis cs:0.802927055726145,2.6)
--(axis cs:0.776167947395,2.6)
--cycle;
\path [draw=white!25.098039215686274!black, fill=color1, opacity=0.5, semithick]
(axis cs:0.776537539180066,3.6)
--(axis cs:0.776537539180066,4.4)
--(axis cs:0.801747945636878,4.4)
--(axis cs:0.801747945636878,3.6)
--(axis cs:0.776537539180066,3.6)
--cycle;
\path [draw=white!25.098039215686274!black, fill=color2, opacity=0.5, semithick]
(axis cs:0.736722053713968,4.6)
--(axis cs:0.736722053713968,5.4)
--(axis cs:0.814510722335005,5.4)
--(axis cs:0.814510722335005,4.6)
--(axis cs:0.736722053713968,4.6)
--cycle;
\path [draw=white!25.098039215686274!black, fill=color3, opacity=0.5, semithick]
(axis cs:0.724336098100569,5.6)
--(axis cs:0.724336098100569,6.4)
--(axis cs:0.799592986487375,6.4)
--(axis cs:0.799592986487375,5.6)
--(axis cs:0.724336098100569,5.6)
--cycle;
\addplot [thick, black, dashed]
table {%
0.7 6.5
0.7 -0.5
};
\addlegendentry{reference}
\addplot [thick, red, dash pattern=on 1pt off 3pt on 3pt off 3pt]
table {%
0.724558827029191 6.5
0.724558827029191 -0.5
};
\addlegendentry{samples}
\addplot [very thin, black, forget plot]
table {%
0.435252329469937 3.5
1.14063034363784 3.5
};
\addplot [very thin, black, forget plot]
table {%
0.435252329469937 4.5
1.14063034363784 4.5
};
\addplot [semithick, white!25.098039215686274!black, forget plot]
table {%
0.661748665957165 0
0.467314966477569 0
};
\addplot [semithick, white!25.098039215686274!black, forget plot]
table {%
0.895199059414287 0
1.10856770663021 0
};
\addplot [semithick, white!25.098039215686274!black, forget plot]
table {%
0.467314966477569 -0.2
0.467314966477569 0.2
};
\addplot [semithick, white!25.098039215686274!black, forget plot]
table {%
1.10856770663021 -0.2
1.10856770663021 0.2
};
\addplot [semithick, white!25.098039215686274!black, forget plot]
table {%
0.702881026376353 1
0.555212832806473 1
};
\addplot [semithick, white!25.098039215686274!black, forget plot]
table {%
0.867531318111347 1
1.02900094985026 1
};
\addplot [semithick, white!25.098039215686274!black, forget plot]
table {%
0.555212832806473 0.8
0.555212832806473 1.2
};
\addplot [semithick, white!25.098039215686274!black, forget plot]
table {%
1.02900094985026 0.8
1.02900094985026 1.2
};
\addplot [semithick, white!25.098039215686274!black, forget plot]
table {%
0.742904662879729 2
0.647974692649679 2
};
\addplot [semithick, white!25.098039215686274!black, forget plot]
table {%
0.839398832484503 2
0.934032552091688 2
};
\addplot [semithick, white!25.098039215686274!black, forget plot]
table {%
0.647974692649679 1.8
0.647974692649679 2.2
};
\addplot [semithick, white!25.098039215686274!black, forget plot]
table {%
0.934032552091688 1.8
0.934032552091688 2.2
};
\addplot [semithick, white!25.098039215686274!black, forget plot]
table {%
0.776167947395 3
0.747416053888716 3
};
\addplot [semithick, white!25.098039215686274!black, forget plot]
table {%
0.802927055726145 3
0.829971706717116 3
};
\addplot [semithick, white!25.098039215686274!black, forget plot]
table {%
0.747416053888716 2.8
0.747416053888716 3.2
};
\addplot [semithick, white!25.098039215686274!black, forget plot]
table {%
0.829971706717116 2.8
0.829971706717116 3.2
};
\addplot [semithick, white!25.098039215686274!black, forget plot]
table {%
0.776537539180066 4
0.748818791483963 4
};
\addplot [semithick, white!25.098039215686274!black, forget plot]
table {%
0.801747945636878 4
0.830671175650064 4
};
\addplot [semithick, white!25.098039215686274!black, forget plot]
table {%
0.748818791483963 3.8
0.748818791483963 4.2
};
\addplot [semithick, white!25.098039215686274!black, forget plot]
table {%
0.830671175650064 3.8
0.830671175650064 4.2
};
\addplot [semithick, white!25.098039215686274!black, forget plot]
table {%
0.736722053713968 5
0.627772444156228 5
};
\addplot [semithick, white!25.098039215686274!black, forget plot]
table {%
0.814510722335005 5
0.930055494112673 5
};
\addplot [semithick, white!25.098039215686274!black, forget plot]
table {%
0.627772444156228 4.8
0.627772444156228 5.2
};
\addplot [semithick, white!25.098039215686274!black, forget plot]
table {%
0.930055494112673 4.8
0.930055494112673 5.2
};
\addplot [semithick, white!25.098039215686274!black, forget plot]
table {%
0.724336098100569 6
0.613435096625014 6
};
\addplot [semithick, white!25.098039215686274!black, forget plot]
table {%
0.799592986487375 6
0.912032056968479 6
};
\addplot [semithick, white!25.098039215686274!black, forget plot]
table {%
0.613435096625014 5.8
0.613435096625014 6.2
};
\addplot [semithick, white!25.098039215686274!black, forget plot]
table {%
0.912032056968479 5.8
0.912032056968479 6.2
};
\addplot [semithick, white!25.098039215686274!black, forget plot]
table {%
0.780738267115723 -0.4
0.780738267115723 0.4
};
\addplot [semithick, white!25.098039215686274!black, forget plot]
table {%
0.787759908620813 0.6
0.787759908620813 1.4
};
\addplot [semithick, white!25.098039215686274!black, forget plot]
table {%
0.792520314936043 1.6
0.792520314936043 2.4
};
\addplot [semithick, white!25.098039215686274!black, forget plot]
table {%
0.789505105500501 2.6
0.789505105500501 3.4
};
\addplot [semithick, white!25.098039215686274!black, forget plot]
table {%
0.788801117972485 3.6
0.788801117972485 4.4
};
\addplot [semithick, white!25.098039215686274!black, forget plot]
table {%
0.774656407340232 4.6
0.774656407340232 5.4
};
\addplot [semithick, white!25.098039215686274!black, forget plot]
table {%
0.759092815460897 5.6
0.759092815460897 6.4
};
\end{axis}

\end{tikzpicture}
	\end{subfigure}\\
	\begin{subfigure}{0.9\textwidth}
		\centering
		\setlength\figureheight{0.225\textheight} 
		\setlength\figurewidth{1\textwidth}
\begin{tikzpicture}

\definecolor{color0}{rgb}{0.901982184391939,0.507089934991626,0.577991751552348}
\definecolor{color1}{rgb}{0.732074288267711,0.54808023353899,0.271747550010192}
\definecolor{color2}{rgb}{0.548486066615094,0.585804228120484,0.249549387016444}
\definecolor{color3}{rgb}{0.260050252892592,0.633337204647202,0.411321010000812}

\begin{axis}[
axis line style={white!15.0!black},
height=\figureheight,
legend cell align={left},
legend style={at={(0.97,0.03)}, anchor=south east, draw=white!80.0!black},
tick align=outside,
tick pos=left,
width=\figurewidth,
x grid style={white!80.0!black},
xlabel={\(\displaystyle m_{T}\)},
xmin=1.65012768556398, xmax=3.84920610706529,
xtick style={color=white!15.0!black},
y grid style={white!80.0!black},
ymin=-0.5, ymax=6.5,
ytick style={color=white!15.0!black},
ytick={0,1,2,3,4,5,6},
y dir=reverse,
yticklabels={ABC(\(\displaystyle \delta=7.5\)),ABC(\(\displaystyle \delta=5.5\)),ABC(\(\displaystyle \delta=3.5\)),ABC(\(\displaystyle \delta=1.5\)),SMC ABC,MCMC,MCMC(MAP)}
]
\path [draw=white!25.098039215686274!black, fill=color0, opacity=0.5, semithick]
(axis cs:2.17977519899349,-0.4)
--(axis cs:2.17977519899349,0.4)
--(axis cs:3.10813331849339,0.4)
--(axis cs:3.10813331849339,-0.4)
--(axis cs:2.17977519899349,-0.4)
--cycle;
\path [draw=white!25.098039215686274!black, fill=color0, opacity=0.5, semithick]
(axis cs:2.17161197261838,0.6)
--(axis cs:2.17161197261838,1.4)
--(axis cs:2.93207062417272,1.4)
--(axis cs:2.93207062417272,0.6)
--(axis cs:2.17161197261838,0.6)
--cycle;
\path [draw=white!25.098039215686274!black, fill=color0, opacity=0.5, semithick]
(axis cs:2.1404749216307,1.6)
--(axis cs:2.1404749216307,2.4)
--(axis cs:2.6330148786051,2.4)
--(axis cs:2.6330148786051,1.6)
--(axis cs:2.1404749216307,1.6)
--cycle;
\path [draw=white!25.098039215686274!black, fill=color0, opacity=0.5, semithick]
(axis cs:2.29385747818613,2.6)
--(axis cs:2.29385747818613,3.4)
--(axis cs:2.43267429250282,3.4)
--(axis cs:2.43267429250282,2.6)
--(axis cs:2.29385747818613,2.6)
--cycle;
\path [draw=white!25.098039215686274!black, fill=color1, opacity=0.5, semithick]
(axis cs:2.29968802287778,3.6)
--(axis cs:2.29968802287778,4.4)
--(axis cs:2.43909262549719,4.4)
--(axis cs:2.43909262549719,3.6)
--(axis cs:2.29968802287778,3.6)
--cycle;
\path [draw=white!25.098039215686274!black, fill=color2, opacity=0.5, semithick]
(axis cs:2.34343745420294,4.6)
--(axis cs:2.34343745420294,5.4)
--(axis cs:2.37998204747456,5.4)
--(axis cs:2.37998204747456,4.6)
--(axis cs:2.34343745420294,4.6)
--cycle;
\path [draw=white!25.098039215686274!black, fill=color3, opacity=0.5, semithick]
(axis cs:2.34764112799702,5.6)
--(axis cs:2.34764112799702,6.4)
--(axis cs:2.38278187687102,6.4)
--(axis cs:2.38278187687102,5.6)
--(axis cs:2.34764112799702,5.6)
--cycle;
\addplot [thick, black, dashed]
table {%
2.5 6.5
2.5 -0.5
};
\addlegendentry{reference}
\addplot [thick, red, dash pattern=on 1pt off 3pt on 3pt off 3pt]
table {%
2.47097080833574 6.5
2.47097080833574 -0.5
};
\addlegendentry{samples}
\addplot [very thin, black, forget plot]
table {%
1.65012768556398 3.5
3.84920610706529 3.5
};
\addplot [very thin, black, forget plot]
table {%
1.65012768556398 4.5
3.84920610706529 4.5
};
\addplot [semithick, white!25.098039215686274!black, forget plot]
table {%
2.17977519899349 0
1.7522284527927 0
};
\addplot [semithick, white!25.098039215686274!black, forget plot]
table {%
3.10813331849339 0
3.74924799699705 0
};
\addplot [semithick, white!25.098039215686274!black, forget plot]
table {%
1.7522284527927 -0.2
1.7522284527927 0.2
};
\addplot [semithick, white!25.098039215686274!black, forget plot]
table {%
3.74924799699705 -0.2
3.74924799699705 0.2
};
\addplot [semithick, white!25.098039215686274!black, forget plot]
table {%
2.17161197261838 1
1.75103899887897 1
};
\addplot [semithick, white!25.098039215686274!black, forget plot]
table {%
2.93207062417272 1
3.74001185226571 1
};
\addplot [semithick, white!25.098039215686274!black, forget plot]
table {%
1.75103899887897 0.8
1.75103899887897 1.2
};
\addplot [semithick, white!25.098039215686274!black, forget plot]
table {%
3.74001185226571 0.8
3.74001185226571 1.2
};
\addplot [semithick, white!25.098039215686274!black, forget plot]
table {%
2.1404749216307 2
1.75008579563222 2
};
\addplot [semithick, white!25.098039215686274!black, forget plot]
table {%
2.6330148786051 2
3.1462667013891 2
};
\addplot [semithick, white!25.098039215686274!black, forget plot]
table {%
1.75008579563222 1.8
1.75008579563222 2.2
};
\addplot [semithick, white!25.098039215686274!black, forget plot]
table {%
3.1462667013891 1.8
3.1462667013891 2.2
};
\addplot [semithick, white!25.098039215686274!black, forget plot]
table {%
2.29385747818613 3
2.14256672010947 3
};
\addplot [semithick, white!25.098039215686274!black, forget plot]
table {%
2.43267429250282 3
2.57982524615957 3
};
\addplot [semithick, white!25.098039215686274!black, forget plot]
table {%
2.14256672010947 2.8
2.14256672010947 3.2
};
\addplot [semithick, white!25.098039215686274!black, forget plot]
table {%
2.57982524615957 2.8
2.57982524615957 3.2
};
\addplot [semithick, white!25.098039215686274!black, forget plot]
table {%
2.29968802287778 4
2.15882702153804 4
};
\addplot [semithick, white!25.098039215686274!black, forget plot]
table {%
2.43909262549719 4
2.57642086559621 4
};
\addplot [semithick, white!25.098039215686274!black, forget plot]
table {%
2.15882702153804 3.8
2.15882702153804 4.2
};
\addplot [semithick, white!25.098039215686274!black, forget plot]
table {%
2.57642086559621 3.8
2.57642086559621 4.2
};
\addplot [semithick, white!25.098039215686274!black, forget plot]
table {%
2.34343745420294 5
2.28939836241511 5
};
\addplot [semithick, white!25.098039215686274!black, forget plot]
table {%
2.37998204747456 5
2.43478137627722 5
};
\addplot [semithick, white!25.098039215686274!black, forget plot]
table {%
2.28939836241511 4.8
2.28939836241511 5.2
};
\addplot [semithick, white!25.098039215686274!black, forget plot]
table {%
2.43478137627722 4.8
2.43478137627722 5.2
};
\addplot [semithick, white!25.098039215686274!black, forget plot]
table {%
2.34764112799702 6
2.2953328496495 6
};
\addplot [semithick, white!25.098039215686274!black, forget plot]
table {%
2.38278187687102 6
2.43506863759944 6
};
\addplot [semithick, white!25.098039215686274!black, forget plot]
table {%
2.2953328496495 5.8
2.2953328496495 6.2
};
\addplot [semithick, white!25.098039215686274!black, forget plot]
table {%
2.43506863759944 5.8
2.43506863759944 6.2
};
\addplot [semithick, white!25.098039215686274!black, forget plot]
table {%
2.61520977615033 -0.4
2.61520977615033 0.4
};
\addplot [semithick, white!25.098039215686274!black, forget plot]
table {%
2.52915019854115 0.6
2.52915019854115 1.4
};
\addplot [semithick, white!25.098039215686274!black, forget plot]
table {%
2.39362184399738 1.6
2.39362184399738 2.4
};
\addplot [semithick, white!25.098039215686274!black, forget plot]
table {%
2.36343568009505 2.6
2.36343568009505 3.4
};
\addplot [semithick, white!25.098039215686274!black, forget plot]
table {%
2.37279322354536 3.6
2.37279322354536 4.4
};
\addplot [semithick, white!25.098039215686274!black, forget plot]
table {%
2.36217839828779 4.6
2.36217839828779 5.4
};
\addplot [semithick, white!25.098039215686274!black, forget plot]
table {%
2.365184551296 5.6
2.365184551296 6.4
};
\end{axis}

\end{tikzpicture}
	\end{subfigure}\\
	\begin{subfigure}{0.9\textwidth}
		\centering
		\setlength\figureheight{0.225\textheight} 
		\setlength\figurewidth{1\textwidth}
\begin{tikzpicture}

\definecolor{color0}{rgb}{0.901982184391939,0.507089934991626,0.577991751552348}
\definecolor{color1}{rgb}{0.732074288267711,0.54808023353899,0.271747550010192}
\definecolor{color2}{rgb}{0.548486066615094,0.585804228120484,0.249549387016444}
\definecolor{color3}{rgb}{0.260050252892592,0.633337204647202,0.411321010000812}

\begin{axis}[
axis line style={white!15.0!black},
height=\figureheight,
legend cell align={left},
legend style={at={(0.97,0.03)}, anchor=south east, draw=white!80.0!black},
tick align=outside,
tick pos=left,
width=\figurewidth,
x grid style={white!80.0!black},
xlabel={\(\displaystyle \sigma_{T}\)},
xmin=0.0300256763191912, xmax=0.469679033443805,
xtick style={color=white!15.0!black},
y grid style={white!80.0!black},
ymin=-0.5, ymax=6.5,
ytick style={color=white!15.0!black},
ytick={0,1,2,3,4,5,6},
y dir=reverse,
yticklabels={ABC(\(\displaystyle \delta=7.5\)),ABC(\(\displaystyle \delta=5.5\)),ABC(\(\displaystyle \delta=3.5\)),ABC(\(\displaystyle \delta=1.5\)),SMC ABC,MCMC,MCMC(MAP)}
]
\path [draw=white!25.098039215686274!black, fill=color0, opacity=0.5, semithick]
(axis cs:0.148102852838094,-0.4)
--(axis cs:0.148102852838094,0.4)
--(axis cs:0.327179565596602,0.4)
--(axis cs:0.327179565596602,-0.4)
--(axis cs:0.148102852838094,-0.4)
--cycle;
\path [draw=white!25.098039215686274!black, fill=color0, opacity=0.5, semithick]
(axis cs:0.142495315619796,0.6)
--(axis cs:0.142495315619796,1.4)
--(axis cs:0.299943928212054,1.4)
--(axis cs:0.299943928212054,0.6)
--(axis cs:0.142495315619796,0.6)
--cycle;
\path [draw=white!25.098039215686274!black, fill=color0, opacity=0.5, semithick]
(axis cs:0.144958112016719,1.6)
--(axis cs:0.144958112016719,2.4)
--(axis cs:0.266618507797087,2.4)
--(axis cs:0.266618507797087,1.6)
--(axis cs:0.144958112016719,1.6)
--cycle;
\path [draw=white!25.098039215686274!black, fill=color0, opacity=0.5, semithick]
(axis cs:0.21018573021215,2.6)
--(axis cs:0.21018573021215,3.4)
--(axis cs:0.240897997089402,3.4)
--(axis cs:0.240897997089402,2.6)
--(axis cs:0.21018573021215,2.6)
--cycle;
\path [draw=white!25.098039215686274!black, fill=color1, opacity=0.5, semithick]
(axis cs:0.211151785200163,3.6)
--(axis cs:0.211151785200163,4.4)
--(axis cs:0.240563167858911,4.4)
--(axis cs:0.240563167858911,3.6)
--(axis cs:0.211151785200163,3.6)
--cycle;
\path [draw=white!25.098039215686274!black, fill=color2, opacity=0.5, semithick]
(axis cs:0.249167095616308,4.6)
--(axis cs:0.249167095616308,5.4)
--(axis cs:0.277540712505797,5.4)
--(axis cs:0.277540712505797,4.6)
--(axis cs:0.249167095616308,4.6)
--cycle;
\path [draw=white!25.098039215686274!black, fill=color3, opacity=0.5, semithick]
(axis cs:0.241401427236189,5.6)
--(axis cs:0.241401427236189,6.4)
--(axis cs:0.265103718188364,6.4)
--(axis cs:0.265103718188364,5.6)
--(axis cs:0.241401427236189,5.6)
--cycle;
\addplot [thick, black, dashed]
table {%
0.2 6.5
0.2 -0.5
};
\addlegendentry{reference}
\addplot [thick, red, dash pattern=on 1pt off 3pt on 3pt off 3pt]
table {%
0.217078557114774 6.5
0.217078557114774 -0.5
};
\addlegendentry{samples}
\addplot [very thin, black, forget plot]
table {%
0.0300256763191911 3.5
0.469679033443805 3.5
};
\addplot [very thin, black, forget plot]
table {%
0.0300256763191911 4.5
0.469679033443805 4.5
};
\addplot [semithick, white!25.098039215686274!black, forget plot]
table {%
0.148102852838094 0
0.0500847709418427 0
};
\addplot [semithick, white!25.098039215686274!black, forget plot]
table {%
0.327179565596602 0
0.449647737736358 0
};
\addplot [semithick, white!25.098039215686274!black, forget plot]
table {%
0.0500847709418427 -0.2
0.0500847709418427 0.2
};
\addplot [semithick, white!25.098039215686274!black, forget plot]
table {%
0.449647737736358 -0.2
0.449647737736358 0.2
};
\addplot [semithick, white!25.098039215686274!black, forget plot]
table {%
0.142495315619796 1
0.0500099198248554 1
};
\addplot [semithick, white!25.098039215686274!black, forget plot]
table {%
0.299943928212054 1
0.44969478993814 1
};
\addplot [semithick, white!25.098039215686274!black, forget plot]
table {%
0.0500099198248554 0.8
0.0500099198248554 1.2
};
\addplot [semithick, white!25.098039215686274!black, forget plot]
table {%
0.44969478993814 0.8
0.44969478993814 1.2
};
\addplot [semithick, white!25.098039215686274!black, forget plot]
table {%
0.144958112016719 2
0.0500099198248554 2
};
\addplot [semithick, white!25.098039215686274!black, forget plot]
table {%
0.266618507797087 2
0.372815484376668 2
};
\addplot [semithick, white!25.098039215686274!black, forget plot]
table {%
0.0500099198248554 1.8
0.0500099198248554 2.2
};
\addplot [semithick, white!25.098039215686274!black, forget plot]
table {%
0.372815484376668 1.8
0.372815484376668 2.2
};
\addplot [semithick, white!25.098039215686274!black, forget plot]
table {%
0.21018573021215 3
0.179000834736631 3
};
\addplot [semithick, white!25.098039215686274!black, forget plot]
table {%
0.240897997089402 3
0.27020125005186 3
};
\addplot [semithick, white!25.098039215686274!black, forget plot]
table {%
0.179000834736631 2.8
0.179000834736631 3.2
};
\addplot [semithick, white!25.098039215686274!black, forget plot]
table {%
0.27020125005186 2.8
0.27020125005186 3.2
};
\addplot [semithick, white!25.098039215686274!black, forget plot]
table {%
0.211151785200163 4
0.178450728321466 4
};
\addplot [semithick, white!25.098039215686274!black, forget plot]
table {%
0.240563167858911 4
0.271240638010153 4
};
\addplot [semithick, white!25.098039215686274!black, forget plot]
table {%
0.178450728321466 3.8
0.178450728321466 4.2
};
\addplot [semithick, white!25.098039215686274!black, forget plot]
table {%
0.271240638010153 3.8
0.271240638010153 4.2
};
\addplot [semithick, white!25.098039215686274!black, forget plot]
table {%
0.249167095616308 5
0.207947442931345 5
};
\addplot [semithick, white!25.098039215686274!black, forget plot]
table {%
0.277540712505797 5
0.319620636101703 5
};
\addplot [semithick, white!25.098039215686274!black, forget plot]
table {%
0.207947442931345 4.8
0.207947442931345 5.2
};
\addplot [semithick, white!25.098039215686274!black, forget plot]
table {%
0.319620636101703 4.8
0.319620636101703 5.2
};
\addplot [semithick, white!25.098039215686274!black, forget plot]
table {%
0.241401427236189 6
0.207285139031036 6
};
\addplot [semithick, white!25.098039215686274!black, forget plot]
table {%
0.265103718188364 6
0.300547577483981 6
};
\addplot [semithick, white!25.098039215686274!black, forget plot]
table {%
0.207285139031036 5.8
0.207285139031036 6.2
};
\addplot [semithick, white!25.098039215686274!black, forget plot]
table {%
0.300547577483981 5.8
0.300547577483981 6.2
};
\addplot [semithick, white!25.098039215686274!black, forget plot]
table {%
0.237434458577811 -0.4
0.237434458577811 0.4
};
\addplot [semithick, white!25.098039215686274!black, forget plot]
table {%
0.219210948396069 0.6
0.219210948396069 1.4
};
\addplot [semithick, white!25.098039215686274!black, forget plot]
table {%
0.208411206777365 1.6
0.208411206777365 2.4
};
\addplot [semithick, white!25.098039215686274!black, forget plot]
table {%
0.226406592048518 2.6
0.226406592048518 3.4
};
\addplot [semithick, white!25.098039215686274!black, forget plot]
table {%
0.225421563979391 3.6
0.225421563979391 4.4
};
\addplot [semithick, white!25.098039215686274!black, forget plot]
table {%
0.262978334570878 4.6
0.262978334570878 5.4
};
\addplot [semithick, white!25.098039215686274!black, forget plot]
table {%
0.252704136301206 5.6
0.252704136301206 6.4
};
\end{axis}

\end{tikzpicture}
	\end{subfigure}%
	\caption[Marginal posterior distributions]{
		Boxplots of the samples approximating the hyper-parameters marginal posterior distributions in the test bench case obtained with ABC, SMC ABC, MCMC and MCMC(MAP) in comparison to the reference values and sample mean and standard deviation.}
	\label{fig:posterior_testbench}
\end{figure}

With almost the same methods setting as in the synthetic case (see Section~\ref{ssec:synthetic}) Figure~\ref{fig:posterior_testbench} presents boxplots of the marginal posterior distributions. Only difference is that for ABC and SMC ABC  $\delta \in \{7.5,\, 5.5,\, 3.5,\, 1.5\}$. 
%
Table~\ref{tab:runtimes_testbench} displays the runtimes, number of proposals and samples. SMC ABC takes less than half of the time than MCMC(MAP).
%
\begin{table}[htbp]
	\centering\small
	\begin{tabular}{llll}
\toprule
{} & runtime [s] & samples $m$ & proposals $M$ \\
\midrule
ABC($\delta=7.5$) &           4 &       1,500 &         6,000 \\
ABC($\delta=5.5$) &           9 &       1,500 &        13,000 \\
ABC($\delta=3.5$) &          41 &       1,500 &        66,000 \\
ABC($\delta=1.5$) &       7,472 &       1,500 &    11,968,000 \\
SMC ABC           &         315 &       1,500 &       484,000 \\
MCMC              &         710 &   $n_{eff}$ &       3*11000 \\
MCMC(MAP)         &         700 &   $n_{eff}$ &       3*11000 \\
\bottomrule
\end{tabular}

	\caption{Runtimes in seconds, number of samples $m$ and proposals $M$ of the considered methods. For MCMC the number of effective samples $n_{eff}$ is given in detail in Table~\ref{tab:mcmc_gelman_rubin_neff_tb}.}
	\label{tab:runtimes_testbench}
\end{table}

The results are comparable to the synthetic case, however there is an offset to the reference values. This is most likely due to the varying noise structure and model error.
Further we notice a difference in the SMC ABC and MCMC(MAP) posterior distributions for $\sigma_V$ and $\sigma_{T}$. This is due to the estimation of the noise standard deviations $\sigma_{I}, \sigma_{\omega}$ and due to the additive structure in the summary statistic (see Equation~\eqref{eq:sumstat_std}).
Figure~\ref{fig:MCMC_noise_tb} shows the marginal MCMC and MCMC(MAP) samples for the noise standard deviations $\sigma^{I}$ and $\sigma^{\omega}$. Note that there is no reference value available. The structure of the noise is not as regular as in the artificial 
setting and can not be perfectly described by the assumption of independent and 
identically distributed noise in time and for every measurement. The MCMC 
estimates are higher than the median of the estimates $\overline{\sigma_i^{(\cdot)}}$ obtained from the stationary time domain, which are used for ABC and SMC ABC. This leads to the difference in the posterior distributions.
In order to improve the MCMC estimation one could introduce a more complex noise model, which would lead to a higher number of parameters and consequently to even more difficulties in sampling.
%
\begin{figure}[htbp]
	\flushleft
	\begin{subfigure}{0.9\textwidth}
		\flushright
		\setlength\figureheight{0.2\textheight} 
		\setlength\figurewidth{0.9\textwidth}
		\input{Plots/testbench/511_posterior_noise_sigma_I.tex}
	\end{subfigure}\\
	\begin{subfigure}{0.9\textwidth}
		\flushright
		\setlength\figureheight{0.2\textheight} 
		\setlength\figurewidth{0.9\textwidth}
		\input{Plots/testbench/511_posterior_noise_sigma_omega.tex}
	\end{subfigure}
	\caption[MCMC posterior for noise standard deviations]{This figure shows the marginal MCMC and MCMC(MAP) samples for the noise standard deviations $\sigma^{I}$ and $\sigma^{\omega}$ for the test bench. Also a histogram of the estimates 
		$\overline{\sigma^{(\cdot)}} := [\overline{\sigma_i^{(\cdot)}}]_{i=1,\dots,N}$ 
		via Equation~\eqref{eq:noise_est} and the corresponding mean and median are displayed.}
	\label{fig:MCMC_noise_tb}
\end{figure}

\begin{figure}[htbp]
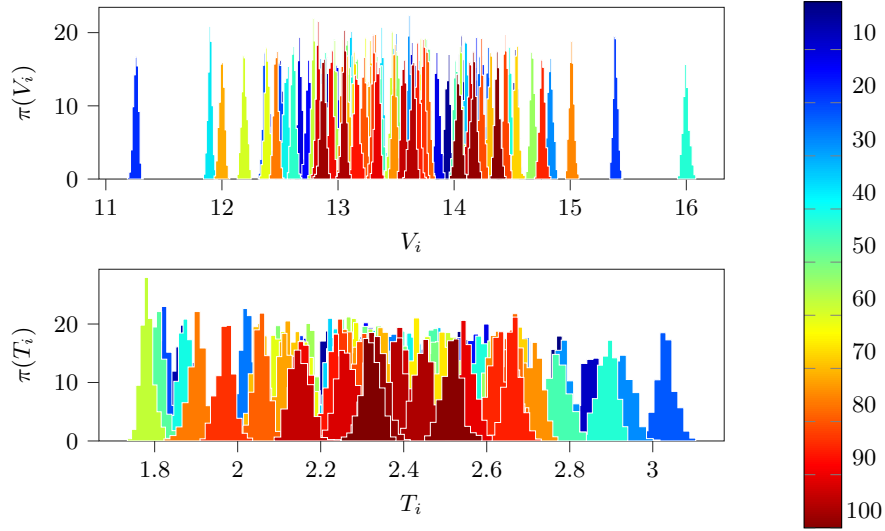

	\centering
	\begin{tabular}[c]{cc}
		\begin{tabular}[c]{c}
			\smallskip
			\begin{subfigure}[c]{0.85\textwidth}
				\centering
				\setlength\figureheight{0.2\textheight} 
				\setlength\figurewidth{0.96\textwidth}
				\input{Plots/testbench/511_posterior_100_U.tex}
			\end{subfigure}\\
			\begin{subfigure}[c]{0.85\textwidth}
				\centering
				\setlength\figureheight{0.2\textheight} 
				\setlength\figurewidth{0.96\textwidth}
				\input{Plots/testbench/511_posterior_100_load.tex}
			\end{subfigure}
		\end{tabular}
		&
		\begin{subfigure}[c]{0.15\textwidth}
			\setlength\figureheight{0.2\textheight} 
			\setlength\figurewidth{0.89\textwidth}
			\begin{tikzpicture}
\begin{axis}[
    hide axis,
    scale only axis,
    height=\figureheight,
    width=\figurewidth,
    colormap/jet,
    colorbar horizontal,
    point meta min=1,
    point meta max=100,
    colorbar style={
        xtick={10,20,30,...,100},
        xticklabel pos=upper,
        rotate=270,
        width=7cm,
        x tick label style={xshift=0.3cm}
    }]
    \addplot [draw=none] coordinates {(0,0)};
\end{axis}
\end{tikzpicture}
		\end{subfigure}
	\end{tabular}
	\caption[MCMC(MAP) estimates for the samples $V_i$ and ${T}_i$ based on test bench data]{This figure shows the marginal MCMC(MAP) samples for $V_i$ and 
		${T}_i$ for $i=1,...,100$ based on test bench data. The color bar indicates the index $i$.}
	\label{fig:MCMC_samples_100_tb}
\end{figure}

Marginal posterior distributions of $V_i$ and $T_{i}$ for $i=1,\dots,N$ obtained via MCMC(MAP) are visualized in Figure \ref{fig:MCMC_samples_100_tb}. 
For $T_i$ they are wider as in the synthetic case which is due to larger noise in the data. This slightly simplifies for MCMC and MCMC(MAP) to generate samples, which can also be observed for the sampler efficiency statistics in Table~\ref{tab:mcmc_gelman_rubin_neff_tb}. Overall we see again a good improvement from MCMC to MCMC(MAP), but even for MCMC(MAP) $n_{eff}$ is still below 15\%. The statistics for MCMC recommend to increase the sample size.
\begin{table}[htbp]
	\centering\small
	\begin{tabular}{lrrrrrrrr}
	\toprule
	{} & \multicolumn{2}{l}{mean($\hat{R}$)} & \multicolumn{6}{c}{$\hat{R}$} \\
	{} &             $V$ &   $T$ &     $m_V$ & $\sigma_V$ & $m_{T}$ & $\sigma_{T}$ & $\sigma_{I}$ & $\sigma_{\omega}$ \\
	\midrule
	MCMC      &           2.257 & 1.835 &     1.003 &      1.033 &   1.000 &        1.169 &        1.270 &             1.219 \\
	MCMC(MAP) &           1.036 & 1.036 &     1.000 &      1.001 &   1.003 &        1.002 &        1.001 &             1.001 \\
\midrule
	{} & \multicolumn{2}{l}{mean($n_{eff}$)} & \multicolumn{6}{c}{$n_{eff}$} \\
	{} &             $V$ & $T$ &     $m_V$ & $\sigma_V$ & $m_{T}$ & $\sigma_{T}$ & $\sigma_{I}$ & $\sigma_{\omega}$ \\
	\midrule
	MCMC      &               4 &   5 &     2,852 &        132 &   1,667 &            7 &            5 &                 6 \\
	MCMC(MAP) &              45 &  49 &     2,085 &      2,138 &   1,804 &        1,985 &        1,297 &             1,237 \\
	\bottomrule
\end{tabular}

	\caption{Gelman-rubin statistics $\hat{R}$ and number of effective samples $n_{eff}$ for MCMC and MCMC(MAP), based on the last 5000 samples of 3 parallel sampled chains, each of total length 11.000, for $N=100$. Note that the values presented for $V$ and $T$ are 
	$mean(\hat{R}(V)) := \frac{1}{N} \sum_{i=1}^{N} \hat{R}(V_i)$, 
	respectively for $T$ (analog for $n_{eff}$).}
	\label{tab:mcmc_gelman_rubin_neff_tb}
\end{table}

%
%

\section{Conclusions}\label{ch:conlusion}


This work presents methods for inference of parameter distributions from noisy time series data. The introduced methods are a hierarchical surrogate-based MCMC approach and a hierarchical surrogate-based ABC approach with summary statistics. For both methods an important ingredient is a PCE surrogate, which allows sampling in the first place by drastically speeding up the model evaluations.
Additionally, ABC speeds up the inference further by exploiting parametric assumptions on the unknown parameter distributions and using summary statistics together with sparse grid quadrature. This is in particular effective for large data and consequently high dimensional parameter spaces for our examples.

Application to a complex, real world industrial example, i.e.\ an electric motor, show the effectiveness and robustness of the methods. We first analyze the methods in detail for synthetic data with a basic model and then test them for robustness with real world data from an electric motor test bench hardware. 

With Theorem~\ref{thm:consistency} we show posterior consistency for the surrogate-based ABC method and also illustrate this with numerical results in the synthetic case. Of course, plain rejection ABC with small threshold $\delta$ is very inefficient, but nevertheless obtains a decent result. In order to get a rough overview on posterior distribution, ABC with moderate $\delta$ is extremely fast. This fact is further exploited by using a population based SMC ABC method with a sequence of decreasing $\delta$'s, where we obtain results comparable to plain rejection ABC with small $\delta$ with roughly 90\% less samples. 

For an increase of data size and consequently increasing number of parameters in our examples the ABC method performs better for almost constant computational cost. This is because the summary statistics and noise standard deviation estimation quality improve, which improves ABC sampling. For MCMC it is vice versa: performance decreases and also computational costs raise, due to the high dimensions and the concentration effect. The MCMC performance can be improved by providing curvature information and appropriate starting values (with MAP and inverse Hessian initialization), but still the samples are highly autocorrelated.
Even without further improvements, the surrogate-based ABC approach with summary statistics is significantly faster than the MCMC approach and shows results of similar accuracy.
Additionally, ABC is embarrassingly parallel, thus further multiprocessing would easily yield additional efficiency gains, whereas potential efficiency gain for MCMC is lower. Here other gradient free sampling methods might yield advantages.

In the real world case we observe similar results, however due to the more complex noise structure and due to model error a deviation of the posterior from reference values can be observed. 
Future work might include a correction for model discrepancy in the real world application to correct the inference of the parameter distributions, as it was done for example in \cite{john2018emclpp} for one deterministic parameter. However, this results in additional identification problems leading to a far more difficult problem to solve. Allowing for larger classes of distributions as well as for more realistic noise models, the resulting inverse problem becomes much more complex. However, the presented method showed a significant speed-up compared to other state-of-the-art methods and is a promising direction also for distribution robust approaches.

The inference results of the ABC method are strongly influenced by the choice of the 
summary statistics as well as by the distance function. With multiple input distributions and multiple output quantities of interest scaling of the summary statistics is important. E.g. in \cite{jung2011choice} they  showed that appropriate weighting improves posterior approximation. In our example the outputs are much more sensitive to variations of the voltage than to variations of the mechanical loading. This issue could be addressed in future work with a  sensitivity analysis in order to choose an appropriate distance function. For example, Sobol sensitivity indices \cite{sobol2001global,saltelli2008global} could be used to weight the distance function pro-rata.

In this work we used the mean and standard deviation as summary statistics. However, the applied measurements are only roughly Gaussian distributed for every considered 
point in time. In future work, other statistics could be included to the summary 
statistics in order to increase the information content and further improve the ABC results. However, the selection of summary statistics is non-trivial as already discussed in the introduction. An optimal selection of summary statistics based on minimum entropy  \cite{Nunes} was computationally not feasible for this work.

Similarly, we estimated only the first and second moment of the unknown aleatoric parameters. Future work might address higher moment approximation to infer a broader class 
of probability distributions.

In summary, albeit the presented surrogate-based ABC method with summary statistics has still some open points to work on, it already provides results that are in terms of accuracy comparable to those of the hierarchical surrogate-based MCMC method in much less time, in particular for large data sets and high dimensional state spaces. We showed that in cases where the likelihood is expensive to evaluate or even not available, ABC methods are an efficient way to obtain approximate posterior distributions with comparable quality. Further ABC with summary statistics are an important tool to deal with high dimensions.  

\section*{Acknowledgments}
We thank Bosch Research for partially funding this work. ClS would like to thank the Isaac Newton Institute for Mathematical Sciences for support and hospitality during the program \textit{Uncertainty quantification for complex systems: theory and methodologies} when work on this paper was undertaken. This work was supported by: EPSRC grant numbers EP/K032208/1 and EP/R014604/1.
Further, we would like to thank Philipp Glaser for providing the test bench hardware and Daniel Schwarzer for valuable comments during proof reading.


\bibliography{mybibfile}

\begin{thebibliography}{10}
\expandafter\ifx\csname url\endcsname\relax
  \def\url#1{\texttt{#1}}\fi
\expandafter\ifx\csname urlprefix\endcsname\relax\def\urlprefix{URL }\fi
\expandafter\ifx\csname href\endcsname\relax
  \def\href#1#2{#2} \def\path#1{#1}\fi

\bibitem{Stuart.2010}
A.~M. Stuart, {Inverse problems: A Bayesian perspective}, {Acta Numerica} 19
  (2010) 451--559.
\newblock \href {http://dx.doi.org/10.1017/S0962492910000061}
  {\path{doi:10.1017/S0962492910000061}}.

\bibitem{Dashti.2017}
M.~Dashti, A.~M. Stuart,
  \href{{https://doi.org/10.1007/978-3-319-12385-1_7}}{{The Bayesian Approach
  to Inverse Problems}}, in: R.~Ghanem, D.~Higdon, H.~Owhadi (Eds.), {Handbook
  of Uncertainty Quantification}, {Springer International Publishing}, Cham,
  2017, pp. 311--428.
\newblock \href {http://dx.doi.org/10.1007/978-3-319-12385-1_7}
  {\path{doi:10.1007/978-3-319-12385-1_7}}.
\newline\urlprefix\url{{https://doi.org/10.1007/978-3-319-12385-1_7}}

\bibitem{Kaipio.2005}
J.~Kaipio, E.~Somersalo, {Statistical and Computational Inverse Problems},
  {Applied Mathematical Sciences}, {Springer Science+Business Media, Inc}, New
  York, NY, 2005.

\bibitem{Robert:2005:MCS:1051451}
C.~P. Robert, G.~Casella, Monte Carlo Statistical Methods (Springer Texts in
  Statistics), Springer-Verlag, Berlin, Heidelberg, 2005.

\bibitem{Wilkinson}
R.~D. Wilkinson, {Approximate Bayesian computation (ABC) gives exact results
  under the assumption of model error}, Statistical Applications in Genetics
  and Molecular Biology 12~(2) (2013) 129--141.
\newblock \href {http://dx.doi.org/10.1515/sagmb-2013-0010}
  {\path{doi:10.1515/sagmb-2013-0010}}.

\bibitem{Tavare505}
S.~Tavar{\'e}, D.~J. Balding, R.~C. Griffiths, P.~Donnelly,
  \href{http://www.genetics.org/content/145/2/505}{{Inferring Coalescence Times
  From DNA Sequence Data}}, Genetics 145~(2) (1997) 505--518.
\newline\urlprefix\url{http://www.genetics.org/content/145/2/505}

\bibitem{Marjoram15324}
P.~Marjoram, J.~Molitor, V.~Plagnol, S.~Tavar{\'e},
  \href{https://www.pnas.org/content/100/26/15324}{{Markov chain Monte Carlo
  without likelihoods}}, Proceedings of the National Academy of Sciences
  100~(26) (2003) 15324--15328.
\newblock \href {http://dx.doi.org/10.1073/pnas.0306899100}
  {\path{doi:10.1073/pnas.0306899100}}.
\newline\urlprefix\url{https://www.pnas.org/content/100/26/15324}

\bibitem{Sisson1760}
S.~A. Sisson, Y.~Fan, M.~M. Tanaka,
  \href{https://www.pnas.org/content/104/6/1760}{{Sequential Monte Carlo
  without likelihoods}}, Proceedings of the National Academy of Sciences
  104~(6) (2007) 1760--1765.
\newblock \href {http://dx.doi.org/10.1073/pnas.0607208104}
  {\path{doi:10.1073/pnas.0607208104}}.
\newline\urlprefix\url{https://www.pnas.org/content/104/6/1760}

\bibitem{10.1093/sysbio/syq054}
R.~D. Wilkinson, M.~E. Steiper, C.~Soligo, R.~D. Martin, Z.~Yang, S.~Tavaré,
  \href{https://dx.doi.org/10.1093/sysbio/syq054}{{Dating Primate Divergences
  through an Integrated Analysis of Palaeontological and Molecular Data}},
  Systematic Biology 60~(1) (2010) 16--31.
\newblock \href {http://dx.doi.org/10.1093/sysbio/syq054}
  {\path{doi:10.1093/sysbio/syq054}}.
\newline\urlprefix\url{https://dx.doi.org/10.1093/sysbio/syq054}

\bibitem{Najm2016}
H.~Najm, K.~Chowdhary,
  \href{https://doi.org/10.1007/978-3-319-11259-6_68-1}{Inference Given Summary
  Statistics}, Springer International Publishing, Cham, 2016, pp. 1--35.
\newblock \href {http://dx.doi.org/10.1007/978-3-319-11259-6_68-1}
  {\path{doi:10.1007/978-3-319-11259-6_68-1}}.
\newline\urlprefix\url{https://doi.org/10.1007/978-3-319-11259-6_68-1}

\bibitem{Prangle12}
P.~Fearnhead, D.~Prangle,
  \href{https://doi.org/10.1111/j.1467-9868.2011.01010.x}{{Constructing summary
  statistics for approximate Bayesian computation: semi-automatic approximate
  Bayesian computation}}, Journal of the Royal Statistical Society Series B
  74~(3) (2012) 419--474.
\newblock \href {http://dx.doi.org/10.1111/j.1467-9868.2011.01010.x}
  {\path{doi:10.1111/j.1467-9868.2011.01010.x}}.
\newline\urlprefix\url{https://doi.org/10.1111/j.1467-9868.2011.01010.x}

\bibitem{Cam}
L.~L. Cam, \href{http://www.jstor.org/stable/2238284}{{Sufficiency and
  Approximate Sufficiency}}, The Annals of Mathematical Statistics 35~(4)
  (1964) 1419--1455.
\newline\urlprefix\url{http://www.jstor.org/stable/2238284}

\bibitem{Nunes}
M.~A. Nunes, D.~J. Balding, {On Optimal Selection of Summary Statistics for
  Approximate Bayesian Computation}, Statistical Applications in Genetics and
  Molecular Biology 9~(1).
\newblock \href {http://dx.doi.org/10.2202/1544-6115.1576}
  {\path{doi:10.2202/1544-6115.1576}}.

\bibitem{Barnes}
C.~P. Barnes, S.~Filippi, M.~P. Stumpf, T.~Thorne,
  \href{http://dx.doi.org/10.1007/s11222-012-9335-7}{Considerate approaches to
  constructing summary statistics for {ABC} model selection}, Statistics and
  Computing 22~(6) (2012) 1181--1197.
\newblock \href {http://dx.doi.org/10.1007/s11222-012-9335-7}
  {\path{doi:10.1007/s11222-012-9335-7}}.
\newline\urlprefix\url{http://dx.doi.org/10.1007/s11222-012-9335-7}

\bibitem{Prangle2014}
D.~Prangle, P.~Fearnhead, M.~P. Cox, P.~J. Biggs, N.~P. French, Semi-automatic
  selection of summary statistics for {ABC} model choice, Statistical
  Applications in Genetics and Molecular Biology 13~(1) (2014) 67--82.
\newblock \href {http://dx.doi.org/10.1515/sagmb-2013-0012}
  {\path{doi:10.1515/sagmb-2013-0012}}.

\bibitem{blum2013}
M.~G.~B. Blum, M.~A. Nunes, D.~Prangle, S.~A. Sisson,
  \href{https://doi.org/10.1214/12-STS406}{{A Comparative Review of Dimension
  Reduction Methods in Approximate Bayesian Computation}}, Statist. Sci. 28~(2)
  (2013) 189--208.
\newblock \href {http://dx.doi.org/10.1214/12-STS406}
  {\path{doi:10.1214/12-STS406}}.
\newline\urlprefix\url{https://doi.org/10.1214/12-STS406}

\bibitem{sisson2018handbook}
S.~A. Sisson, Y.~Fan, M.~Beaumont, {Handbook of approximate Bayesian
  computation}, Chapman and Hall/CRC, 2018.

\bibitem{Beaumont2025}
M.~A. Beaumont, W.~Zhang, D.~J. Balding,
  \href{http://www.genetics.org/content/162/4/2025}{{Approximate Bayesian
  Computation in Population Genetics}}, Genetics 162~(4) (2002) 2025--2035.
\newline\urlprefix\url{http://www.genetics.org/content/162/4/2025}

\bibitem{Ohm.1826}
G.~Ohm, Bestimmung des Gesetzes, nach welchem Metalle die
  Contaktelektricit{\"a}t leiten: nebst einem Entwurfe zur einer Theorie des
  Voltaischen Apparates und des Schweiggerschen Multiplicators, Journal f{\"u}r
  Chemie und Physik, 1826.

\bibitem{Toliyat.2004}
H.~A. Toliyat, G.~B. Kliman, Handbook of electric motors, 2nd Edition, Vol. 120
  of {Electrical and computer engineering}, CRC Press, New York and Basel,
  2004.

\bibitem{Glaser.2016}
P.~Glaser, M.~Schick, K.~Petridis, V.~Heuveline,
  \href{https://www.eccomas2016.org/proceedings/pdf/10011.pdf}{{Comparison
  beween a Polynomial Chaos Surrogate Model and Markov Chain Monte Carlo for
  Inverse Uncertainty Quantification based on an Electric Drive Test Bench}},
  {ECOMAS Congress} 19 (2016) 8809--8826.
\newline\urlprefix\url{https://www.eccomas2016.org/proceedings/pdf/10011.pdf}

\bibitem{Glaser.diss}
P.~Glaser, {Uncertainty Quantiﬁcation for Complex Engineering Systems},
  Dissertation, {Heidelberg University} (2021).

\bibitem{butterworth1930theory}
S.~Butterworth, On the theory of filter amplifiers, Wireless Engineer 7~(6)
  (1930) 536--541.

\bibitem{tuzlukov2018signal}
V.~Tuzlukov, Signal processing noise, CRC Press, 2002.

\bibitem{John.diss}
D.~John, {Uncertainty Quantiﬁcation for an electric motor inverse problem -
  tackling the model discrepancy challenge}, Dissertation, {Heidelberg
  University} (2021).

\bibitem{Sunnaker}
M.~Sunnaker, A.~G. Busetto, E.~Numminen, J.~Corander, M.~Foll, C.~Dessimoz,
  \href{https://doi.org/10.1371/journal.pcbi.1002803}{{Approximate Bayesian
  Computation}}, PLOS Computational Biology 9~(1) (2013) 1--10.
\newblock \href {http://dx.doi.org/10.1371/journal.pcbi.1002803}
  {\path{doi:10.1371/journal.pcbi.1002803}}.
\newline\urlprefix\url{https://doi.org/10.1371/journal.pcbi.1002803}

\bibitem{barber2015}
S.~Barber, J.~Voss, M.~Webster, \href{https://doi.org/10.1214/15-EJS988}{{The
  rate of convergence for approximate Bayesian computation}}, Electron. J.
  Statist. 9~(1) (2015) 80--105.
\newblock \href {http://dx.doi.org/10.1214/15-EJS988}
  {\path{doi:10.1214/15-EJS988}}.
\newline\urlprefix\url{https://doi.org/10.1214/15-EJS988}

\bibitem{Beaumont2009}
M.~A. Beaumont, J.-M. Cornuet, J.-M. Marin, C.~P. Robert,
  \href{http://dx.doi.org/10.1093/biomet/asp052}{{Adaptive approximate Bayesian
  computation}}, Biometrika 96~(4) (2009) 983–--990.
\newblock \href {http://dx.doi.org/10.1093/biomet/asp052}
  {\path{doi:10.1093/biomet/asp052}}.
\newline\urlprefix\url{http://dx.doi.org/10.1093/biomet/asp052}

\bibitem{Marin2012}
J.-M. Marin, P.~Pudlo, C.~P. Robert, R.~J. Ryder,
  \href{https://doi.org/10.1007/s11222-011-9288-2}{{Approximate Bayesian
  computational methods}}, Statistics and Computing 22~(6) (2012) 1167--1180.
\newblock \href {http://dx.doi.org/10.1007/s11222-011-9288-2}
  {\path{doi:10.1007/s11222-011-9288-2}}.
\newline\urlprefix\url{https://doi.org/10.1007/s11222-011-9288-2}

\bibitem{Lintusaari2016}
J.~Lintusaari, M.~U. Gutmann, R.~Dutta, S.~Kaski, J.~Corander,
  \href{https://doi.org/10.1093/sysbio/syw077}{{Fundamentals and Recent
  Developments in Approximate Bayesian Computation}}, Systematic Biology 66~(1)
  (2016) e66--e82.
\newblock \href {http://dx.doi.org/10.1093/sysbio/syw077}
  {\path{doi:10.1093/sysbio/syw077}}.
\newline\urlprefix\url{https://doi.org/10.1093/sysbio/syw077}

\bibitem{wiener1938}
N.~Wiener, \href{http://www.jstor.org/stable/2371268}{The homogeneous chaos},
  American Journal of Mathematics 60~(4) (1938) 897--936.
\newline\urlprefix\url{http://www.jstor.org/stable/2371268}

\bibitem{cameron_martin}
R.~H. Cameron, W.~T. Martin, \href{http://www.jstor.org/stable/1969178}{The
  orthogonal development of non-linear functionals in series of fourier-hermite
  functionals}, Annals of Mathematics 48~(2) (1947) 385--392.
\newline\urlprefix\url{http://www.jstor.org/stable/1969178}

\bibitem{xiu_gpx}
D.~Xiu, G.~E. Karniadakis, \href{https://doi.org/10.1137/S1064827501387826}{The
  wiener--askey polynomial chaos for stochastic differential equations}, SIAM
  Journal on Scientific Computing 24~(2) (2002) 619--644.
\newblock \href
  {http://arxiv.org/abs/https://doi.org/10.1137/S1064827501387826}
  {\path{arXiv:https://doi.org/10.1137/S1064827501387826}}, \href
  {http://dx.doi.org/10.1137/S1064827501387826}
  {\path{doi:10.1137/S1064827501387826}}.
\newline\urlprefix\url{https://doi.org/10.1137/S1064827501387826}

\bibitem{Smolyak}
F.~Nobile, R.~Tempone, C.~G. Webster,
  \href{https://doi.org/10.1137/060663660}{A sparse grid stochastic collocation
  method for partial differential equations with random input data}, SIAM
  Journal on Numerical Analysis 46~(5) (2008) 2309--2345.
\newblock \href {http://arxiv.org/abs/https://doi.org/10.1137/060663660}
  {\path{arXiv:https://doi.org/10.1137/060663660}}, \href
  {http://dx.doi.org/10.1137/060663660} {\path{doi:10.1137/060663660}}.
\newline\urlprefix\url{https://doi.org/10.1137/060663660}

\bibitem{Constantine}
P.~G. Constantine, M.~S. Eldred, E.~T. Phipps,
  \href{http://dx.doi.org/10.1016/j.cma.2012.03.019}{Sparse pseudospectral
  approximation method}, Computer Methods in Applied Mechanics and Engineering
  229-232 (2012) 1–12.
\newblock \href {http://dx.doi.org/10.1016/j.cma.2012.03.019}
  {\path{doi:10.1016/j.cma.2012.03.019}}.
\newline\urlprefix\url{http://dx.doi.org/10.1016/j.cma.2012.03.019}

\bibitem{Dashti2013}
M.~Dashti, K.~J.~H. Law, A.~M. Stuart, J.~Voss,
  \href{https://doi.org/10.1088%2F0266-5611%2F29%2F9%2F095017}{{MAP estimators
  and their consistency in Bayesian nonparametric inverse problems}}, Inverse
  Problems 29~(9) (2013) 095017.
\newblock \href {http://dx.doi.org/10.1088/0266-5611/29/9/095017}
  {\path{doi:10.1088/0266-5611/29/9/095017}}.
\newline\urlprefix\url{https://doi.org/10.1088%2F0266-5611%2F29%2F9%2F095017}

\bibitem{Helin_2015}
T.~Helin, M.~Burger,
  \href{https://doi.org/10.1088%2F0266-5611%2F31%2F8%2F085009}{{Maximum a
  posteriori probability estimates in infinite-dimensional Bayesian inverse
  problems}}, Inverse Problems 31~(8) (2015) 085009.
\newblock \href {http://dx.doi.org/10.1088/0266-5611/31/8/085009}
  {\path{doi:10.1088/0266-5611/31/8/085009}}.
\newline\urlprefix\url{https://doi.org/10.1088%2F0266-5611%2F31%2F8%2F085009}

\bibitem{schillings2019convergence}
C.~Schillings, B.~Sprungk, P.~Wacker,
  \href{https://arxiv.org/abs/1901.03958}{{On the Convergence of the Laplace
  Approximation and Noise-Level-Robustness of Laplace-based Monte Carlo Methods
  for Bayesian Inverse Problems}}, arXiv preprint arXiv:1901.03958.
\newline\urlprefix\url{https://arxiv.org/abs/1901.03958}

\bibitem{refId0}
C.~Schillings, C.~Schwab, \href{https://doi.org/10.1051/m2an/2016005}{{Scaling
  limits in computational Bayesian inversion}}, ESAIM: Mathematical Modelling
  and Numerical Analysis 50~(6) (2016) 1825--1856.
\newblock \href {http://dx.doi.org/10.1051/m2an/2016005}
  {\path{doi:10.1051/m2an/2016005}}.
\newline\urlprefix\url{https://doi.org/10.1051/m2an/2016005}

\bibitem{Rudolf2018}
D.~Rudolf, B.~Sprungk, \href{https://doi.org/10.1007/s10208-016-9340-x}{{On a
  Generalization of the Preconditioned Crank--Nicolson Metropolis Algorithm}},
  Foundations of Computational Mathematics 18~(2) (2018) 309--343.
\newblock \href {http://dx.doi.org/10.1007/s10208-016-9340-x}
  {\path{doi:10.1007/s10208-016-9340-x}}.
\newline\urlprefix\url{https://doi.org/10.1007/s10208-016-9340-x}

\bibitem{Sprungk.2018}
B.~Sprungk, {Numerical Methods for Bayesian Inference in Hilbert Spaces}, 1st
  Edition, {Universit{\"a}tsverlag der TU Chemnitz}, Chemnitz, 2018.

\bibitem{Hu2017adaptive_pCN}
Z.~Hu, Z.~Yao, J.~Li,
  \href{http://www.sciencedirect.com/science/article/pii/S0021999116306106}{{On
  an adaptive preconditioned Crank–Nicolson MCMC algorithm for infinite
  dimensional Bayesian inference}}, Journal of Computational Physics 332 (2017)
  492 -- 503.
\newblock \href {http://dx.doi.org/https://doi.org/10.1016/j.jcp.2016.11.024}
  {\path{doi:https://doi.org/10.1016/j.jcp.2016.11.024}}.
\newline\urlprefix\url{http://www.sciencedirect.com/science/article/pii/S0021999116306106}

\bibitem{Bishop.2006}
C.~M. Bishop, Pattern Recognition and Machine Learning (Information Science and
  Statistics), Springer-Verlag, Berlin, Heidelberg, 2006.

\bibitem{robert2007bayesian}
C.~Robert, {The Bayesian choice: from decision-theoretic foundations to
  computational implementation}, Springer Science \& Business Media, 2007.

\bibitem{sraj2016coordinate}
I.~Sraj, O.~P. Le~Ma{\^\i}tre, O.~M. Knio, I.~Hoteit, {Coordinate
  transformation and Polynomial Chaos for the Bayesian inference of a Gaussian
  process with parametrized prior covariance function}, Computer Methods in
  Applied Mechanics and Engineering 298 (2016) 205--228.

\bibitem{dunlop2017hierarchical}
M.~M. Dunlop, M.~A. Iglesias, A.~M. Stuart, {Hierarchical Bayesian level set
  inversion}, Statistics and Computing 27~(6) (2017) 1555--1584.

\bibitem{roininen2019hyperpriors}
L.~Roininen, M.~Girolami, S.~Lasanen, M.~Markkanen, Hyperpriors for mat{\'e}rn
  fields with applications in bayesian inversion, Inverse Problems \& Imaging
  13~(1) (2019) 1--29.

\bibitem{latz2019fast}
J.~Latz, M.~Eisenberger, E.~Ullmann, {Fast sampling of parameterised Gaussian
  random fields}, Computer Methods in Applied Mechanics and Engineering 348
  (2019) 978--1012.

\bibitem{glaser2017modeling}
P.~Glaser, P.~Kosmas, M.~Schick, V.~Heuveline, {Modeling of a Likelihood
  Function based on a Global Sensitivity Analysis}, PAMM 17~(1) (2017)
  719--720.

\bibitem{MARZOUK2009}
Y.~M. Marzouk, H.~N. Najm,
  \href{http://www.sciencedirect.com/science/article/pii/S0021999108006062}{{Dimensionality
  reduction and polynomial chaos acceleration of Bayesian inference in inverse
  problems}}, Journal of Computational Physics 228~(6) (2009) 1862 -- 1902.
\newblock \href {http://dx.doi.org/https://doi.org/10.1016/j.jcp.2008.11.024}
  {\path{doi:https://doi.org/10.1016/j.jcp.2008.11.024}}.
\newline\urlprefix\url{http://www.sciencedirect.com/science/article/pii/S0021999108006062}

\bibitem{lemaitre2010spectral}
O.~Le~Ma{\^\i}tre, O.~M. Knio, Spectral methods for uncertainty quantification:
  with applications to computational fluid dynamics, Springer Science \&
  Business Media, 2010.

\bibitem{LI2015KL}
J.~Li,
  \href{http://www.sciencedirect.com/science/article/pii/S0167715215002291}{{A
  note on the Karhunen–Lo\`{e}ve expansions for infinite-dimensional Bayesian
  inverse problems}}, Statistics \& Probability Letters 106 (2015) 1 -- 4.
\newblock \href {http://dx.doi.org/https://doi.org/10.1016/j.spl.2015.06.025}
  {\path{doi:https://doi.org/10.1016/j.spl.2015.06.025}}.
\newline\urlprefix\url{http://www.sciencedirect.com/science/article/pii/S0167715215002291}

\bibitem{URIBE2020}
F.~Uribe, I.~Papaioannou, W.~Betz, D.~Straub,
  \href{http://www.sciencedirect.com/science/article/pii/S004578251930516X}{{Bayesian
  inference of random fields represented with the Karhunen–Lo\`{e}ve
  expansion}}, Computer Methods in Applied Mechanics and Engineering 358 (2020)
  112632.
\newblock \href {http://dx.doi.org/https://doi.org/10.1016/j.cma.2019.112632}
  {\path{doi:https://doi.org/10.1016/j.cma.2019.112632}}.
\newline\urlprefix\url{http://www.sciencedirect.com/science/article/pii/S004578251930516X}

\bibitem{turner2014hierarchical}
B.~M. Turner, T.~Van~Zandt, Hierarchical approximate bayesian computation,
  Psychometrika 79~(2) (2014) 185--209.

\bibitem{Elfi}
J.~Lintusaari, H.~Vuollekoski, A.~Kangasrääsiö, K.~Skytén, M.~Järvenpää,
  P.~Marttinen, M.~Gutmann, A.~Vehtari, J.~Corander, S.~Kaski,
  \href{https://arxiv.org/abs/1708.00707}{{ELFI: Engine for Likelihood Free
  Inference}} (2018).
\newline\urlprefix\url{https://arxiv.org/abs/1708.00707}

\bibitem{salvatier2016probabilistic}
J.~Salvatier, T.~V. Wiecki, C.~Fonnesbeck,
  \href{https://peerj.com/articles/cs-55/}{{Probabilistic programming in Python
  using PyMC3}}, PeerJ Computer Science 2 (2016) e55.
\newline\urlprefix\url{https://peerj.com/articles/cs-55/}

\bibitem{john2018emclpp}
D.~John, M.~Schick, V.~Heuveline,
  \href{https://journals.ub.uni-heidelberg.de/index.php/emcl-pp/article/view/51320}{{Learning
  model discrepancy of an electric motor with Bayesian inference}}, Preprint
  Series of the Engineering Mathematics and Computing Lab~(01).
\newblock \href {http://dx.doi.org/http://doi.org/10.11588/emclpp.2018.1.51320}
  {\path{doi:http://doi.org/10.11588/emclpp.2018.1.51320}}.
\newline\urlprefix\url{https://journals.ub.uni-heidelberg.de/index.php/emcl-pp/article/view/51320}

\bibitem{jung2011choice}
H.~Jung, P.~Marjoram, {Choice of summary statistic weights in approximate
  Bayesian computation}, Statistical applications in genetics and molecular
  biology 10~(1).

\bibitem{sobol2001global}
I.~M. Sobol,
  \href{http://citeseerx.ist.psu.edu/viewdoc/download?doi=10.1.1.466.9144&rep=rep1&type=pdf}{{Global
  sensitivity indices for nonlinear mathematical models and their Monte Carlo
  estimates}}, Mathematics and computers in simulation 55~(1-3) (2001)
  271--280.
\newline\urlprefix\url{http://citeseerx.ist.psu.edu/viewdoc/download?doi=10.1.1.466.9144&rep=rep1&type=pdf}

\bibitem{saltelli2008global}
A.~Saltelli, M.~Ratto, T.~Andres, F.~Campolongo, J.~Cariboni, D.~Gatelli,
  M.~Saisana, S.~Tarantola, Global sensitivity analysis: the primer, John Wiley
  \& Sons, 2008.

\end{thebibliography}

\end{document}